\documentclass[paperpaper,twocolumn,english,superscriptaddress,aps,prb,floatfix,amsfonts,amssymb]{revtex4}
\usepackage{float}
\usepackage{epsfig}
\usepackage{graphicx}
\usepackage{dcolumn}
\usepackage{bm}
\usepackage{appendix}
\usepackage{subfigure}
\usepackage{color}
\usepackage{natbib}

\begin{document}


\title{Effects of finite-range interactions on the one-electron spectral properties of 
one-dimensional metals: Application to Bi/InSb(001)}
\author{Jos\'e M. P. Carmelo}
\affiliation{Boston University, Department of Physics, 590 Commonwealth Ave, Boston, Massachusetts 02215, USA}
\affiliation{Massachusetts Institute of Technology, Department of Physics, Cambridge, Massachusetts 02139, USA}
\affiliation{Center of Physics of University of Minho and University of Porto, P-4169-007 Oporto, Portugal}
\affiliation{Department of Physics, University of Minho, Campus Gualtar, P-4710-057 Braga, Portugal}
\author{Tilen \v{C}ade\v{z}}
\affiliation{Beijing Computational Science Research Center, Beijing 100193, China}
\affiliation{Center of Physics of University of Minho and University of Porto, P-4169-007 Oporto, Portugal}
\author{Yoshiyuki Ohtsubo}
\affiliation{Graduate School of Frontier Biosciences, Osaka University, Suita 565-0871, Japan}
\affiliation{Department of Physics, Graduate School of Science, Osaka University, Toyonaka 560-0043, Japan}
\author{Shin-ichi Kimura}
\affiliation{Graduate School of Frontier Biosciences, Osaka University, Suita 565-0871, Japan}
\affiliation{Department of Physics, Graduate School of Science, Osaka University, Toyonaka 560-0043, Japan}
\author{David K. Campbell}
\affiliation{Boston University, Department of Physics, 590 Commonwealth Ave, Boston, MA 02215, USA}

\date{1 April 2019}


\begin{abstract}
We study the one-electron spectral properties of one-dimensional interacting electron systems in which the interactions have finite range. 
We employ a mobile quantum impurity scheme that describes the interactions of the fractionalized excitations at energies above the 
standard Tomonga-Luttinger liquid limit and show that the phase shifts induced by the impurity describe universal properties of the 
one-particle spectral function. We find the explicit forms in terms of these phase shifts for the momentum dependent exponents that 
control the behavior of the spectral function near and at the $(k,\omega)$-plane singularities where most of the spectral weight is located. 
The universality arises because the line shape near the singularities is independent of the short-distance part of the interaction potentials. 
For the class of potentials considered here, the charge fractionalized particles have screened Coulomb interactions that decay with a 
power-law exponent $l>5$. We apply the theory to the angle-resolved photo-electron spectroscopy (ARPES) in the highly one-dimensional
bismuth-induced anisotropic structure on indium antimonide Bi/InSb(001). Our theoretical predictions agree quantitatively with both (i) the 
experimental value found in Bi/InSb(001) for the exponent $\alpha$ that controls the suppression of the density of states at very small 
excitation energy $\omega$ and (ii) the location in the $(k,\omega)$ plane of the experimentally observed high-energy peaks in the 
ARPES momentum and energy distributions. We conclude with a discussion of experimental properties beyond the range of our present 
theoretical framework and further open questions regarding the one-electron spectral properties of Bi/InSb(001). 
\end{abstract}


\maketitle

\section{Introduction}
\label{INT}

One-dimensional (1D) interacting systems are characterized by a breakdown of the basic Fermi liquid quasiparticle picture. 
Indeed, no quasiparticles with the same quantum numbers as the electrons exist when the motion is restricted to a single 
spatial dimension. Rather, in a 1D lattice, correlated electrons split into basic fractionalized charge-only and spin-only 
particles \cite{Blumenstein_11,Voit_95}. Hence the removal or addition of electrons generates an energy continuum of 
excitations described by these exotic fractionalized particles which are not adiabatically connected to free electrons. 
Hence they must be described using a different language. 

These models share common low-energy properties associated with the universal class of the Tomonaga-Luttinger liquid (TLL) \cite{Blumenstein_11,Voit_95,Tomonaga_50,Luttinger_63}. To access their high-energy dynamical correlation functions 
beyond the low-energy TLL limit, approaches such as the pseudofermion dynamical theory (PDT) \cite{Carmelo_05}
or the mobile quantum impurity model (MQIM) \cite{Imambekov_09,Imambekov_12} must be used. Those approaches 
incorporate nonlinearities in the dispersion relations of the fractionalized particles. 

An important low-energy TLL property of 1D correlated electronic metallic systems is the universal power-law scaling of the 
spectral intensity $I (\omega,T)$ such that $I (0,T) \propto T^{\alpha}$ and $I (\omega,0)\propto\vert\omega\vert^{\alpha}$.
Here the exponent $\alpha$ controls the suppression of the density of states (SDS) and $\omega$ is a small excitation 
energy near the ground-state level. The value SDS exponent $\alpha = (1-K_c)^2/(4K_c)$ is determined by that of the 
TLL charge parameter $K_c$ \cite{Blumenstein_11,Voit_95,Schulz_90}. Importantly, this exponent provides useful 
information about the range of the underlying electron interactions.

In the case of integrable 1D models solvable by the Bethe ansatz \cite{Bethe_31} (such as the 1D Hubbard model 
\cite{Lieb_68,Martins_98}), the PDT and MQIM describe the same mechanisms and lead to the same expressions for 
the dynamical correlation functions \cite{Carmelo_18}. The advantage of the MQIM is that it also applies to non-integrable 
systems \cite{Imambekov_12}. The exponents characterizing the singularities in these systems differ significantly from 
the predictions of the linear TLL theory, except in the low-energy limit where the latter is valid.

For integrable 1D lattice electronic models with only onsite repulsion (such as the Hubbard model), the TLL charge 
parameter $K_c $ is larger than $1/2$ and thus the SDS exponent $\alpha = (1-K_c)^2/(4K_c)$ is smaller than $1/8$. 
In non-integrable systems a SDS exponent $\alpha$ larger than $1/8$ stems from finite-range interactions \cite{Schulz_90}.

In fact, as shown in Table \ref{table1}, for the metallic states of both 1D and 
quasi-1D electronic systems, the SDS exponent $\alpha$ frequently has  experimental values in the range $0.5-0.8$
\cite{Blumenstein_11,Voit_95,Schulz_90,Claessen_02,Kim_06,Schoenhammer_93,Ma_17,Ohtsubo_15,Ohtsubo_17}. 
In actual materials, a finite  effective range interaction \cite{Bethe_49,Blatt_49,Joachain_75,Preston_75,Flambaum_99}
generally results from screened long-range Coulomb interactions with potentials 
vanishing as an inverse power of the separation with an exponent larger than one. 
In general, such finite-range interactions in 1D lattice systems represent a complex 
and unsolved quantum problem involving non-perturbative  microscopic electronic processes. 
Indeed, as originally formulated, the MQIM does not apply to lattice electronic systems with finite-range 
interactions whose screened Coulomb potentials vanish as an inverse power of the electron distance.

Recently, the MQIM has been extended to a class of electronic systems with effective interaction ranges of 
about one lattice spacing, compatible with the high-energy one-electron spectral properties observed in twin grain 
boundaries of molybdenum diselenide MoSe$_2$ \cite{Ma_17,Cadez_19}. This has been achieved by suitable 
renormalization of the phase shifts of the charge fractionalized particles. That theoretical scheme, called here 
``MQIM-LO'', accounts for the effects of only the {\it leading order} (LO) in the effective range expansion 
\cite{Bethe_49,Blatt_49} of such phase shifts.

In this paper we consider a bismuth-induced anisotropic structure on indium antimonide which we henceforth call 
Bi/InSb(001) \cite{Ohtsubo_15}. Experimentally, strong evidence has been found that Bi/InSb(001) exhibits 1D 
physics \cite{Ohtsubo_15,Ohtsubo_17}. However, a detailed understanding of the exotic one-electron spectral 
properties revealed by its angle resolved photo-emission spectroscopy (ARPES) \cite{Ohtsubo_15,Ohtsubo_17} 
at energy scales beyond the TLL has remained elusive. In particular, the predictions of the MQIM-LO for the 
location in the $(k,\omega)$ plane of the experimentally observed high-energy peaks in the ARPES momentum 
distribution curves (MDC) and energy distribution curves (EDC) of Bi/InSb(001) do not lead to the same quantitative 
agreement as for the ARPES in the MoSe$_2$ line defects \cite{Ma_17,Cadez_19}. This raises the important 
question of what additional effects must be included to obtain agreement with the experimental data.

In this paper, we answer this question by extending  the MQIM-LO to a larger class of 1D lattice electronic systems 
with finite-range interactions by accounting for higher-order terms in the effective range expansion  
\cite{Bethe_49,Blatt_49,Landau_65,Kermode_90,Burke_11} of the phase shifts of the fractionalized charged 
particles. While the corresponding {\it higher order} ``MQIM-HO'' corresponds in general to a complicated, 
non-perturbative many-electron problem, we find, unexpectedly, that the interactions of the fractionalized 
charged particles with the charge mobile quantum impurity occur in the unitary limit of (minus) infinite scattering length \cite{Newton_82,Zwerger_12,Horikoshi_17}. In that limit, the separation between the interacting charged 
particles (the inverse density) is much greater than the range of the interactions, and the calculations simplify 
considerably.

The unitary limit plays an important role in the  physics of many physical systems, including the dilute neutron 
matter in shells of neutron stars \cite{Dean_03} and in atomic scattering in systems of trapped cold atoms 
\cite{Zwerger_12,Horikoshi_17}. Our discovery of its relevance in a condensed matter system is new and reveals 
new physics.
\begin{widetext}
\begin{table}
\begin{center}
\begin{tabular}{|c||c|c|c|c|} 
\hline
System & Parameter $K_c$ & SDS exponent $\alpha$ & Technique & Source \\
\hline
(TMTSF)$_2$XX=PF$_6$,AsF$_6$,ClO$_4$ & $0.23$ & $0.64$ & Optical conductivity & from SI of Ref. \onlinecite{Blumenstein_11} \\
\hline
Carbon Nanotubes & $0.28$ & $0.46$ & Photoemission  & from SI of Ref. \onlinecite{Blumenstein_11} \\
\hline
Purple Bronze Li$_{0.9}$Mo$_6$O$_{17}$ & $0.24$ & $0.60$ & ARPES and tunneling spectroscopy & from SI of Ref. \onlinecite{Blumenstein_11} \\
\hline
1D Gated Semiconductors & $0.26-0.28$ & $0.46-0.53\approx 0.5$ & Transport conductivity & from SI of Ref. \onlinecite{Blumenstein_11} \\
\hline
MoSe$_2$ 1D line defects & $0.20-0.22$ & $0.70-0.80$ & ARPES & from Ref. \onlinecite{Ma_17} \\
\hline
Bi/InSb(001) & $0.22-0.24$  & $0.60-0.70$ & ARPES & from Ref. \onlinecite{Ohtsubo_15} \\
\hline
\end{tabular}
\caption{Experimental TLL charge parameter $K_c={\tilde{\xi}}_c^2/2$ and
related SDS exponent $\alpha = (1-K_c)^2/(4K_c)$ (SI stands for Supplementary Information.)}
\label{table1}
\end{center}
\end{table} 
\end{widetext}

The results of the MQIM-HO are consistent with the expectation that the microscopic mechanisms behind the
one-electron spectral properties of Bi/InSb(001) include finite-range interactions. Indeed, accounting for the effective 
range of the corresponding interactions \cite{Joachain_75,Preston_75,Flambaum_99} leads to theoretical 
predictions that quantitatively agree with both (i) the experimental value of the SDS exponent ($\alpha\in [0.6-0.7]$) in 
Bi/InSb(001) observed in $I (\omega,0)\propto\vert\omega\vert^{\alpha}$ and (ii) the location in the $(k,\omega)$ plane 
of the experimentally observed high-energy peaks in the ARPES MDC and EDC. 

Since Bi/InSb(001) is a complex system and the MQIM-HO predictions are limited to the properties (i) and (ii), 
in the discussion section of this paper we consider other possible effects beyond the present theoretical framework 
that might contribute to the microscopic mechanisms determining spectral properties of Bi/InSb(001).

In this paper we employ units of $\hbar =1$ and $k_B =1$. In Sec. \ref{model} we introduce the theoretical scheme 
used in our studies. The effective-range expansion of the phase shift associated with the interactions of the charge 
fractionalized particles and charge hole mobile impurity, the corresponding unitary limit, and the scattering lengths 
are all issues we address in Sec. \ref{EREUL}. In Sec. \ref{ER} the effective range expression is derived and 
expressed in terms of the ratio of the renormalized and bare scattering lengths. In Sec. \ref{ARPES} we show how 
our approach predicts the location in the $(k,\omega)$ plane of the experimentally observed high-energy Bi/InSb(001) 
ARPES MDC and EDC peaks. In Sec. \ref{DISCONCL}, we discuss our results and experimental properties outside 
the present theoretical framework, mention open questions on the Bi/InSb(001) spectral properties, and offer concluding 
remarks.  

\section{The model}
\label{model}

The 1D model Hamiltonian associated with the MQIM-HO for electronic density $n_e\in ]0,1[$ is given by,
\begin{eqnarray}
{\hat{H}} & = & t\,\hat{T} + \hat{V}\hspace{0.20cm}{\rm where}
\nonumber \\
\hat{T} & = & - \sum_{\sigma=\uparrow,\downarrow }\sum_{j=1}^{L}\left(c_{j,\sigma}^{\dag}\,
c_{j+1,\sigma} + c_{j+1,\sigma}^{\dag}\,c_{j,\sigma}\right)
\nonumber \\
\hat{V} & = & \sum_{r=0}^{L/2-1}V_e (r)
\sum_{\sigma=\uparrow,\downarrow}\sum_{\sigma'=\uparrow,\downarrow}\sum_{j=1}^{L}\hat{\rho}_{j,\sigma}\hat{\rho}_{j+r,\sigma'} \, .
\label{equ1}
\end{eqnarray}
Here $\hat{\rho}_{j,\sigma} = \left(c_{j,\sigma}^{\dag}\,c_{j,\sigma} - {1\over 2}\right)$,
$V_e (0) = U/2$, $V_e (r) = U\,F_e (r)/r$ for $r>0$, and $F_e (r)$ is a continuous 
screening function such that $F_e (r)\leq 1/4$, which at large $r$ vanishes as some inverse 
power of $r$ whose exponent is larger than one, so that $\lim_{r\rightarrow\infty}F_e (r)=0$. 

We use a representation of the fractionalized $c$ (charge) and $s$ (spin) particles 
that also naturally emerges in the MQIM-LO \cite{Ma_17}. For simplicity, in general in this
paper they are called $c$ particles and $s$ particles, respectively. They occupy a $c$ band and
an $s$ band whose momentum values $q_j$ and $q_j'$, respectively, are such that
$q_{j+1}-q_j = 2\pi/L$ and $q_{j+1}'-q_j' = 2\pi/L$. In the thermodynamic limit one often
uses a continuum representation in terms of corresponding $c$ band momentum
variables $q$ and $s$ band momentum variables $q'$ with ground-state occupancies
$q \in [-2k_F,2k_F]$ and $q' \in [-k_F,k_F]$, respectively, where $2k_F=\pi n_e$.
The energy dispersions for $c$ and $s$ particles, ${\tilde{\varepsilon}}_c (q)$ and ${\tilde{\varepsilon}}_s (q')$,
respectively, are defined for these momentum intervals
in Eqs. (\ref{equA2}) and (\ref{equA4})-(\ref{equA10}) of Appendix \ref{APA}.

\begin{figure}
\begin{center}
\centerline{\includegraphics[width=8.0cm]{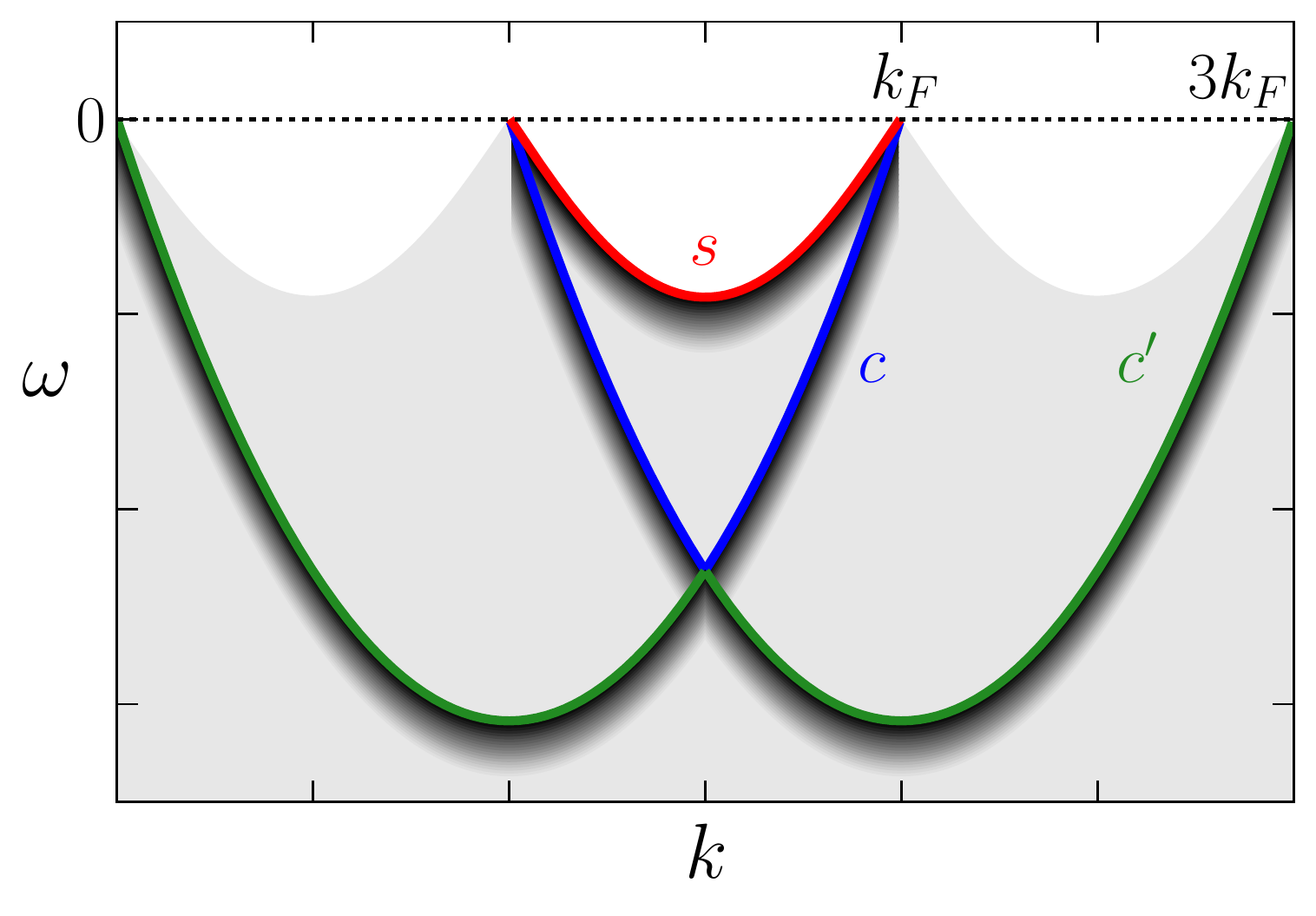}}
\caption{Sketch of the $s$ (spin) and $c$ and $c'$ (charge) branch lines in the one-electron removal spectral function of the lattice 
electronic correlated models discussed in this paper. The soft grey region refers to the small spectral-weight distribution 
continuum whereas the darker grey regions below the branch lines typically display more weight. In the actual spectral function, 
Eq. (\ref{equ2}), this applies to $k$ subdomains for which the exponents that control the line shape near those lines are negative.
The lack of spectral weight in some of the figure $(k,\omega)$-plane regions is imposed by kinematical constraints.}
\label{figure1}
\end{center}
\end{figure}

Most of the weight of the one-electron spectral function is generated by transitions to excited states 
involving creation of one hole in the $c$ band, one hole in the $s$ band, plus 
low-energy particle-hole processes in such bands. Processes where
both holes are created away from the $c$ band and $s$ band Fermi points $\pm 2k_F$ and $\pm k_F$, respectively,
contribute to the spectral-function continuum. Processes where the $c$ band hole is created at momentum values 
spanning its band interval $q\in ]-2k_F,2k_F[$ and the $s$ hole (spinon)
is created near one of its Fermi points $\pm k_F$ contribute to the $c$ and $c'$ branch lines whose spectra 
run from $k\in ]-k_F,k_F[$ and $k \in ]-3k_F,3k_F[$, respectively. Since in such processes
the $c$ band hole is created away from the $c$ band Fermi points, we call it 
a {\it $c$ (charge) hole mobile impurity}. Finally, processes where the $s$ band hole is created at momentum 
values in the interval $q'\in ]-k_F,k_F[$ and the $c$ hole (holon)
is created near one of its Fermi points $\pm 2k_F$ contribute to the $s$ branch line whose spectrum 
runs from $k\in ]-k_F,k_F[$. In the case of these processes it is the $s$ band hole that is created away from the 
corresponding $s$ band Fermi points. Hence we call it {\it $s$ (spin) hole mobile impurity}. 
See a sketch of such spectra in Fig. \ref{figure1}. In the remainder of this paper the charge (and spin) hole 
mobile impurity is merely called $c$ (and $s$) impurity. 

The one-electron operators matrix elements between energy eigenstates
in the expressions for the spectral function involve phase shifts and the charge parameter ${\tilde{\xi}}_c=\sqrt{2K_c}$ 
whose value is determined by them. Its range for the present lattice systems is ${\tilde{\xi}}_c=\sqrt{2K_c}\in ]1/2,\xi_c]$, 
where the bare parameter $\xi_c \in ]1,\sqrt{2}[$ defined by Eq. (\ref{equA16}) of Appendix \ref{APA} refers to the
1D Hubbard model. Note that the model in Eq. (\ref{equ1}), 
becomes the 1D Hubbard model at the bare charge parameter value, ${\tilde{\xi}}_c =\xi_c$. In this limit, 
the SDS exponent reads $\alpha_0 = (2 - \xi_c^2)^2/(8\xi_c^2)\in [0,1/8]$ with
$\alpha_0 =0$ and $\alpha_0 =1/8$ for $u\rightarrow 0$ and $u\rightarrow\infty$, respectively.
For $n_e\in ]0,1[$ there is a $\xi_c\rightarrow {\tilde{\xi}}_c$ transformation  
\cite{Ma_17} for each fixed value of $\xi_c$ and ${\tilde{\xi}}_c<\xi_c$ such that
$\xi_c \in ]1,\sqrt{2}[$ and ${\tilde{\xi}}_c\in ]1/2,1[\,;]1,\xi_c]$. This maps the 
1D Hubbard model onto the model, Eq. (\ref{equ1}), upon gently turning on $F_e (r)$. Consistent with this result,
$\lim_{{\tilde{\xi}}_c\rightarrow\xi_c}F_e (r)\rightarrow 0$ for $r\in [0,\infty]$.
For ${\tilde{\xi}}_c<\xi_c$ the corresponding SDS exponent intervals are 
$\alpha = (2 - {\tilde{\xi}}_c^2)^2/(8{\tilde{\xi}}_c^2)\in [\alpha_0,1/8[\,;]1/8,49/32[$.
<
The phase shifts in the one-electron matrix elements play a major role in our study 
by appearing  explicitly in the expressions of the momentum-dependent exponents of the 
one-electron removal spectral function. These phases shifts are $2\pi{\tilde{\Phi}}_{c,s}(\pm 2k_F,q')$ and 
$2\pi{\tilde{\Phi}}_{c,c}(\pm 2k_F,q)$. Specifically, $-2\pi{\tilde{\Phi}}_{c,s}(\pm 2k_F,q')$ and 
$-2\pi{\tilde{\Phi}}_{c,c}(\pm 2k_F,q)$ are the phase shifts, respectively, imposed on a $c$ particle of $c$ band 
momentum $\pm 2k_F$ by a $s$ and $c$ impurity created at momentum $q' \in  [-k_F,k_F]$ and 
$q \in  [-2k_F,2k_F]$. (Their explicit expressions are given below.) 
The charge parameter ${\tilde{\xi}}_c$ is given by a superposition of charge-charge phase shifts,
\begin{eqnarray}
{\tilde{\xi}}_c = 1 + \lim_{q\rightarrow 2k_F}\{{\tilde{\Phi}}_{c,c}(+2k_F,q)+{\tilde{\Phi}}_{c,c}(-2k_F,q)\} \, .
\nonumber
\end{eqnarray}

The expressions for the exponents of spectral functions also involve the phase shifts 
$2\pi{\tilde{\Phi}}_{s,c}(\pm k_F,q) = \mp {\pi\over\sqrt{2}}$ and
$2\pi{\tilde{\Phi}}_{s,s} (\pm k_F,q') = \pm (\sqrt{2}-1)(\sqrt{2} + (-1)^{\delta_{q',\pm k_F}}){\pi\over\sqrt{2}}$
induced on a $s$ particle of $s$ band momentum $\pm k_F$ by a 
$c$ and $s$ impurity created at momentum $q \in  [-2k_F,2k_F]$ and $q' \in  [-k_F,k_F]$, 
respectively. Their simple expressions are invariant under the 
$\xi_c\rightarrow {\tilde{\xi}}_c$ transformation and, due to the spin $SU(2)$ symmetry, are interaction, density, and 
momentum independent. (Except for $(-1)^{\delta_{q',\pm k_F}}$ in the
$2\pi{\tilde{\Phi}}_{s,s} (\pm k_F,q')$ expression at $q'=\pm k_F$.)

For small energy deviations $({\tilde{\omega}}_{\beta} (k)-\omega)>0$ and $({\tilde{\omega}}_{s} (k)-\omega)>0$  
near the $\beta =c,c'$ branch lines and $s$ branch line, the spectral function behaves as,
\begin{eqnarray}
{\tilde{B}} (k,\omega) & \approx & \sum_{\iota=\pm 1}C_{\beta,\iota}
{\rm Im}\left\{\left({(\iota)\over{\tilde{\omega}}_{\beta} (k)-\omega - {i\over 2\tau_{\beta} (k)}}\right)^{-{\tilde{\zeta}}_{\beta} (k)}\right\} 
\nonumber \\
{\tilde{B}} (k,\omega) & = & C_{s} ({\tilde{\omega}}_{s} (k)-\omega)^{{\tilde{\zeta}}_{s} (k)} \, ,
\label{equ2}
\end{eqnarray}
respectively. Here $C_{\beta,\iota}$ and $C_{s}$ are $n_e$, $u=U/4t$, and ${\tilde{\xi}}_c$ dependent constants
for energy and momentum values corresponding to the small energy deviations $({\tilde{\omega}}_{\beta} (k)-\omega)>0$ 
and $({\tilde{\omega}}_{s} (k)-\omega)>0$, respectively, and $\omega<0$ are high energies beyond those of the TLL. 

The upper bounds of the constants $C_{c,\iota}$, $C_{c',\iota}$, and $C_{s}$ in Eq. (\ref{equ2}) 
are known from matrix elements and sum rules for spectral weights, but their precise values remain in general
an unsolved problem. The expressions for the $\gamma =c,c',s$ spectra 
${\tilde{\omega}}_{\gamma} (k)$ and exponents ${\tilde{\zeta}}_{\gamma} (k)$ are given 
in Eqs. (\ref{equA1}) and (\ref{equA3}) of Appendix \ref{APA}, respectively. As
discussed in Appendix \ref{RTLL}, the MQIM-HO also applies to the 
low-energy TLL limit in which such exponents have different expressions. For the present high-energy regime,
they have the same expressions as for the MQIM-LO except that 
the phase shift $2\pi{\tilde{\Phi}}_{c,c}(\pm 2k_F,q)$ in that of the spectral function
exponents ${\tilde{\zeta}}_{c} (k)$ and ${\tilde{\zeta}}_{c'} (k)$ has MQIM-HO additional terms.

That the $s$ branch line coincides with the edge of the support for the spectral function 
ensures that near it the line shape is power-law like, as given in Eq. (\ref{equ2}). 
For the $c,c'$ branch likes, which run within the spectral weight continuum, 
the $\beta =c,c'$ lifetime $\tau_{\beta} (k)$ in Eq. (\ref{equ2}) is very large for the interval ${\tilde{\xi}}_c \in ]1,\xi_c[$,
so that the expression given in that equation is {\it nearly} 
power-law like, ${\tilde{B}} (k,\omega) \propto \left({\tilde{\omega}}_{\beta} (k)-\omega\right)^{{\tilde{\zeta}}_{\beta} (k)}$.
The finite-range interaction effects increase upon decreasing ${\tilde{\xi}}_c$ in the
interval ${\tilde{\xi}}_c\in [{\tilde{\xi}}_c^{\oslash},1[$ where ${\tilde{\xi}}_c^{\oslash} = 1/\xi_c$.
In it the corresponding $c$ impurity relaxation 
processes associated with large lifetimes $\tau_{c} (k)$ and $\tau_{c'} (k)$ in Eq. (\ref{equ2}) 
for the $k$ intervals for which ${\tilde{\zeta}}_{c} (k)<0$ and ${\tilde{\zeta}}_{c'} (k)<0$, respectively,
start transforming the power-law singularities into broadened peaks with small widths. Such effects become more pronounced 
upon further decreasing ${\tilde{\xi}}_c$ in the interval ${\tilde{\xi}}_c\in ]{\tilde{\xi}}_c^{\oslash},1]$. 
As discussed in more detail below in Sec. \ref{Relaxation}, for $k$ ranges for which the 
exponents ${\tilde{\zeta}}_{c} (k)$ and ${\tilde{\zeta}}_{c'} (k)$ become positive upon 
decreasing ${\tilde{\xi}}_c$, the relaxation processes wash out the peaks entirely.

\section{The effective-range expansion and the unitary limit}
\label{EREUL}

\subsection{The effective-range expansion}
\label{ERAE}

As we shall establish in detail below, the finite-range electron interactions have their strongest effects in 
the charge-charge interaction channel. In contrast, for the charge-spin channel, 
the renormalization factor of the phase shift,
\begin{equation}
2\pi{\tilde{\Phi}}_{c,s} (\pm 2k_F,q')={{\tilde{\xi}}_c\over\xi_c}\,2\pi\Phi_{c,s} (\pm 2k_F,q') \, ,
\label{equ3}
\end{equation}
remains that of the MQIM-LO. 

For small relative momentum $k_r = q \mp 2k_F$ of the $c$ impurity 
of momentum $q$ and $c$ particle of momentum $\pm 2k_F$ the phase shift 
${\tilde{\Phi}}_c (k_r) = -2\pi{\tilde{\Phi}}_{c,c} (\pm 2k_F,\pm 2k_F + k_r)$
associated with the charge-charge channel obeys an effective range expansion,
\begin{eqnarray}
\cot ({\tilde{\Phi}}_c (k_r)) = {-1\over {\tilde{a}}\,k_r} + {1\over 2}\,R_{\rm eff}\,k_r 
- P_{\rm eff}\,R_{\rm eff}^3\,k_r^3 + {\cal{O}} (k_r^5) \, .
\label{equ4}
\end{eqnarray}
This equation is the same as for three-dimensional (3D) s-wave scattering problems if $k_r$
is replaced by $\vert k_r\vert$ \cite{Bethe_49,Blatt_49}. The first and second terms involve the 
scattering length ${\tilde{a}}$ and effective range $R_{\rm eff}$, respectively. The third and higher 
terms are negligible and involve the shape parameters
\cite{Bethe_49,Blatt_49,Burke_11,Kermode_90,Landau_65}. 

One finds that in the bare charge parameter limit, ${\tilde{\xi}}_c = \xi_c$, the effective range expansion reads
$\cot (\Phi_c (k_r)) = -1/(a\,k_r)$ where $\Phi_c (k_r) = -2\pi\Phi_{c,c} (\pm 2k_F,\pm 2k_F + k_r)$, $2\pi\Phi_{c,c} (\pm 2k_F,q)$ is the bare
phase shift defined in Eqs. (\ref{equA11})-(\ref{equA15}) of Appendix \ref{APA}, and
$a=\lim_{{\tilde{\xi}}_c\rightarrow\xi_c}{\tilde{a}}$ is the bare scattering length.

Due to the 1D charge-spin separation at all MQIM energy scales,
the repulsive electronic potential $V_e (r)$ gives rise to an attractive potential $V_c (x)$ associated with the 
interaction of the $c$ particle and $c$ impurity at a distance $x$. To go beyond the MQIM-LO, we 
must explicitly account for the general properties of $V_c (x)$ whose form is determined by that of $V_e (r)$. 
The corresponding relation between the electron and $c$ particle representations is discussed see Appendix \ref{RECP}. The 
attractive potential $V_c (x)$ is negative for $x>x_0$ where $x_0$ is a non-universal distance
that either vanishes or is much smaller than 
the lattice spacing $a_0$. Moreover, for the present class of systems $V_c (x)$ vanishes for {\it large $x$} as,
\begin{eqnarray}
V_c^{\rm asy} (x) & = & - {\gamma_c\over x^l} = - {C_c\over (x/2r_l)^l}\hspace{0.2cm}{\rm where}
\nonumber \\
C_c & = & {1\over (2r_l)^2\mu}\hspace{0.2cm}{\rm and}\hspace{0.2cm}\gamma_c = {(2r_l)^{l-2}\over \mu} \, .
\label{equ5}
\end{eqnarray}
Here $\mu$ is a non-universal reduced mass,
$l$ is an integer determined by the large-$r$ behavior of $V_e (r)$,
and $2r_l$ is a length scale whose $l$ dependence for ${\tilde{\xi}}_c <1$ is given below in Sec. \ref{ER}. 
(And is twice the van der Waals length at $l=6$). 

Since $V_c (x)$ has asymptotic behavior $1/x^l$, the scattering length, effective range, and shape parameter terms 
in Eq. (\ref{equ4}) only converge if $l > 5$, $l > 7$, and $l > 9$, respectively \cite{Burke_11}. 
We shall find that agreement with the experimental results is achieved provided that
the effective range studied in Sec. \ref{ER} is finite and this requires that $l > 5$ in Eq. (\ref{equ5}).

Similarly to the potentials considered in Refs. \onlinecite{Flambaum_99} and \onlinecite{Gribakin_93},
the class of potentials with large-distance behavior, Eq. (\ref{equ5}) and whose depth
is larger that the scattering energy of the corresponding interactions considered here are such that the 
positive ``momentum'' $\sqrt{2\mu (-V_c (x))}$ obeys a sum rule of general form,
\begin{eqnarray}
& & \int_{x_0}^{\infty}dx\sqrt{2\mu (-V_c (x))} = \Phi + {\theta_c\pi\over 2(l-2)}\hspace{0.20cm}{\rm where}
\nonumber \\
& & \tan (\Phi) = - {\Delta a\over{\tilde{a}}}\cot \left({\pi\over l-2}\right)\hspace{0.20cm}{\rm and}\hspace{0.20cm}{\rm thus}
\nonumber \\
&& {a\over {\tilde{a}}} = 1 - \tan\left({\pi\over l-2}\right)\tan(\Phi) \, .
\label{equ6}
\end{eqnarray}
Here $\Delta a/{\tilde{a}}$ where $\Delta a = a - {\tilde{a}}$ 
is a relative fluctuation that involves two uniquely defined yet non-universal scattering lengths, 
$a$ and ${\tilde{a}}$. As justified in Sec. \ref{ER}, in the present unitary-limit case discussed in
Sec. \ref{UL}, they are the bare and renormalized scattering lengths, respectively,
defined in that section. The physically important renormalized charge parameter range is
${\tilde{\xi}}_c\in ]1/2,1[$ for which $\alpha >1/8$. The term $\theta_c\pi/[2(l-2)]$ in Eq. (\ref{equ6}) refers to a 
potential boundary condition\cite{Flambaum_99,Gribakin_93} with $\theta_c = 1$ for that interval.
(In that regime, the expressions in Eq. (\ref{equ6}) are similar to those in
Eqs. (4) and (6) of Ref. \onlinecite{Flambaum_99} with $a$, ${\tilde{a}}$, $l$, and $\Phi$ corresponding
to $a$, ${\bar{a}}$, $n$, and $\Phi - \pi/[2(n-2)]$, respectively.)
A function $\theta_c =\sqrt{(\xi_c^4 - {\tilde{\xi}}_c^4)/(\xi_c^4 - 1)}$ 
for the interval ${\tilde{\xi}}_c\in ]1,\xi_c[$ for which $\alpha <1/8$ merely ensures that the sum rule
in Eq. (\ref{equ6}) continuously vanishes as ${\tilde{\xi}}_c\rightarrow\xi_c$. 

Our choice of potentials with large-$x$ behavior given in Eq. (\ref{equ5}) is such that the 
sum rule, Eq. (\ref{equ6}), is obeyed yet for small $x$ the form of $V_c (x)$ is not 
universal and is determined by the specific small-$r$ form of $V_e (r)$ itself. The 
zero-energy phase $\Phi$ in Eq. (\ref{equ6}) whose physics is further clarified below 
can be expressed as,
\begin{equation}
\Phi = \int_{x_0}^{x_2}dx\sqrt{2\mu (-V_c (x))} 
\hspace{0.20cm}{\rm where}\hspace{0.20cm} x_2 = 2r_l\left({4\sqrt{2}\over\pi\theta_c}\right)^{2\over l-2} \, .
\label{equ7}
\end{equation} 
Indeed, $V_c (x)= V_c^{\rm asy} (x)$ for $x>x_2$ and 
$\int_{x_2}^{\infty}dx\sqrt{2\mu (-V_c^{\rm asy} (x))} = {\theta_c\pi\over 2(l-2)}$.
Here $x_2 \approx 2r_l$ for ${\tilde{\xi}}_c\in ]1/2,1[$
with the ratio $x_2/2r_l$ decreasing from $1.342$ at $l=6$ to $1$ at $l=\infty$.
For ${\tilde{\xi}}_c\in ]1,\xi_c]$ it is an increasing function of ${\tilde{\xi}}_c$ 
such that $\lim_{{\tilde{\xi}}_c\rightarrow\xi_c}x_2/2r_l = \infty$ for $l$ finite. 

The universal form of the spectral function near the singularities, Eq. (\ref{equ2}), 
is determined by the large $x$ behavior of $V_c (x)$, Eq. (\ref{equ5}), and sum rules, 
Eqs. (\ref{equ6}) and (\ref{equ7}). In the spectral-weight continuum, its form is not universal, as it depends on the 
specific small $x$ form of $V_c (x)$ determined by $V_e (r)$.

The length scale $2r_l$ in Eq. (\ref{equ5}) is found below in Sec. \ref{ER} to obey the
inequality $\sqrt{2}\,(2r_l)^{l-2\over 2}=\sqrt{2\mu\gamma_c}\gg 1$ in units of $a_0 =1$.  
($\sqrt{2\mu\gamma_c}$ in such units corresponds to the important parameter 
$\gamma = \sqrt{2\,\mu\,\alpha}/\hbar$ of Ref. \onlinecite{Flambaum_99} in units of Bohr 
radius $a_0 = 0.529177$\,\textrm{\AA} with $\mu$ and $\alpha$ corresponding to $\mu$ and $\gamma_c$, 
respectively.) This inequality justifies why $V_c (x)= V_c^{\rm asy} (x)$ for $x>x_2$  and implies that 
$\int_{x_0}^{x_2}dx\sqrt{2\mu (-V_c (x))}$ in Eq. (\ref{equ7})
has for ${\tilde{\xi}}_c\in ]1/2,1[$ large values. This is consistent with the above
mentioned requirement of the scattering energy of the residual interactions 
of the $c$ particles and $c$ impurity being 
smaller than the depth $-V_c (x_1)$ of the potential $V_c (x)$ well, which since 
$\int_{x_0}^{\infty}dx\sqrt{2\mu (-V_c (x))}/\pi\gg 1$ must be large. 
Here $x_1$ is a small non-universal potential-dependent $x$ value such that $x_0<x_1<a_0$ at 
which $\partial V_c (x)/\partial x =0$ and $-V_c (x)$ reaches its maximum value. 

\subsection{The unitary limit and the scattering lengths}
\label{UL}

As confirmed below in Sec. \ref{ER}, the expression for the phase shift in the thermodynamic limit, 
\begin{equation}
-2\pi{\tilde{\Phi}}_{c,c} (\pm 2k_F, \pm 2k_F + k_r)\vert_{k_r = \mp {2\pi\over L}}
=\mp {({\tilde{\xi}}_c -1)^2\pi\over {\tilde{\xi}}_c} \, ,
\label{equ8}
\end{equation}
for $\lim_{k_r\rightarrow0}{\tilde{\Phi}}_c (k_r)$
remains the same as for the MQIM-LO. Its use along with that of
$\mp (\xi_c -1)^2\pi/\xi_c$ for the bare phase shift 
$\lim_{k_r\rightarrow0}\cot (\Phi_c (k_r))$
in the leading term of the corresponding effective-range expansions gives
the scattering lengths. In the thermodynamic limit they read,
\begin{eqnarray}
\tilde{a} & = & - {L\over 2\pi}\tan \left({({\tilde{\xi}}_c -1)^2\pi\over {\tilde{\xi}}_c}\right)
\rightarrow - \infty 
\hspace{0.20cm}{\rm for}\hspace{0.20cm}{\tilde{\xi}}_c\neq 1\hspace{0.20cm}{\rm and}
\nonumber \\
a & = & - {L\over 2\pi}\tan \left({(\xi_c -1)^2\pi\over \xi_c}\right) \rightarrow - \infty 
\hspace{0.20cm}{\rm for}\hspace{0.20cm}\xi_c\neq 1 \, ,
\label{equ9}
\end{eqnarray}
respectively. This is known as the unitary limit \cite{Zwerger_12,Horikoshi_17}. 

The validity of the MQIM-HO refers to this limit, which  occurs provided that $\xi\neq 1$, ${\tilde{\xi}}_c\neq 1$, 
and as confirmed below that ${\tilde{\xi}}_c>1/2$. The
dependence of the bare charge parameter $\xi_c = \xi_c (n_e,u)$ on the density $n_e$ and $u=U/4t$ 
is defined by Eq. (\ref{equA16}) of Appendix \ref{APA}. It is such that $\xi_c=\sqrt{2}$ for $u\rightarrow 0$ and 
$\xi_c=1$ for $u\rightarrow\infty$ for $n_e\in ]0,1[$
and $\xi_c=1$ for $u>0$ and $\xi_c=\sqrt{2}$ at $u=0$ for both $n_e\rightarrow 0$ and
$n_e\rightarrow 1$. This implies that $a=-\infty$ provided that the relative momentum obeys the inequality 
$\vert k_r\vert\ll {\tan (\pi\,n_e)\over 4u}$. This excludes electronic densities very near $n_e=0$ and $n_e=1$
for all $u$ values and excludes large $u$ values for the remaining electronic densities.

The phase shifts ${\tilde{\Phi}}_c  = -2\pi{\tilde{\Phi}}_{c,c} (\pm 2k_F,q)$
incurred by the $c$ particles from their interactions with the $c$ impurity
created at momentum $q\in [-2k_F+k_{Fc}^0,2k_F-k_{Fc}^0[$ 
have $c$ band momenta in two small intervals $[-2k_F,-2k_F+k_{Fc}^0]$ and
$[2k_F-k_{Fc}^0,2k_F]$ near the $c$ band Fermi points $-2k_F$ and $2k_F$, respectively.
As discussed in Appendix \ref{RTLL}, the creation of an impurity in the 
$c$ band intervals $q\in [-2k_F,-2k_F+k_{Fc}^0]$ and $q\in [2k_F-k_{Fc}^0,2k_F]$ 
refers to the low-energy TLL regime. Its velocity 
becomes that of the low-energy particle-hole excitations near $-2k_F$
and $2k_F$, respectively. In this regime, the physics is different, as the
impurity loses its identity, since it cannot be distinguished from the $c$ band holes 
(TLL holons) in such excitations. 

The small momentum $k_{Fc}^0$ can be written as $k_{Fc}^0 = \pi n_{Fc}^0$.  
The unitary limit refers to the corresponding low-density $n_{Fc}^0$ of $c$ particle scatterers with phase shift 
${\tilde{\Phi}}_c  = -2\pi{\tilde{\Phi}}_{c,c} (\pm 2k_F,q)$
and $c$ band momentum values $[-2k_F,-2k_F+k_{Fc}^0]$ and $[2k_F-k_{Fc}^0,2k_F]$ near
$-2k_F$ and $2k_F$, respectively. They, plus the single $c$ impurity 
constitute the usual dilute quantum liquid of the unitary limit whose density is thus $n_{Fc}^0$.
$k_{Fc}^0$ is such that $k_{Fc}^0\,\vert{\tilde{a}}\vert = {1\over 2}N_{Fc}^0\tan([({\tilde{\xi}}_c -1)^2/{\tilde{\xi}}_c]\pi)$ 
and $k_{Fc}^0\,\vert a\vert = {1\over 2}N_{Fc}^0\tan([(\xi_c -1)^2/\xi_c]\pi)$ for ${\tilde{\xi}}_c=\xi_c$. 
Here $N_{Fc}^0$ is the number of $c$ particle scatterers in $n_{Fc}^0=N_{Fc}^0/L$.

In the thermodynamic limit one has that $n_{Fc}^0$ is very small or even such that
$\lim_{L\rightarrow\infty}n_{Fc}^0\rightarrow 0$. Consistent with this result, the following relations of the 
usual unitary limit of the dilute quantum liquid unitary limit  
\cite{Zwerger_12}, $R_{\rm eff}\ll 1/k_{Fc}^0\ll\vert{\tilde{a}}\vert$ and
$0\ll 1/k_{Fc}^0\ll \vert a\vert$ hold. The effective range $R_{\rm eff}$ derived 
in Sec. \ref{ER} is such that $R_{\rm eff}\rightarrow\infty$ as ${\tilde{\xi}}_c\rightarrow 1/2$. The unitary
limit requirement that $R_{\rm eff}\ll 1/k_{Fc}^0$ in the thermodynamic limit is the reason that
the value ${\tilde{\xi}}_c =1/2$ is excluded from the regime in which the MQIM-HO is valid.
\begin{figure}
\begin{center}
\centerline{\includegraphics[width=8.0cm]{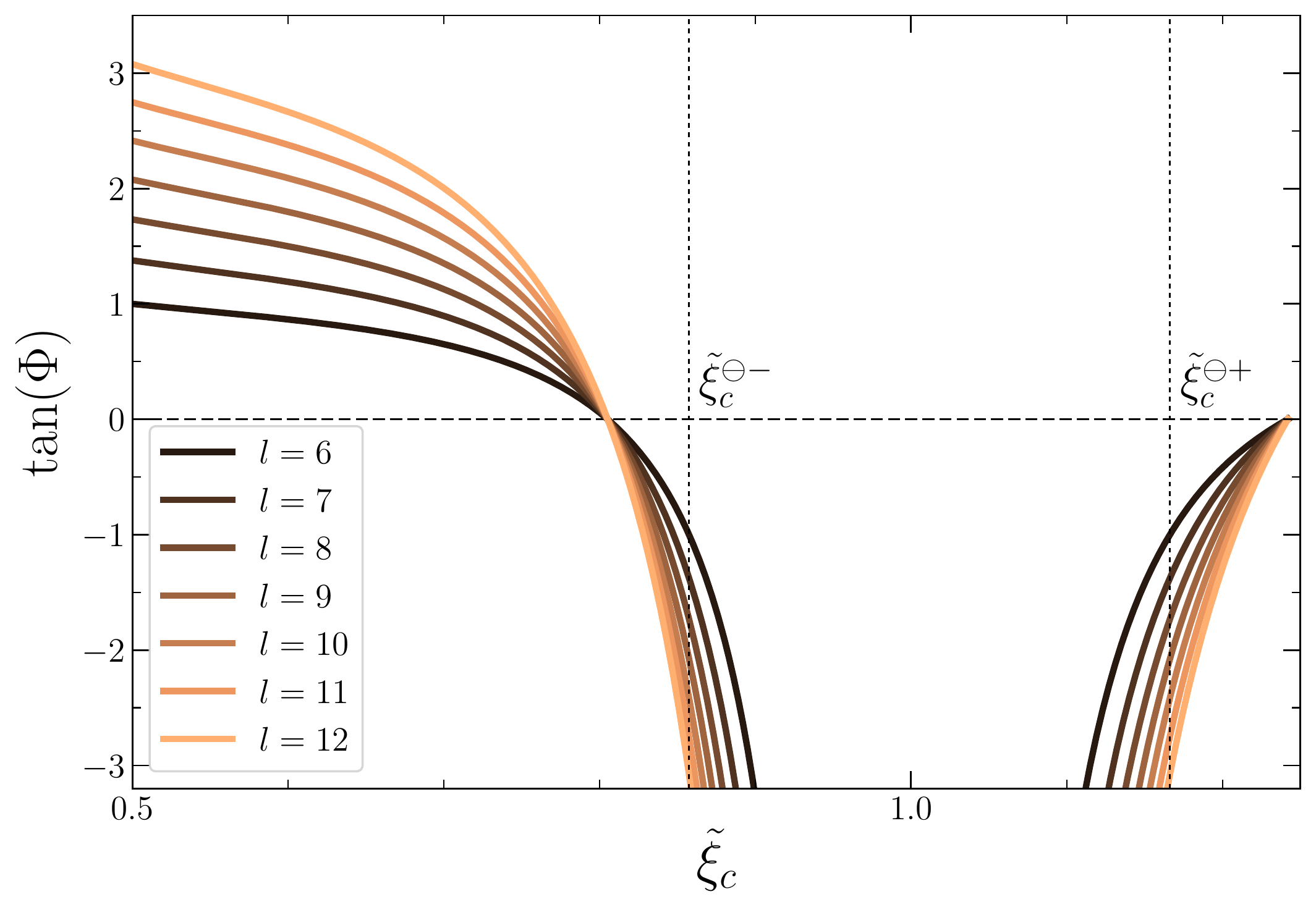}}
\caption{$\tan (\Phi) = - (\Delta a/{\tilde{a}})\cot (\pi/[l-2])$, Eqs. (\ref{equ6}) and (\ref{equ11}), as a function of 
the renormalized charge parameter ${\tilde{\xi}}_c$ for the electronic density $n_e = 0.176$, 
interaction $u=U/4t=0.30$, and integer quantum numbers $l=6-12$ used in Sec. \ref{ARPES} for Bi/InSb(001). 
For ${\tilde{\xi}}_c\rightarrow 1/2$ and at ${\tilde{\xi}}_c={\tilde{\xi}}_c^{\oslash}=1/\xi_c=0.805$,
$\tan (\Phi)$ reads $\cot (\pi/(l-2))$ and $0$, respectively, and at both ${\tilde{\xi}}_c={\tilde{\xi}}_c^{\ominus-}=0.857$ and 
${\tilde{\xi}}_c={\tilde{\xi}}_c^{\ominus+}=1.166$ it is given by $-\cot (\pi/(l-2))$. The MQIM-HO is not valid
near ${\tilde{\xi}}_c= 1$ at which $\Delta a/{\tilde{a}}=\infty$ and the corresponding scattering problem does not
refer to the unitary limit. The finite-range effects are more pronounced 
for ${\tilde{\xi}}_c \in ]1/2,{\tilde{\xi}}_c^{\oslash}[$ when $\Delta a/{\tilde{a}}<0$ and $\tan (\Phi)>0$.}
\label{figure2}
\end{center}
\end{figure}

Importantly, although both $a^{-1}=0$ and ${\tilde{a}}^{-1}=0$, the ratio $\tilde{a}/a$ is finite. 
Since below in Sec. \ref{ER} we confirm that $a$ and ${\tilde{a}}^{-1}$ are in Eq. (\ref{equ6})
the scattering lengths given by Eq. (\ref{equ9}), the value of ${\tilde{\xi}}$ in their ratio $\tilde{a}/a$ 
expression is found to be controlled by the potential $V_c (x)$ though $\tan (\Phi)$ in the 
sum rules provided in Eqs. (\ref{equ6}) and (\ref{equ7}) as,
\begin{equation}
{{\tilde{a}}\over a} = {\tan (\pi({\tilde{\xi}}_c -1)^2/{\tilde{\xi}}_c)\over\tan (\pi(\xi_c -1)^2/\xi_c)}
= {1\over 1 - \tan\left({\pi\over l-2}\right)\tan(\Phi)} \, .
\label{equ10}
\end{equation}
The first expression on the right-hand side of this equation is specific to the present 1D quantum problem and
follows directly from Eq. (\ref{equ9}). Hence in the present case $\tan (\Phi)$ in Eq. (\ref{equ6}) can 
be expressed as,
\begin{equation}
\tan (\Phi) = - {\sin\left({(\xi_c - {\tilde{\xi}}_c)(\xi_c{\tilde{\xi}}_c-1)\pi\over \xi_c{\tilde{\xi}}_c}\right)
\cot\left({\pi\over l-2}\right)
\over \sin\left({({\tilde{\xi}}_c -1)^2\over {\tilde{\xi}}_c}\pi\right)
\cos\left({(\xi_c -1)^2\over\xi_c}\pi\right)} \, .
\label{equ11}
\end{equation}
One finds from the use of Eq. (\ref{equ10}) that effects of the finite-range interactions controlled by relative fluctuation $\Delta a/{\tilde{a}}$ 
in $\tan (\Phi)=- {\Delta a\over{\tilde{a}}}\cot \left({\pi\over l-2}\right)$, Eq. (\ref{equ6}),
are stronger for ${\tilde{\xi}}_c\in ]1/2,{\tilde{\xi}}_c^{\oslash}]= ]1/2,1/\xi_c]$ when 
$\Delta a/{\tilde{a}}<0$, ${\tilde{a}}/a>1$, and $\tan (\Phi)>0$ in Fig. \ref{figure2}. 
Upon increasing ${\tilde{\xi}}_c$ within the intervals ${\tilde{\xi}}_c \in ]1/2,1[$ and ${\tilde{\xi}}_c \in ]1,\xi_c]$, the 
relative fluctuation increases from $\Delta a/{\tilde{a}}=-1$ for 
${\tilde{\xi}}_c\rightarrow 1/2$ to $\Delta a/{\tilde{a}}=\infty$ for ${\tilde{\xi}}_c\rightarrow 1$, crossing $0$ and $1$ at 
${\tilde{\xi}}_c={\tilde{\xi}}_c^{\oslash}=1/\xi_c$ and ${\tilde{\xi}}_c={\tilde{\xi}}_c^{\ominus-}$,
respectively. Upon further increasing ${\tilde{\xi}}_c$, the ratio decreases from $\Delta a/{\tilde{a}}=\infty$ to $\Delta a/{\tilde{a}}=0$ at
${\tilde{\xi}}_c=\xi_c$, crossing $1$ at ${\tilde{\xi}}_c={\tilde{\xi}}_c^{\ominus+}$. Here ${\tilde{\xi}}_c^{\ominus -}\in ]0.778,1[$
and ${\tilde{\xi}}_c^{\ominus -}\in ]1,1.284[$ are given by Eqs. (\ref{equ12}) and (\ref{equ13})
with $\eta_c (\xi_c,\Phi,l) = 1 + {1\over 2\pi}\arctan \left({\vert a\vert\,\pi\over L}\right)$ where $a$ is the bare scattering length
in Eq. (\ref{equ9}). For the electronic density $n_e = 0.176$ and interaction $u=U/4t=0.30$ (the values used in Sec. \ref{ARPES} 
for Bi/InSb(001)), ${\tilde{\xi}}_c^{\oslash}=1/\xi_c = 0.805$, ${\tilde{\xi}}_c^{\ominus-} = 0.857$, and ${\tilde{\xi}}_c^{\ominus+} = 1.166$.

The renormalized charge parameter intervals ${\tilde{\xi}}_c \in ]1/2,1[$ for which $\alpha >1/8$ and ${\tilde{\xi}}_c \in ]1,\xi_c]$
for which $\alpha <1/8$ refer to two qualitatively different problems.
Importantly, the ${\tilde{\xi}}_c$ value in the $\xi_c\rightarrow {\tilde{\xi}}_c$ transformation
is uniquely defined for {\it each} of these two intervals solely by the bare charge parameter $\xi_c = \xi_c (n_e,u)$, 
Eq. (\ref{equA16}) of Appendix \ref{APA}, the integer quantum number $l$ in the potential $V_c  (x)$ large-$x$ expression, 
Eq. (\ref{equ5}), and its sum rule zero-energy phase $\Phi $, Eq. (\ref{equ7}), as follows:
\begin{eqnarray}
{\tilde{\xi}}_c & = & \eta_c (\xi_c,\Phi,l) \left(1 - \sqrt{1 - {1\over \eta_c^2 (\xi_c,\Phi,l)}}\right) \in ]1/2,1[ 
\nonumber \\
& = & \eta_c (\xi_c,\Phi,l)\left(1 + \sqrt{1 - {1\over \eta_c^2 (\xi_c,\Phi,l)}}\right) \in ]1,\xi_c] 
\label{equ12}
\end{eqnarray}
where,
\begin{equation}
\eta_c (\xi_c,\Phi,l) = 1 + {1\over 2\pi}\arctan\left({\tan \left({(\xi_c -1)^2\pi\over \xi_c}\right)\over
1 + \tan \left({\pi\over l-2}\right) \tan (\Phi) }\right)  \, .
\label{equ13}
\end{equation}

\section{The effective range}
\label{ER}

\subsection{The effective-range general problem and cancellation of its unwanted terms}
\label{ERGP}

The MQIM-HO accounts for the higher terms in the effective range expansion, Eq. (\ref{equ4}),
so that as anticipated the phase shift $2\pi{\tilde{\Phi}}_{c,c} (\pm 2k_F,q)$ acquires an additional term, 
$2\pi{\tilde{\Phi}}_{c,c}^{R_{\rm eff}} (k_r)$, relative to the MQIM-LO, 
\begin{eqnarray}
&& 2\pi{\tilde{\Phi}}_{c,c} (\pm 2k_F,q) = 2\pi{\tilde{\Phi}}_{c,c}^{{\tilde{a}}} (\pm 2k_F,q) + 2\pi{\tilde{\Phi}}_{c,c}^{R_{\rm eff}} (k_r) 
\nonumber \\
&& 2\pi{\tilde{\Phi}}_{c,c}^{{\tilde{a}}} (\pm 2k_F,q) = {\xi_c\over {\tilde{\xi}}_c}{({\tilde{\xi}}_c -1)^2\over (\xi_c -1)^2}\,2\pi\Phi_{c,c} (\pm 2k_F,q)
\nonumber \\
& & \hspace{2.6cm} = {\arctan\left({{\tilde{a}}\over L}\,2\pi\right)\over\arctan\left({a\over L}\,2\pi\right)}\,2\pi\Phi_{c,c} (\pm 2k_F,q)
\nonumber \\
& & 2\pi{\tilde{\Phi}}_{c,c}^{R_{\rm eff}} (k_r) =
\nonumber \\ 
&& \arctan\left({1\over 2}R_{\rm eff}\,k_r\sin^2 \left({({\tilde{\xi}}_c -1)^2\over  {\tilde{\xi}}_c}\pi\right) + P_c (k_r)\right) .
\label{equ14}
\end{eqnarray}

The second term in the expression for the phase shift $2\pi{\tilde{\Phi}}_{c,c}^{{\tilde{a}}} (\pm 2k_F,q)$ 
reveals that its renormalization is controlled by the scattering lengths associated with the leading  term in the effective 
range expansion. The ${\tilde{\xi}}_c =\xi_c$ bare phase shift $2\pi\Phi_{c,c} (\pm 2k_F,q)$ in that 
expression is defined in Eqs. (\ref{equA11})-(\ref{equA15}) of Appendix \ref{APA}. 
Furthermore, the function $P_c (k_r)$ in the expression of $2\pi{\tilde{\Phi}}_{c,c}^{R_{\rm eff}} (k_r)$
vanishes for $l<8$ and is such that its use in the term on the left-hand side of Eq. (\ref{equ4}), 
$\cot ({\tilde{\Phi}}_c (k_r))=\cot(-2\pi{\tilde{\Phi}}_{c,c}^{{\tilde{a}}} (\pm 2k_F,q) - 2\pi{\tilde{\Phi}}_{c,c}^{R_{\rm eff}} (k_r))$,
gives rise to all the shape parameter terms in the expansion, Eq. (\ref{equ4}), beyond
the two leading terms, ${-1\over {\tilde{a}}\,k_r} + {1\over 2}\,R_{\rm eff}\,k_r$.

Fortunately, in the unitary limit all properties that are characterized by these higher-order terms 
become irrelevant also for $l>7$. Hence $2\pi{\tilde{\Phi}}_{c,c}^{R_{\rm eff}} (k_r)$ is given by
$\arctan\left({1\over 2}R_{\rm eff}\,k_r\sin^2 ([({\tilde{\xi}}_c -1)^2/{\tilde{\xi}}_c]\pi)\right)$, which
gives $\cot ({\tilde{\Phi}}_c (k_r))={-1\over {\tilde{a}}\,k_r} + {1\over 2}\,R_{\rm eff}\,k_r$
at small $k_r$. (That $2\pi{\tilde{\Phi}}_{c,c}^{R_{\rm eff}} (\mp 2\pi/L)$ vanishes
in the thermodynamic limit confirms that at $k_r = \mp 2\pi/L$ 
the phase shift $2\pi{\tilde{\Phi}}_{c,c} (\pm 2k_F,\pm 2k_F + k_r)$
has the same value $\pm ({\tilde{\xi}}_c -1)^2\pi/{\tilde{\xi}}_c$,
Eq. (\ref{equ8}), as for the MQIM-LO.)

Both the unitary limit and the fact that for ${\tilde{\xi}}_c\in ]1/2,1[$ the scattering energy of the residual interactions 
of the $c$ particles and $c$ impurity are much smaller than the depth $-V_c (x_1)$
of the potential $V_c (x)$ will play  important roles in the following derivations of   
the effective range $R_{\rm eff}$ in the expression of $2\pi{\tilde{\Phi}}_{c,c}^{R_{\rm eff}} (k_r)$,
Eq.(\ref{equ14}). 

First, note that the phase shift term $-2\pi{\tilde{\Phi}}_{c,c}^{{\tilde{a}}} (\pm 2k_F,\pm 2k_F + k_r)$
(see Eq. (\ref{equ14})) of ${\tilde{\Phi}}_c (k_r) = -2\pi{\tilde{\Phi}}_{c,c} (\pm 2k_F,\pm 2k_F + k_r)$ in the
effective range expansion, Eq. (\ref{equ4}), contributes only  to the leading term in that expansion
, ${-1\over {\tilde{a}}\,k_r}$. Thus it does not contribute to the effective range
$R_{\rm eff}$. Indeed, that phase shift term reads $\mp ({\tilde{\xi}}_c -1)^2\pi/{\tilde{\xi}}_c$,
Eq. (\ref{equ8}), at $k_r=\mp 2\pi/L$ whereas it vanishes at $k_r =0$, so that in the thermodynamic 
limit the derivative $-2\pi\partial {\tilde{\Phi}}_{c,c}^{{\tilde{a}}} (\pm 2k_F,\pm 2k_F + k_r)/\partial k_r\vert_{k_r =0}$ 
is ill defined.

For a potential with large-$x$ behavior, $- C_c/(x/2r_l)^l$,
Eq. (\ref{equ5}), the effective range $R_{\rm eff}$ in the phase shift term
$2\pi{\tilde{\Phi}}_{c,c}^{R_{\rm eff}} (k_r)$ of Eq. (\ref{equ14}) follows from 
standard scattering-theory methods, and becomes \cite{Joachain_75,Preston_75,Flambaum_99}
\begin{eqnarray}
R_{\rm eff} = 2\int_0^{\infty} dx\left((\psi_c^0 (x))^2 - (\psi_c (x))^2\right) \, .
\label{equ15}
\end{eqnarray}
This integral converges provided that $l>5$.

The bare limit ${\tilde{\xi}}_c =\xi_c$ boundary condition $V_c (x)=0$ for all $x$ 
corresponds to the wave function $\psi_c^0 (x)$ in Eq. (\ref{equ15}). 
It is the zero-energy solution of the Schr\"odinger equation for the free motion,
\begin{eqnarray}
- {1\over 2\mu}{d^2\psi_c^0 (x)\over dx^2} = 0 \, .
\nonumber 
\end{eqnarray}
Here $\mu$ is the reduced mass of the $c$ particle and $c$ impurity.
The function $\psi_c^0 (x)$ then has the form $\psi_c^0 (x) = 1 - x/a$ for all $x\in [0,\infty]$.

In contrast,  the wave function $\psi_c (x)$ in Eq. (\ref{equ15})
is associated with the potential $V_c (x)$ induced by the potential $V_e (r)$ in Eq. (\ref{equ1}).
The former is associated with the interaction of the $c$ particle and $c$ impurity. 
That wave function is thus the solution of a corresponding 
Schr\"odinger equation at zero energy,
\begin{eqnarray}
- {1\over 2\mu}{d^2\psi_c (x)\over dx^2} + V_c (x)\,\psi_c (x) = 0 \, ,
\label{equ16}
\end{eqnarray}
with the boundary condition $\psi_c (0) = 0$. It is normalized at $x\rightarrow\infty$
as $\psi_c (x) = \psi_c^0 (x) = 1 - x/a$. 

The charge parameter interval ${\tilde{\xi}}_c\in ]1,\xi_c]$ for which $\alpha <1/8$ that corresponds
to the second ${\tilde{\xi}}_c$ expression in Eq. (\ref{equ12}) is of little interest for our studies,
as similar $\alpha$ values are reachable by the 1D Hubbard model.
Two boundary conditions that must be obeyed by $R_{\rm eff}$ in that parameter interval are
$\lim_{{\tilde{\xi}}_c\rightarrow \xi_c} R_{\rm eff}=0$ and $\lim_{{\tilde{\xi}}_c\rightarrow 1} R_{\rm eff}= a_0$.
They are satisfied by the following {\it phenomenological} effective range expression,
\begin{equation}
R_{\rm eff}  = a_0\left(1 - {{\tilde{a}}\over a}\right) < a_0 \, .
\label{equ17}
\end{equation}

In the case of the interval ${\tilde{\xi}}_c\in ]1/2,1[$ for which $\alpha >1/8$
that corresponds to the first ${\tilde{\xi}}_c$ expression in Eq. (\ref{equ12}), the
explicit derivation of the integral in the effective range expression, Eq. (\ref{equ15}), simplifies
because the inequalities $\sqrt{2\mu\gamma_c}\gg 1$ in units of $a_0 =1$ and $\Phi/\pi\gg 1$ are found to  
apply, as reported in Sec. \ref{EREUL}. This ensures that for ${\tilde{\xi}}_c<1$ 
the scattering energy of the residual interactions of the $c$ particles and 
$c$ impurity is much smaller than the depth $-V_c (x_1)$ of the potential $V_c (x)$. 

The following analysis applies to general scattering lengths $a$, finite or infinite, 
and potentials with these general properties. They imply that the large-$x$ function
$\psi_c (x)$ obeying a Schr\"odinger equation, 
\begin{eqnarray}
{d^2\psi_c (x)\over dx^2} + {2(2r_l)^{l-2}\over x^l}\psi_c (x)  = 0 \, ,
\nonumber 
\end{eqnarray}
whose attractive potential is given by its large-distance asymptotic 
behavior $V_c^{\rm asy} (x) = - C_c/(x/2r_l)^l$, Eq. (\ref{equ5}), which has the general form,
\begin{eqnarray}
\psi_c (x) = \sqrt{x}\,\left(B_1\,\phi_{1\over l-2}(x) 
- B_2\,{\phi_{{-1\over l-2}}(x)\over\cos\left({\pi\over l-2}\right)}\right) \, .
\label{equ18}
\end{eqnarray}
This expression  can be used for all $x\in [0,\infty]$ 
{\it provided} that $V_c (x)$ at small distances $x\approx x_1$ where 
it is deep is replaced by a suitable energy-independent boundary condition. 
This is also valid for 3D s-wave scattering problems 
whose potentials have the above general properties and whose scattering 
lengths are parametrically large \cite{Flambaum_99}. 

$B_1$ and $B_2$ are in Eq. (\ref{equ18}) $x$ independent constants and
$\phi_{1\over l-2} (x) = J_{1\over l-2} (y)$ and
$\phi_{{-1\over l-2}} (x) = J_{{-1\over l-2}} (y)$ where
$J_{1\over l-2} (y)$ and $J_{{-1\over l-2}} (y)$ are
Bessel functions of argument,
\begin{eqnarray}
y = {2\sqrt{2}\over (l-2)(x/2r_l)^{l-2\over 2}} \, .
\nonumber 
\end{eqnarray}
From the use in Eq. (\ref{equ18}) of the asymptotic behavior, $J_{\nu} (y) \approx y^{\nu}/(2^{\nu}\Gamma (1+\nu))$, of 
the Bessel functions for $x\gg 1$ and thus $y\ll 1$
one finds that the normalization at $x\rightarrow\infty$ 
as $\psi_c (x) = \psi_c^0 (x) = 1 - x/a$ requires that,
\begin{equation}
B_1 = {1\over\sqrt{2r_l}}\left({l-2\over\sqrt{2}}\right)^{1\over l-2}
\Gamma \left({l-1\over l-2}\right) \, ,
\label{equ19}
\end{equation}
and
\begin{eqnarray}
B_2 & = & B_2^0 = {\bar{a}\over a}\,B_1\hspace{0.20cm}{\rm where}\hspace{0.20cm}{\rm the}
\hspace{0.20cm}{\rm length}\hspace{0.20cm}\bar{a}\hspace{0.20cm}{\rm reads}
\nonumber \\
\bar{a} & = & 2r_l\,\left({\sqrt{2}\over l-2}\right)^{2\over l-2}
{\Gamma \left({l-3\over l-2}\right) \over\Gamma \left({l-1\over l-2}\right)}\cos \left({\pi\over l-2}\right) \, .
\label{equ20}
\end{eqnarray}

It is convenient to write the integrand in Eq. (\ref{equ15}) as
$(\psi_c^0 (x))^2 - \psi_c^2 (x) = g_c^{\rm virtual} (x) + g_c (x)$ where,
\begin{eqnarray}
g_c^{\rm virtual} (x) & = & (\psi_c^0 (x))^2 - f_c (x) \hspace{0.20cm}{\rm and}
\nonumber \\
\psi_c (x) & = & \sqrt{f_c (x) - g_c (x)} \, ,
\label{equ21}
\end{eqnarray}
and the functions $f_c (x)$ and $g_c (x)$ are given by,
\begin{eqnarray}
& & f_c (x) = (2r_l)^2{d\over dx}\{\left({x\over 2r_l}\right)^2[B_1^2\,\phi_{1\over l-2}^2 (x)
-  {B_2\over\cos\left({\pi\over l-2}\right)}
\nonumber \\
& & \times \{B_1\,\phi_{{1\over l-2}}(x) 
- {B_2\over 3\cos\left({\pi\over l-2}\right)}\,\phi_{{-1\over l-2}}(x)\}\,\phi_{{-1\over l-2}}(x)]\} 
\label{equ22}
\end{eqnarray}
and
\begin{eqnarray}
& & g_c (x) = \left({x\over 2r_l}\right)^{-{(l-2)\over 2}+1}4\sqrt{2}\, r_l
\{B_1^2\,\phi_{1\over l-2}(x)\,\phi_{{1\over l-2}+1}(x) 
\nonumber \\
& - & {B_1\,B_2\over 2\cos\left({\pi\over l-2}\right)} [\phi_{1\over l-2}(x)\,\phi_{{-1\over l-2}+1}(x) 
+ \phi_{{-1\over l-2}}(x)\,\phi_{{1\over l-2}+1}(x)]
\nonumber \\
& + & {B_2^2\over 3\cos^2\left({\pi\over l-2}\right)}\,\phi_{{-1\over l-2}}(x)\,\phi_{{-1\over l-2}+1}(x)\} \, ,
\label{equ23}
\end{eqnarray}
respectively.

The separation in Eq. (\ref{equ21}) is convenient because 
the divergences all appear in the functions $(\psi_c^0 (x))^2$ and $f_c (x)$. 
That $B_1$ and $B_2$ have the expressions given in Eqs. (\ref{equ19}) 
and (\ref{equ20}), respectively, ensures that both $2\int_0^{\infty} dx\,(\psi_c^0 (x))^2$ and
$2\int_0^{\infty} dx\,f_c (x)$ read $\lim_{x\rightarrow\infty} 2\left(x - {x^2\over a} + {1\over 3}{x^3\over a^2}\right)$ 
and thus the divergences from $(\psi_c^0 (r))^2$ and $f_c (x)$ exactly cancel each other
under the integration of Eq. (\ref{equ15}). Hence $R_{\rm eff}$ can be expressed as, 
\begin{eqnarray}
R_{\rm eff} = 2\int_0^{\infty} dx\,g_c (x)  \, . 
\label{equ24}
\end{eqnarray}

\subsection{The energy-independent boundary condition}
\label{EIBC}

The only role of $f_c (x)$, Eq. (\ref{equ22}), is to cancel $(\psi_c^0 (x))^2$
within $g_c^{\rm virtual} (x)$, Eq. (\ref{equ21}), under the integration in Eq. (\ref{equ15}).
In the general expression for the effective range given in that equation, 
$\psi_c (x)$ is the solution of Eq. (\ref{equ16}) with
the actual potential $V_c (x)$ defined in its full domain, $x\in [0,\infty]$. 
The alternative use of Eq. (\ref{equ24}), which was derived
by using the function $\psi_c (x)$ large-$x$ expression, Eq. (\ref{equ18}), for the whole domain
$x\in [0,\infty]$, also leads to the effective range, Eq. (\ref{equ15}). This applies provided that $V_c (x)$ is replaced at small 
distances near $x=x_1$, where it is deep, by the energy-independent boundary condition defined below. 
It accounts for the effects from $V_c (x)$ for small distances. 

In the unitary limit the inverse scattering length, $a^{-1}=0$, which appears in the $B_2$
expression, Eq. (\ref{equ20}), is at the middle of negative $a^{-1}<0$ and positive $a^{-1}>0$ values and could
refer to $a = -\infty$ or $a = \infty$. Hence in that limit there is not much difference between 
the repulsive and attractive scattering cases. As discussed in Ref. \onlinecite{Castin_12}
for the case of two particles with a s-wave interaction, the scattering lengths in the attractive $a=-\infty$ 
and repulsive $a=\infty$ cases of the unitary limit merely refer to different states of the {\it same} $a^{-1}=0$ 
scattering problem.

For a potential $V (r)$ with a {\it finite} scattering length $a$ and having the general properties 
reported above, at small distances $r$ where the potential is deep it can be replaced by 
an energy-independent boundary condition such that the ratio
$B_2/B_1 = \bar{a}/a$ reads $\left[1 - \tan\left({\pi\over l-2}\right)\tan ({\bar{\Phi}})\right]^{-1}$ where
${\bar{\Phi}} = \int_{r_0}^{\infty}dr\sqrt{2\mu (-V (r))} - \pi/[2(l-2)]$. Here
${\bar{\Phi}}\gg\pi$ is a potential dependent zero-energy phase, $V (r_0)=0$, and $V (r)<0$ for $r>r_0$.
Moreover, $\tan ({\bar{\Phi}}) = - {\Delta a\over {\bar{a}}}\cos \left({\pi\over l-2}\right)$ where $\Delta a = a - {\bar{a}}$,
$\Delta a/{\bar{a}}$ is a relative fluctuation, and ${\bar{a}}$ given in Eq. (\ref{equ20}) is a mean scattering length determined by the 
asymptotic behavior $\propto 1/r^l$ of the potential $V (r)$ through the integer $l>5$ and
the length scale $2r_l$. For instance, in terms of the constants $A=B_1-B_2$ and $B=-B_2\tan\left({\pi\over l-2}\right)$,
of the scattering problem studied in Ref. \onlinecite{Flambaum_99},
the ratio $\bar{a}/a = B_2/B_1$ on the left hand side of the above boundary condition reads 
$\left[1 - (A/B)\tan\left({\pi\over n-2}\right)\right]^{-1}$ where $n=l$. 

For the present range ${\tilde{\xi}}_c\in ]1/2,1[$ the length scale $2r_l$ whose expression is given below is finite 
in the unitary limit and thus the related length scale ${\bar{a}}$ in Eq. (\ref{equ20}) is also finite.
It follows both that $\bar{a}/a = 0$ and the constant $B_2 = B_2^0$, Eq. (\ref{equ20}), vanishes. This result is
clearly incorrect. The reason is that we have have not yet accounted for the behavior of $V_c (x)$ at small 
distances through a suitable energy-independent boundary condition. In the case of the unitary limit, 
this boundary condition renders both $\bar{a}/a$ and $B_2 = B_2^0$ in Eq. (\ref{equ20}) finite. 
Specifically, the scattering length $a$ is suitably mapped under it into a finite 
scattering length $a_{\rm f}= a\,{\bar{a}\over {\tilde{a}}} $, so that $B_2^0$ is mapped 
onto the following corresponding finite constant $B_2$,
\begin{eqnarray}
B_2 & = & {\bar{a}\over a_{\rm f}}\,B_1 = {{\tilde{a}}\over a}\,B_1 
\hspace{0.20cm}{\rm where}
\nonumber \\
a_{\rm f} & = & a\,{\bar{a}\over {\tilde{a}}} = \bar{a}\left[1 - \tan\left({\pi\over l-2}\right)\tan ({\bar{\Phi}}_{\rm f})\right] \, .
\label{equ25}
\end{eqnarray}
Here $\tan ({\bar{\Phi}}_{\rm f})=\tan (\Phi)$ yet
${\bar{\Phi}}_{\rm f}\gg \pi$ may be different from $\Phi\gg \pi$ in Eq. (\ref{equ10}). Indeed, the relation 
$\tan ({\bar{\Phi}}_{\rm f})=\tan (\Phi)$ is insensitive to such phase differences. In the unitary limit the boundary 
condition is thus equivalent to a transformation $a\rightarrow a_{\rm f}$ such that $\tan ({\bar{\Phi}}_{\rm f})=\tan (\Phi)$.

The energy-independent boundary condition in Eq. (\ref{equ25}) is in terms of the finite scattering length $a_{\rm f}$
such that $B_2/B_1 = \bar{a}/a_{\rm f}$ is given by $\left[1 - \tan\left({\pi\over l-2}\right)\tan ({\bar{\Phi}}_{\rm f})\right]^{-1}$, 
similarly to scattering problems of the same universality class whose scattering lengths are 
parametrically large \cite{Flambaum_99,Gribakin_93}. The positivity of $a_{\rm f} = a\,{\bar{a}\over {\tilde{a}}}$ 
often occurs for potentials that for large distances are attractive \cite{Flambaum_99}. If $a_{\rm f}$ were negative, 
$\bar{a}/a_{\rm f} = - \tilde{a}/a$,
then $\tan ({\bar{\Phi}}_{\rm f})$ would be given by $2\cot\left({\pi\over l-2}\right)-\tan (\Phi)$, which would violate 
both the requirements that $\tan ({\bar{\Phi}}_{\rm f})=\tan (\Phi)$ and  that $\tan ({\bar{\Phi}}_{\rm f})=0$ in the 
bare limit, ${\tilde{\xi}}_c =\xi_c$. 

Importantly, the cancellation  $(\psi_c^0 (x))^2 - f_c (x)=0$ is {\it independent} of the value
of the scattering length in the expressions for $\psi_c^0 (x)$ and $f_c (x)$. Hence all results associated with Eqs. (\ref{equ15})-(\ref{equ24})
remain the same, with $a$ replaced by $a_{\rm f}$. This includes the effective range 
$R_{\rm eff}$, Eq. (\ref{equ24}), remaining determined only by $g_c (x)$.

The main property of the transformation $a\rightarrow a_{\rm f}$ is the corresponding exact equality of the ratios, 
$\bar{a}/a_{\rm f} = \tilde{a}/a$. This actually justifies why the scattering lengths $a$ and $\tilde{a}$, 
Eqs. (\ref{equ9}), can be used in $\tan (\Phi)$ in Eq. (\ref{equ6}). That transformation is also the mechanism 
through which the renormalized scattering length ${\tilde{a}}$ emerges in $\psi_c (x)$.

Hence similarly to finite-$a$ scattering problems of the same universality class, as for instance
those studied in Refs. \onlinecite{Flambaum_99} and \onlinecite{Kishi_17}, the relations 
of general form, Eq. (\ref{equ6}), apply. In the present unitary limit the scattering length ratio $\tilde{a}/a$ 
in them equals the ratio $\bar{a}/a_{\rm f}$ also given by Eq. (\ref{equ10}). The sum rule 
$\Phi = \int_{x_0}^{x_2}dx\sqrt{2\mu (-V_c (x))}$, Eq. (\ref{equ7}), encodes the effects from $V_c (x)$ for 
small distance near $x=x_1$, referring to the interval $x\in [x_0,x_2]$ around $x_1$. 

\subsection{The effective range dependence on the scattering length finite ratio $\tilde{a}/a$}
\label{EISL}

The use of the function $g_c (x)$, Eq. (\ref{equ23}), with the constants $B_1$ and $B_2$ given
in Eqs. (\ref{equ19}) and (\ref{equ25}), respectively, in Eq. (\ref{equ24}) leads for ${\tilde{\xi}}_c\in ]1/2,1[$ to,
\begin{eqnarray} 
& & R_{\rm eff} = 2\sqrt{2}\,\Gamma^2 \left({l-1\over l-2}\right)\,\left({l-2\over\sqrt{2}}\right)^{2\over l-2}
\nonumber \\
& \times & \{\int_0^{\infty} dx\, \left({x\over 2r_l}\right)^{-{l-2\over 2}+1}\,\phi_{1\over l-2}(x)\,\phi_{{-1\over l-2}+1}(x) 
\nonumber \\
& - & \left({{\tilde{a}}\over a}\right)\int_0^{\infty} dx\,\left({x\over 2r_l}\right)^{-{l-2\over 2}+1}
\nonumber \\
& \times & {\phi_{1\over l-2}(x)\,\phi_{{-1\over l-2}+1}(x) 
+ \phi_{{-1\over l-2}}(x)\,\phi_{{1\over l-2}+1}(x)\over 2\cos \left({\pi\over l-2}\right)}
\nonumber \\
& + & \left({{\tilde{a}}\over a}\right)^2\int_0^{\infty} dx\,\left({x\over 2r_l}\right)^{-{l-2\over 2}+1}
{\phi_{{-1\over l-2}}(x)\,\phi_{{-1\over l-2}+1}(x) 
\over 3\cos^2 \left({\pi\over l-2}\right)}\}   
\nonumber
\end{eqnarray}
After performing the integrations, one finally reaches the following expression 
valid for the present charge parameter interval ${\tilde{\xi}}_c \in ]1/2,1[$ and $\alpha >1/8$,
\begin{equation}
R_{\rm eff} = a_0\left(1 - c_1\,\left({{\tilde{a}}\over a}\right) + c_2\,\left({{\tilde{a}}\over a}\right)^2\right) \, .
\label{equ26}
\end{equation}
That here the coefficient,
\begin{equation}
a_0 = 2r_l\left({2\over 3\pi}{\left({2\over (l-2)^2}\right)^{1\over l-2}\over \sin\left({\pi\over l-2}\right)}\right)
{\Gamma \left({1\over l-2}\right)\Gamma \left({4\over l-2}\right)
\over \Gamma^2 \left({2\over l-2}\right)\Gamma \left({3\over l-2}\right)} \, ,
\label{equ27}
\end{equation}
is identified with the lattice spacing results from the imposition of $R_{\rm eff}$ having the same value for 
${\tilde{\xi}}_c\rightarrow 1^-$ and ${\tilde{\xi}}_c\rightarrow 1^+$. The coefficients $c_1$ and $c_2$ in 
Eq. (\ref{equ26}) can be expressed in terms the usual $\Gamma$ function and are given by,
\begin{eqnarray}
c_1 & = & {2\over \cos\left({\pi\over l-2}\right)} {\Gamma \left({2\over l-2}\right)\Gamma \left({l-4\over l-2}\right)
\over \Gamma \left({1\over l-2}\right)\Gamma \left({l-3\over l-2}\right)}\hspace{0.20cm}{\rm and}
\nonumber \\
c_2 & = & {3\,(l+1)\over (l-1)\cos^2\left({\pi\over l-2}\right)}{\Gamma \left({3\over l-2}\right)\Gamma \left(-{l+1\over l-2}\right)
\over \Gamma \left({-1\over l-2}\right)\Gamma \left(-{l-1\over l-2}\right)} \, ,
\label{equ28}
\end{eqnarray}
respectively. They decrease from $c_1=c_2=2$ at $l=6$ to $c_1=1$ and $c_2=1/3$ 
for $l\rightarrow\infty$. 

The effective range $R_{\rm eff}$, Eq. (\ref{equ26}), appears in the expression
of the spectral function exponents ${\tilde{\zeta}}_c (k)$ and ${\tilde{\zeta}}_{c'} (k)$,
Eq. (\ref{equA3}) of Appendix \ref{APA}, through the expression for the phase shift $2\pi{\tilde{\Phi}}_{c,c} (\pm 2k_F,q)$, 
Eq. (\ref{equ14}). $R_{\rm eff}=\infty$ for ${\tilde{\xi}}_c\rightarrow 1/2$ is excluded, as it is outside 
the range of validity of the unitary limit. The $R_{\rm eff}$ values found below 
in Sec. \ref{ARPES} for Bi/InSb(001) are given in Table \ref{table5}. They obey the unitary limit
inequality, $R_{\rm eff}\ll 1/k_{Fc}^0$. 

The effective range, Eq. (\ref{equ26}), can alternatively be expressed in terms of the ratio $\bar{a}/a_{\rm f}$ 
involving the finite scattering lengths $\bar{a}\propto 2r_l$ and $a_{\rm f}$ defined by Eqs. (\ref{equ20}) and (\ref{equ25}), 
respectively. 

The expression for the lattice spacing $a_0$, Eq. (\ref{equ27}), contains important physical information: Its 
inverse gives the following expression valid for ${\tilde{\xi}}_c \in ]1/2,1[$ for the length scale $2r_l$ in the 
potential $V_c^{\rm asy} (x)$ expression, Eq. (\ref{equ5}), and the related length scale $x_2$, Eq. (\ref{equ7}),
\begin{equation}
2r_l = {3\pi a_0\over 2}\sin\left({\pi\over l-2}\right)\left({l-2\over\sqrt{2}}\right)^{2\over l-2}
{\Gamma^2 \left({2\over l-2}\right)\Gamma \left({3\over l-2}\right)\over
\Gamma \left({1\over l-2}\right)\Gamma \left({4\over l-2}\right)}  \, .
\label{equ29}
\end{equation}
Here $2r_l$ is given by $5.95047\,a_0$ at $l=6$, reaches a maximum $6.48960\,a_0$ at $l=10$, and decreases to 
$4.93480\,a_0$ as $l\rightarrow\infty$, so that $\sqrt{2}\,(2r_l)^{l-2\over 2}=\sqrt{2\mu\gamma_c}\gg 1$ 
in units of $a_0 =1$ as given in Table \ref{table5}. Thus $\Phi/\pi\gg 1$ for $l=6-12$. 

As in the case of 3D s-wave atomic scattering problems \cite{Flambaum_99,Gribakin_93}, this shows 
that for ${\tilde{\xi}}_c\in ]1/2,1[$ the scattering energy of the interactions of the $c$ particles and 
$c$ impurity is indeed much smaller than the depth $-V_c (x_1)$ of the potential $V_c (x)$ well.
This confirms the consistency of the derivation of the effective range for ${\tilde{\xi}}_c \in ]1/2,1[$ that 
assumed the validity of such properties. 

The length scales involved in the MQIM-HO description are explicitly defined in Table \ref{table2}.
\begin{widetext}
\begin{table}
\begin{center}
\begin{tabular}{|c||c|} 
\hline
Length scale & Description \\
\hline
$2r_l$ & length scale in the large-$x$ potential decay with exponent $l>5$, $V_c^{\rm asy} (x)\propto (x/2r_l)^{-l}$, 
Eqs. (\ref{equ5}) and (\ref{equ29})  \\
\hline
$a_0$ & lattice spacing related to $2r_l$ (twice the van der Waals length at $l=6$) as given in Eq. (\ref{equ27})\\
\hline
$a$ and $\tilde{a}$ & scattering lengths at $\xi_c$ and ${\tilde{\xi}}_c$, respectively, within the $\xi_c\rightarrow{\tilde{\xi}}_c$ 
transformation, Eqs. (\ref{equ9}) and (\ref{equ10}) \\
\hline
$R_{\rm eff}$ & effective range $R_{\rm eff} = a_0\left(1 - c_1\,\left({{\tilde{a}}\over a}\right) + c_2\,\left({{\tilde{a}}\over a}\right)^2\right)$
for the interval ${\tilde{\xi}}_c<1$ of physical interest, Eqs. (\ref{equ26}) and (\ref{equ28})\\
\hline
\end{tabular}
\caption{Length scales involved in the MQIM-HO description}
\label{table2}
\end{center}
\end{table} 
\end{widetext}

\section{ARPES in Bi/InSb(001)}
\label{ARPES}

\subsection{Brief information on the sample preparation and ARPES experiments}
\label{PREP}

Concerning the preparation of the Bi/InSb(001) surface, a substrate InSb(001) was cleaned by 
repeated cycles of Ar sputtering and annealing up to $680$ K. Bi was evaporated on it up to nominally 
$3$ monolayers (ML): One ML is defined as the atom density of bulk-truncated substrate. Then, the 
substrate was flash-annealed up to $680$ K for $\sim 10$ seconds. The resulting surface showed a 
($1\times 3$) low-energy electron diffraction pattern. 

Although the Bi/InSb(001) surface state is formed by evaporating Bi on the InSb substrate, in addition 
to Bi also In and Sb are found at the surface, modified from their bulk positions by Bi evaporation. 
Hence Bi, In, and Sb can all be significant sources of the surface electronic states. Detailed information 
of the characterization of the Bi/InSb(001) surface sample is provided in Ref. \onlinecite{Ohtsubo_15}.

\begin{figure}[!htb]
\begin{center}
\centerline{\includegraphics[width=8.7cm]{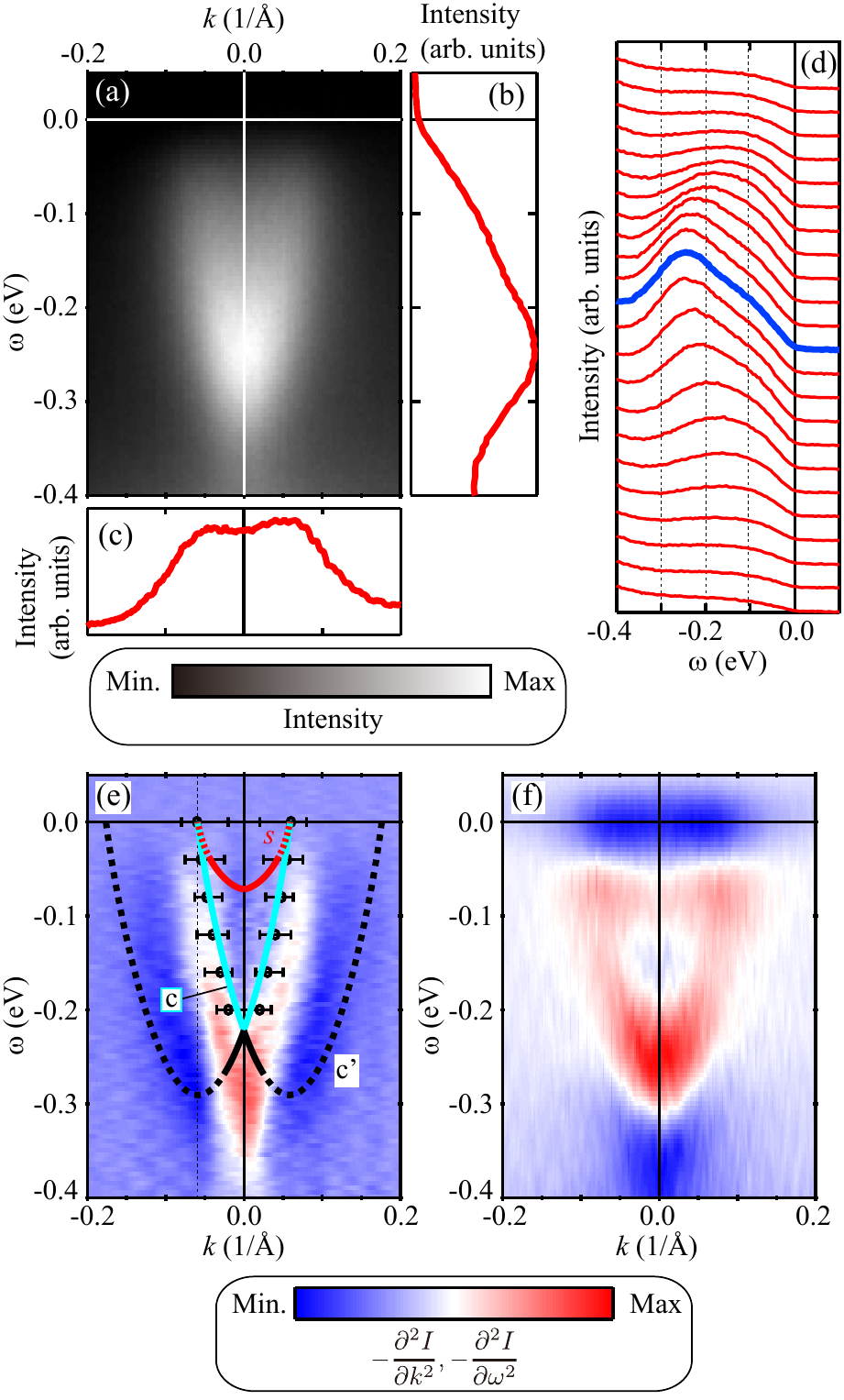}}
\caption{(a) Raw Bi/InSb(001) ARPES data for $\hbar\omega = 15$ eV. (b) An ARPES  
EDC at $k = 0$ (\AA$^{-1}$.) (c) ARPES MDC at $\omega = -0.05$ eV. (d) ARPES EDC
from $k = -0.16$ (bottom) to $+0.16$ (top) (\AA$^{-1}$.) The thick line is the normal-emission spectrum ($k = 0$ (\AA$^{-1}$).) (e) and (f)
Second-derivative ARPES images. Derivation was made along momentum in (e) and energy in (f). Circles and error 
bars in (e) indicate the MDC peak positions. Solid and dashed lines overlaid in (e) are the theoretical $s$ (red), $c$ 
(light blue) and $c$' (black) branch lines for $u = U/4t = 0.30$, $t = 1.22$ eV, and electronic density $n_e = 0.176$. Only 
for the solid-line $k$ ranges in (e) for which the exponents are negative in 
Fig. \ref{figure4} and Figs. \ref{figure5} and \ref{figure6} of Appendix \ref{APA} can they be seen in the ARPES image. 
(ARPES from the same experimental data as in Refs. \onlinecite{Ohtsubo_15} and \onlinecite{Ohtsubo_17}.).}
\label{figure3}
\end{center}
\end{figure}

ARPES measurements were performed at $\hbar\omega=15$ eV and taken at $8$ K in the
CASSIOP\'EE beamline of SOLEIL synchrotron. The photoelectron kinetic energy at $E_F$ and the 
overall energy resolution of the ARPES setup were calibrated by the Fermi edge of the photoelectron 
spectra from Mo foils attached to the sample. The energy resolution was $\sim$20 meV. The ARPES 
taken at $8$ K is shown in Fig. \ref{figure3}. 

The theoretical predictions reported in this paper refer to (i) the $(k,\omega)$-plane location of the 
high-energy Bi/InSb(001) MDC and EDC ARPES peaks and (ii) the value of the power-law SDS 
exponent $\alpha$ associated with the angle integration to detect the low-energy suppression 
of the photoelectron intensity that were performed at $k_y = 0.2\,\textrm{\AA}^{-1}$, near the 
boundary of the ($1 \times 3$) surface Brillouin zone ($0.23\,\textrm{\AA}^{-1}$).

\subsection{Criteria for agreement between ARPES and the present theory}
\label{Criteria}

Refs. \onlinecite{Ohtsubo_15} and \onlinecite{Ohtsubo_17} found strong experimental evidence that Bi/InSb(001) 
at $y$ momentum component $k_y = 0.2\,\textrm{\AA}^{-1}$ and temperature $8$ K displays 1D physics with 
an SDS exponent that for small $\vert\omega\vert < 0.10$ eV has values in the interval $\alpha\in [0.6,0.7]$. 

As discussed and justified below in Sec. \ref{OTHER}, the one-electron spectral properties of Bi/InSb(001)
are expected to be controlled  mainly by the interplay of one dimensionality and finite-range electron interactions, 
despite a likely small level of disorder. Consistent with an SDS exponent $\alpha$ larger than $1/8$ stemming 
from finite-range interactions \cite{Schulz_90}, here we use the MQIM-HO 
to predict one-electron spectral properties of Bi/InSb(001). 

As discussed in Sec. \ref{OTHER}, Bi/InSb(001) is a complex system and some of its experimental properties 
beyond those studied here may involve microscopic processes other than those described by the MQIM-HO 
and the Hamiltonian, Eq. (\ref{equ1}). This includes coupling to two-dimensional (2D) physics if 
$k_y = 0.2\,\textrm{\AA}^{-1}$ is smoothly changed to $k_ y =0$.

As reported in Sec. \ref{ERAE}, the MQIM-HO can describe both the low-energy TLL regime and
the spectral function, Eq. (\ref{equ2}), at high energies near the $(k,\omega)$-plane singularities.
At and in the vicinity of those singularities, the renormalization from its bare ${\tilde{\xi}}_c = \xi_c$ form
is determined by the large $x$ behavior of $V_c (x)$, Eq. (\ref{equ5}), and its sum rules, Eqs. (\ref{equ6}) 
and (\ref{equ7}), which refer to a high energy regime that goes well beyond the TLL limit.

Hence we can predict two properties of the one-electron spectral function : (i) the location in the 
$(k,\omega)$ plane of the experimentally observed high-energy peaks in the ARPES MDC and EDC and
(ii) the value of the low-energy SDS exponent $\alpha$. Our $T=0$ theoretical results describe the former 
high-energy experimental data taken at $8$ K for which the smearing of the spectral function singularities 
by thermal fluctuations is negligible. The quantitative agreement with the corresponding experimental data 
taken at fixed momentum $k_y = 0.2\,\textrm{\AA}^{-1}$ reached below provides further evidence of 1D 
physics and electron finite-range interactions in Bi/InSb(001).
\begin{widetext}
\begin{table}
\begin{center}
\begin{tabular}{|c||c|} 
\hline
Agreement & Description \\
\hline
First type & overall $(k,\omega)$-plane shapes of the theoretical
branch-line spectra, Eq. (\ref{equA1}), versus ARPES experimental spectra \\
\hline
Second type & $(k,\omega)$-plane location of the singularities corresponding
to negative exponents, Eq. (\ref{equA3}), versus ARPES peaks  \\
\hline
Third type & SDS exponent $\alpha$ from the dependence of the exponents, Eq. (\ref{equA3}), 
on ${\tilde{\xi}}_c$ versus its low-$\omega$ experimental value\\
\hline
\end{tabular}
\caption{Types of agreement between theory and experiments}
\label{table3}
\end{center}
\end{table} 
\end{widetext}

A first type of agreement of the theoretical branch-line energy spectra with the $(k,\omega)$-plane
shape of the ARPES image spectra must be reached for well-defined fixed values of electronic
density $n_e$ and interaction $u=U/4t$. Through Eq. (\ref{equA16}) of Appendix (\ref{APA}), these
uniquely determine the value of the bare charge parameter $\xi_c = \xi_c (u,n_e)$ to be used in the
$\xi_c\rightarrow {\tilde{\xi}}_c$ transformations suited to Bi/InSb(001). In addition, that
first type of agreement also determines the value of the transfer integral $t$.

The experimental values of the lattice spacing $a_0$ and of the momentum width of the spectra at $\omega =0$ 
provide the Fermi momentum $k_F = (\pi/2a_0)\,n_e$ and thus the electronic density $n_e$. At the density $n_e$, 
the ratio ${\tilde{W}}_s/{\tilde{W}}_c$ of the experimental energy bandwidths 
${\tilde{W}}_s \equiv \vert{\tilde{\omega}}_{s} (0)\vert$ of the $s$ branch line spectrum and 
${\tilde{W}}_c \equiv \vert{\tilde{\omega}}_c (0)\vert = \vert{\tilde{\omega}}_{c'} (0)\vert$ of the $c$ and $c'$ 
branch line spectra at momentum $k=0$ uniquely determine $u=U/4t$. (See such energy bandwidths in the 
sketch of the theoretical spin $s$ and charge $c$ and $c'$ branch lines in Fig. \ref{figure1}.) Finally, the 
experimental values of ${\tilde{W}}_s$ and ${\tilde{W}}_c$ determine the value of the transfer integral $t$.

As discussed below in Sec. \ref{BiARPES}, from the available experimental data it is not possible to trace the 
energy dispersion of the $s$ branch line. However, combining the experimental data on the EDC with kinematic 
constraints of the MDC provides information about the most probable value of the energy at which its bottom is 
located, which equals the branch line energy bandwidth ${\tilde{W}}_s$.

A second type of agreement is between the momentum interval and corresponding energy interval
for which the exponents ${\tilde{\zeta}}_c (k)$, ${\tilde{\zeta}}_{c'} (k)$, and ${\tilde{\zeta}}_{s} (k)$, Eq. (\ref{equA3}) of Appendix \ref{APA},
are negative and the $(k,\omega)$-plane location of the experimentally observed high-energy ARPES MDC and EDC 
peaks. That agreement must be reached at the fixed $u$ and $n_e$ values and corresponding bare charge parameter 
$\xi_c = \xi_c (u,n_e)$ value obtained from the first type of agreement. This second type of agreement is reached at some values 
of the integer quantum number $l>5$ in the large-$x$ potential $V_c (x)$ expression, Eq. (\ref{equ5}), 
and of the renormalized charge parameter ${\tilde{\xi}}_c$ (and thus of $\tan (\Phi)$, see Eqs. (\ref{equ12}) and (\ref{equ13})).

For the theoretically predicted high-energy ARPES peaks located on the $s$ branch line, there is only limited 
experimental information. Hence we start by finding the ${\tilde{\xi}}_c$ and $l>5$ values 
at which the second type of agreement is reached concerning
the momentum intervals where the exponents ${\tilde{\zeta}}_c (k)$ and ${\tilde{\zeta}}_{c'} (k)$ are negative 
and the corresponding $(k,\omega)$-plane location of the experimentally observed high-energy 
ARPES MDC and EDC peaks. Fortunately, it turns out that the
${\tilde{\xi}}_c$ values lead to a prediction of location in the $(k,\omega)$-plane of the 
high-energy ARPES peaks associated with the $s$ branch line that is consistent with the available experimental EDC data. 

This second type of agreement is reached for specific ${\tilde{\xi}}_c$ values. This then provides a prediction for the SDS exponent
$\alpha = (2-{\tilde{\xi}}_c^2)^2/(8{\tilde{\xi}}_c^2)$ obtained from a different {\it low-energy} experiment
that detects the suppression of the photoelectron intensity. That the SDS exponent $\alpha$ determined by 
the ${\tilde{\xi}}_c$ values for which the second type of agreement is reached 
is also that measured within the low-energy angle integrated photoemission intensity then becomes the 
required third type of agreement.

In the Lehmann representation of the spectral function, the first and second types of agreement correspond 
to the energy spectra and the overlaps of the one-electron matrix elements, respectively. 
The exponents in Eq. (\ref{equA3}) of Appendix \ref{APA} involved in the second type of agreement
depend {\it both} on ${\tilde{\xi}}_c$ and momentum-dependent phase shifts
${\tilde{\Phi}}_{c,c}(\pm 2k_F,q)$  and ${\tilde{\Phi}}_{c,s}(\pm 2k_F,q')$. 
There is no apparent direct relation between the high-energy ARPES MDC peaks and the low-energy SDS.
That the MQIM-HO describes the main microscopic mechanisms behind the specific one-electron
spectral properties of Bi/InSb(001) then requires that the third type of agreement is 
fulfilled. 

The three types of agreement between theory and experiment are explicitly described in Table \ref{table3}.

\subsection{Searching for agreement between theory and experiments}
\label{BiARPES}

\subsubsection{First type of agreement}
\label{BiARPES1}

The MDC spectral shape plotted in Fig. \ref{figure3}(c) displays two peaks centered at well defined 
Fermi momentum values $-k_F = -0.06$\,$\textrm{\AA}^{-1}$ and $k_F = 0.06$\,$\textrm{\AA}^{-1}$, respectively.
Furthermore, the experimental circles (with error bars) in Fig. \ref{figure3}(e) clearly indicate that the MDC peaks are
located on two lines that in the limit of zero energy start at such two Fermi momenta.
Since the experimental data lead to $\pi/a_0\approx 0.68\,\textrm{\AA}^{-1}$, one finds from
$k_F = (\pi/2a_0)\,n_e\approx 0.06\,\textrm{\AA}^{-1}$ a small electronic density, $n_e \approx 0.176$. 

The experimental value of the $c$ branch line energy bandwidth 
${\tilde{W}}_c$ is directly extracted from analysis of the experimental MDC data provided in Fig. \ref{figure3}(e). 
From analysis of the EDCs in Fig. \ref{figure3}(d) alone one finds
that there is a uncertainty $0.05\pm 0.05$ eV concerning the energy at which the bottom of the $s$ branch line 
is located. It is clear that in this energy region there is a hump that cannot be explained by assuming the single 
peak at $0.25$ eV, which refers to the bottom of the $c$ branch line. 

The zero-energy level of the theoretically predicted downward-convex parabolic-like dispersion of the 
$s$ branch line plotted in Fig. \ref{figure3}(e) (see also sketch depicted in Fig. \ref{figure1}) refers to the 
Fermi level. Hence the $s$ branch line energy bandwidth ${\tilde{W}}_s$ equals that of its bottom.
While the energy range uncertainty of that bottom energy is experimentally rather wide, 
one can lessen it by combining the experimental ARPES MDC intensity distribution 
shown in Fig. \ref{figure3}(c) with its kinematical constraints, which follow
from the finite-energy bandwidth of the theoretical $s$ branch line. One then finds that
the most probable value of the $s$ branch line bottom energy and thus
of ${\tilde{W}}_s$ is between $0.05$ and $0.10$ eV.

The maximum momentum width of the ARPES MDC intensity distribution shown in Fig. \ref{figure3}(c) for energy 
$\vert\omega\vert = 0.05$ eV allowed by such kinematic constraints involves the superposition of two maximum 
momentum widths $\Delta k$, centered at $-k_F$ and $k_F$, respectively. Within the MQIM-HO, these kinematical 
constraints explain the lack of spectral weight in well-defined $(k,\omega)$-plane regions shown in Fig. \ref{figure1}. 
Fortunately, the lines that limit such regions without spectral weight only involve the $s$ band dispersion spectrum.

In the case of the spectral weight centered at $-k_F$ and $k_F$, respectively, such kinematical constraints imply 
that for each energy value $\vert\omega\vert = -\omega$ the corresponding maximum 
momentum width reads,
\begin{eqnarray}
\Delta k & = & 2(k_F - k) \hspace{0.20cm}{\rm for}\hspace{0.20cm}
\vert\omega\vert = \vert{\tilde{\omega}}_{s} (k)\vert \hspace{0.20cm}{\rm and}\hspace{0.20cm}k \in ]0,k_F[
\nonumber \\
& = & 2(k_F + k) \hspace{0.20cm}{\rm for}\hspace{0.20cm}
\vert\omega\vert = \vert{\tilde{\omega}}_{s} (k)\vert \hspace{0.20cm}{\rm and}\hspace{0.20cm}k \in ]-k_F,0[
\nonumber \\
{\rm no} & & {\rm constraints}\hspace{0.20cm}{\rm for}\hspace{0.20cm}
\vert\omega\vert > {\tilde{W}}_s\hspace{0.20cm}{\rm and}\hspace{0.20cm}
\vert k\vert > k_F 
\label{equ30}
\end{eqnarray}
where ${\tilde{W}}_s = \vert{\tilde{\omega}}_{s} (0)\vert$ and the $s$ band dispersion spectrum ${\tilde{\omega}}_{s} (k)$ 
is given in Eq. (\ref{equA1}) of Appendix \ref{APA}. 

For $\vert\omega\vert\ll {\tilde{W}}_s$ the kinematical constraints, Eq. (\ref{equ30}), are those of a TLL, 
$\Delta k = 2\vert\omega\vert/v_s (k_F)$, consistent with $v_s (k_F) ={\rm min}(v_s (k_F),v_c (2k_F))$ \cite{Orgad_01}.
However, for energy $\vert\omega\vert = -\omega$ larger than the $s$ branch 
line energy bandwidth ${\tilde{W}}_s = \vert{\tilde{\omega}}_{s} (0)\vert$, which is that at which the $s$ branch line bottom is located 
in the experimental data, there are {\it no} kinematical constraints.

The absolute value of the derivative with respect to $k$ of the ARPES MDC intensity plotted in Fig. \ref{figure3}(c) 
increases in a $\vert k\vert$ interval $\vert k\vert\in [k_F,k_F+k_{\rm MDC}]$ and decreases for 
$\vert k\vert >k_{\rm MDC}$. Theoretically, the ARPES MDC intensity should be symmetrical 
around $k=0$. Its actual experimental shape then introduces a small uncertainty in the value
of $k_{\rm MDC}$. The relatively large intensity in the tails located at the momentum region
$\vert k\vert > k_{\rm MDC}$ is explained by the larger uncertainty in the $s$ branch line bottom energy
${\tilde{W}}_s$. Indeed, the ARPES MDC under consideration 
refers to an energy $\vert\omega\vert = 0.05$ eV within that uncertainty. And, as given in
Eq. (\ref{equ30}), there are {\it no} kinematic constraints for $\vert\omega\vert > {\tilde{W}}_s$.

One can then identify the most probable value of ${\tilde{W}}_s$ within its uncertainty interval as that for which
at the energy $\vert\omega\vert = 0.05$ eV the kinematic constraints would limit the ARPES MDC 
intensity to momentum values within the interval $\vert k\vert \leq k_{\rm MDC}$. The corresponding momenta $k=\pm k_{\rm MDC}$ 
are the inflection points at which the derivative of the variation of the ARPES MDC intensity with respect to 
$k$ changes sign in Fig. \ref{figure3}(c). The momentum width associated with $\vert k\vert \leq k_{\rm MDC}$ 
is thus that of the ARPES MDC shown in that figure if one excludes the tails. 

The corresponding maximum momentum width $\Delta k$, Eq. (\ref{equ30}), of the two overlapping spectral weights
centered at $k_F$ and $-k_F$, respectively, that at $\vert\omega\vert = 0.05$ eV would lead to the
kinematic constraint  $\Delta k = 2(k_{\rm MDC} - k_F)$, so that $\pm (k_F + \Delta k/2)
= \pm k_{\rm MDC}$. According to the kinematic constraints in Eq. (\ref{equ30}), this is fulfilled when at 
$k =\pm (k_F - \Delta k/2)=\pm (2k_F - k_{\rm MDC})$ so that the $s$ branch line energy spectrum reads 
$\vert{\tilde{\omega}}_{s} (k)\vert = - {\tilde{\omega}}_{s} (k) = 0.05$ eV.
Accounting for the combined $k_{\rm MDC}$ and ${\tilde{W}}_s$ uncertainties, the 
most probable value of the energy bandwidth ${\tilde{W}}_s$ is larger 
than $0.05$ eV and smaller than $0.10$ eV, as that of the theoretical $s$ branch line plotted in Fig. \ref{figure3}(e).

At electronic density $n_e = 0.176$ the best second type of agreement between theory and experiments 
discussed in the following is reached within that combined uncertainty by the $u=U/4t$ and $t$ values that
are associated with the energy bandwidth ${\tilde{W}}_s$ of such a theoretical $s$ branch line. They read $u= 0.30$ and $t = 1.22$ eV,
as determined from the corresponding ratio ${\tilde{W}}_s/{\tilde{W}}_c$ and experimental ${\tilde{W}}_c$ value in Fig. \ref{figure3}(e). 
Hence within the MQIM-HO the first type of agreement with the ARPES spectra is reached by choosing these
parameter values for the electronic density $n_e = 0.176$. 

\subsubsection{Second type of agreement}
\label{BiARPES2}

The second type of agreement involves the theoretical $\gamma = c,c',s$ exponents ${\tilde{\zeta}}_{\gamma} (k)$, 
Eq. (\ref{equA3}) of Appendix \ref{APA}. They are plotted for $u=0.30$ and $n_e=0.176$ as a function of the 
momentum $k$ in Fig. \ref{figure4}(a) for $l=6$ and in Fig. \ref{figure4}(b) for $l=12$. In Appendix \ref{APA}, they 
are plotted as a function of $k$ for several additional values of $l$.
\begin{figure}[!htb]
\begin{center}
\subfigure{\includegraphics[width=8.25cm]{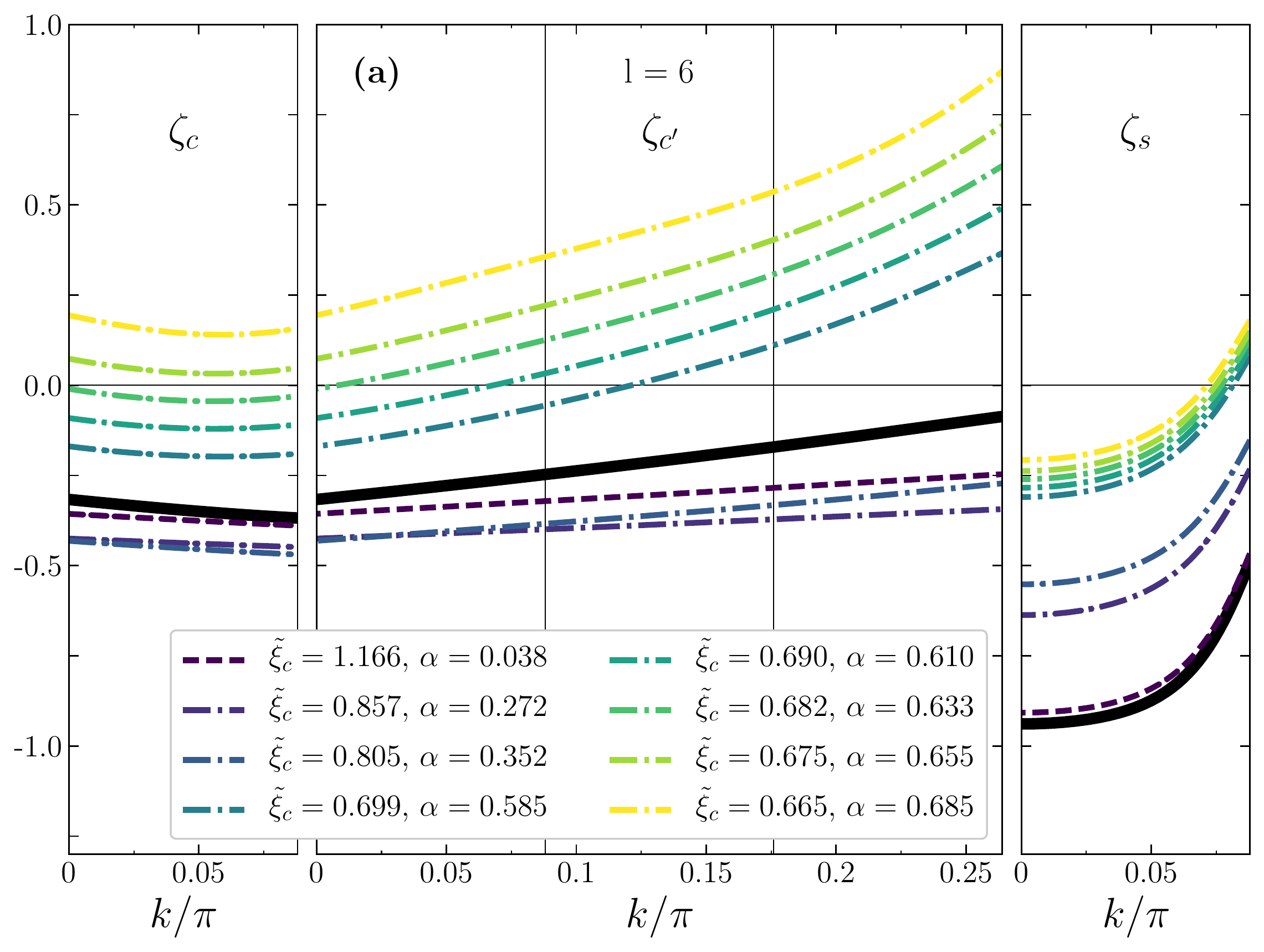}}
\subfigure{\includegraphics[width=8.25cm]{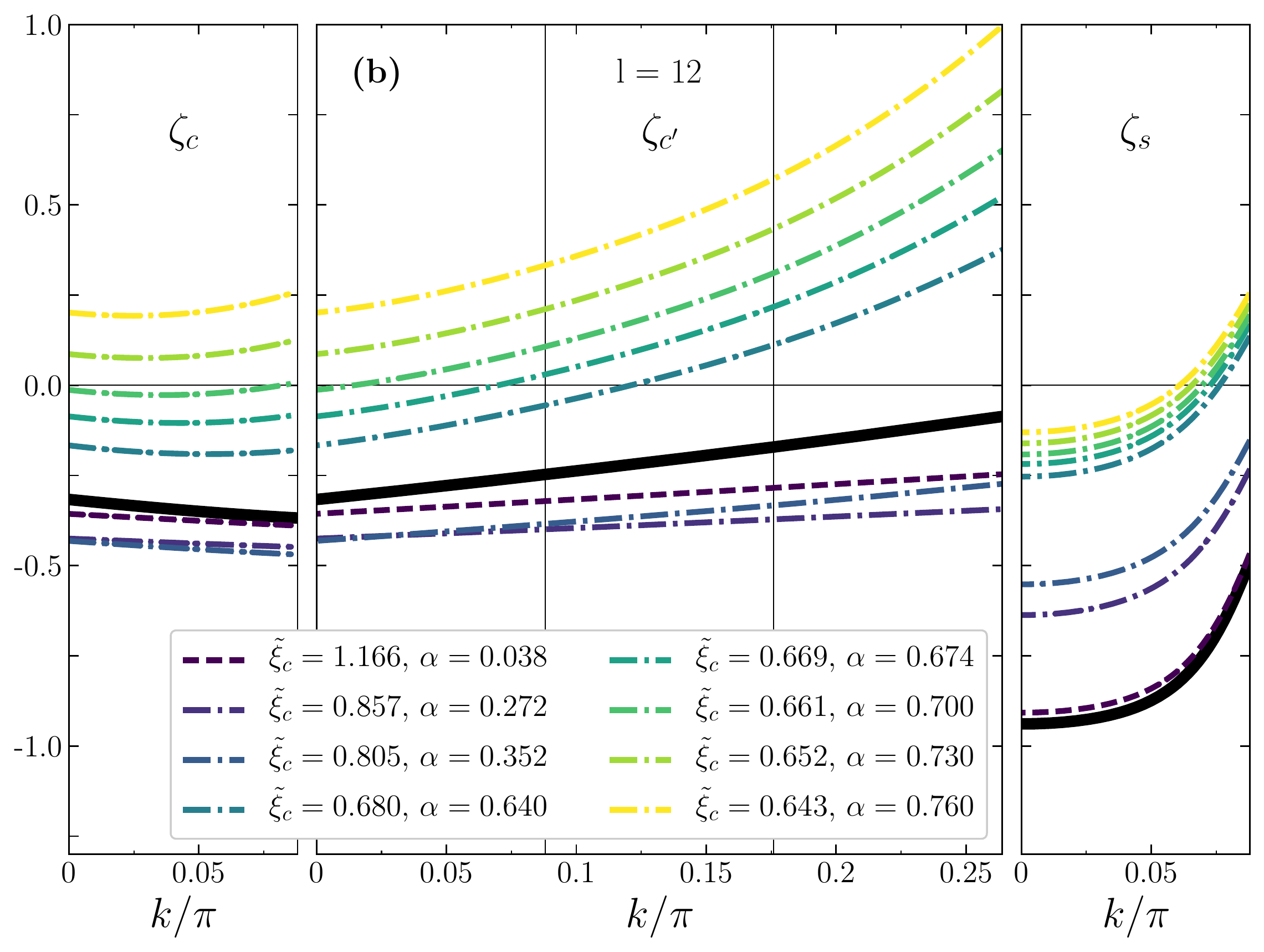}}
\caption{The exponents, Eq. (\ref{equA3}) of Appendix \ref{APA}, in the
spectral function, Eq. (\ref{equ2}), that control the line shape
near the theoretical $c$, $c'$, and $s$ branch lines in Fig. \ref{figure3}(e), respectively,
associated with the experimentally observed high-energy Bi/InSb(001) ARPES MDC and EDC 
peaks \cite{Ohtsubo_15,Ohtsubo_17}.
They are here plotted as a function of the momentum $k$ within the 
MQIM-HO for (a) $l=6$ and (b) $l=12$ at $u=0.30$, $n_e=0.176$, and 
several ${\tilde{\xi}}_c$ and $\alpha = (2-{\tilde{\xi}}_c^2)^2/(8{\tilde{\xi}}_c^2)$ values. The black solid lines
refer to the bare limit, ${\tilde{\xi}}_c =\xi_c$ ($\xi_c = 1.242$ and $\alpha_0= 0.017$).
The black dashed and the dashed-dotted lines correspond to $\alpha<1/8$ and $\alpha>1/8$ values, respectively.
Moreover, ${\tilde{\xi}}_c=0.805$, $0.857$, and $1.166$ refer to ${\tilde{\xi}}_c^{\oslash}=1/\xi_c$, 
${\tilde{\xi}}_c^{\ominus-}$, and ${\tilde{\xi}}_c^{\ominus+}$, respectively. The ${\tilde{\xi}}_c$
values of the lines whose negative exponents ranges agree with the experimentally observed high-energy ARPES $(k,\omega)$-plane
MDC and EDC peaks in Fig. \ref{figure3}(e) are those for which the $c'$ branch-line exponent crosses zero
between $k/\pi =0$ and $k/\pi \approx 0.07$. 
For $l=6$ and $l=12$ this refers to the small ${\tilde{\xi}}_c$ subintervals 
$\alpha = 0.610-0.633$ and $\alpha = 0.674-0.700$, respectively. (Such
limiting values are given in Tables \ref{table4} and \ref{table5} for all $l=6-12$
integers.)}
\label{figure4}
\end{center}
\end{figure}

The different curves in each figure are associated with different values of the charge parameter ${\tilde{\xi}}_c$ 
and thus of the SDS exponent $\alpha = (2-{\tilde{\xi}}_c^2)^2/(8{\tilde{\xi}}_c^2)$ and effective range $R_{\rm eff}$. 
The black solid lines refer to the bare charge parameter limit, ${\tilde{\xi}}_c=\xi_c=1.242$. The values
${\tilde{\xi}}_c=0.805$, $0.857$, and $1.166$ correspond to ${\tilde{\xi}}_c^{\oslash}=1/\xi_c$, 
${\tilde{\xi}}_c^{\ominus-}$, and ${\tilde{\xi}}_c^{\ominus+}$, respectively.

As justified in Sec. \ref{Criteria}, we start by finding the ${\tilde{\xi}}_c$ and $l>5$ values at which the second type of agreement 
is reached. It refers to the momentum intervals (and corresponding energy ranges) at which the exponents 
${\tilde{\zeta}}_c (k)$ and ${\tilde{\zeta}}_{c'} (k)$ are negative. Those are required to agree with 
the corresponding $(k,\omega)$-plane location of the experimentally observed high-energy ARPES MDC and EDC peaks 
in Figs. \ref{figure3}(e) and (f), respectively. This reveals that the integers $l>5$ and the values of the charge 
parameter ${\tilde{\xi}}_c$ and corresponding SDS exponent 
$\alpha = (2-{\tilde{\xi}}_c^2)^2/(8{\tilde{\xi}}_c^2)$ for which agreement is reached are those for which the 
exponents ${\tilde{\zeta}}_c (k)$ and ${\tilde{\zeta}}_{c'} (k)$, Eq. (\ref{equA3}) of Appendix \ref{APA},
in the spectral-function expression near the $c$ and $c'$ branch lines, Eq. (\ref{equ2}), are negative for 
$k \in [-2k_F + k_{Fc}^{\rm ex}, 2k_F - k_{Fc}^{\rm ex}]$ and $k \in [- k_{c'}^{\rm ex}, k_{c'}^{\rm ex}]$, respectively.

In the case of the exponent ${\tilde{\zeta}}_c (k)$, the momentum $k_{Fc}^{\rm ex}$  appearing in the interval
$k \in [-2k_F + k_{Fc}^{\rm ex}, 2k_F - k_{Fc}^{\rm ex}]$ is such that $k_{Fc}^{\rm ex}/k_F$ is vanishing or very small
in the thermodynamic limit. It is the experimental value of the small theoretical 
$c$ band momentum $k_{Fc}^0 = \pi n_{Fc}^0$ associated with the low density $n_{Fc}^0$ of $c$ particle 
scatterers near the $c$ band Fermi points $-2k_F$ and $2k_F$ considered in Sec. \ref{UL}.

Concerning the momentum interval $k \in [- k_{c'}^{\rm ex}, k_{c'}^{\rm ex}]$ for which the 
exponent ${\tilde{\zeta}}_{c'} (k)$ must be negative, there is a small uncertainty in
the value of $k_{c'}^{\rm ex}$. It is such that $k_{c'}^{\rm ex}\in [0,\delta k_0]$ where
$2\delta k_0 \approx 0.10\,\textrm{\AA}^{-1}$ in Fig. \ref{figure3}(e) is the momentum width of
the ARPES image crossed by the $c'$ branch line. 

This small uncertainty, which in the units used in the figures corresponds to $\delta k_0\in [0,0.07\pi]$, 
implies corresponding small uncertainties in the ${\tilde{\xi}}_c$ and $\alpha$ values at which for 
each $l$ agreement with the experimentally observed high-energy ARPES MDC and EDC
peaks is reached. The corresponding two limiting values of such ${\tilde{\xi}}_c$ and $\alpha$ uncertainties 
at which the exponent ${\tilde{\zeta}}_{c'} (k)$ in Fig. \ref{figure4} and in Figs. \ref{figure5} and \ref{figure6} of Appendix \ref{APA} 
crosses zero at $k\approx 0$ and $k\approx 0.07\pi$, respectively, are given in Table \ref{table4} for each integer $l=6-12$.

Following the direct relation between the $c$ and $c'$ branch lines spectra, that $\delta k_0\in [0,0.07\pi]$ ensures that 
the exponent ${\tilde{\zeta}}_c (k)$ is indeed negative for $k$ intervals $k \in [-2k_F + k_{Fc}^{\rm ex}, 2k_F - k_{Fc}^{\rm ex}]$ 
where $k_{Fc}^{\rm ex}/k_F\ll 1$, as also required for the second type of agreement to be reached.

Hence regarding the $c$ and $c'$ branch lines, agreement between theory and experiments is reached by the
${\tilde{\xi}}_c$ and $l>5$ values that in Fig. \ref{figure4} and in Figs. \ref{figure5} and \ref{figure6} of Appendix \ref{APA}
correspond to the $c'$ branch line exponent curves crossing zero between $k\approx 0$ and $k \approx \delta k_0\approx 0.07\pi$.
(In such figures only the two corresponding $c'$ branch line exponent curves crossing zero at $k\approx 0$ and
$k \approx \delta k_0\approx 0.07\pi$, respectively, are plotted.)
\begin{table}
\begin{tabular}{|c||c|c||c|c|} 
\hline
$l$ & ${\tilde{\xi}}_c$ & $\alpha$ & ${\tilde{\xi}}_c$ & $\alpha$ \\
\hline
& (${\tilde{\zeta}}_{c'} (0)=0$) & (${\tilde{\zeta}}_{c'} (0)=0$) & (${\tilde{\zeta}}_{c'} ({7\pi\over 100})=0$) & (${\tilde{\zeta}}_{c'} ({7\pi\over 100})=0$) \\
\hline
\hline
$6$ & $0.682$ & $0.633$ & $0.690$ & $0.610$ \\
\hline
$7$ & $0.673$ & $0.662$ & $0.680$ & $0.640$ \\
\hline
$8$ & $0.667$ & $0.680$ & $0.673$ & $0.660$ \\
\hline
$9$ & $0.663$ & $0.691$ & $0.672$ & $0.665$ \\
\hline
$10$ & $0.662$ & $0.695$ & $0.67$ & $0.670$ \\
\hline
$11$ & $0.661$ & $0.699$ & $0.669$ & $0.672$ \\
\hline
$12$ & $0.661$ & $0.700$ & $0.669$ & $0.674$ \\
\hline
\end{tabular}
\caption{The two values of the charge parameter ${\tilde{\xi}}_c$ and corresponding
SDS exponent $\alpha = (2-{\tilde{\xi}}_c^2)^2/(8{\tilde{\xi}}_c^2)$ that at each integer $l=6-12$ 
are those at which the exponent ${\tilde{\zeta}}_{c'} (k)$ plotted as a function of $k$ in 
Fig. \ref{figure4}(a) for $l=6$ and in Fig. \ref{figure4}(b) for $l=12$ crosses zero at $k\approx 0$ and $k\approx 0.07\pi$, respectively.
The same applies to the exponent ${\tilde{\zeta}}_{c'} (k)$ plotted for $l=7-11$ 
in Figs. \ref{figure5}(a) and (b) and \ref{figure6}(a)-(c) of Appendix \ref{APA}.}
\label{table4}
\end{table} 

The theoretical $s$ branch line exponent ${\tilde{\zeta}}_{s} (k)$, Eq. (\ref{equA3}) of Appendix \ref{APA}, does not
depend on the integer quantum number $l>5$. For the ${\tilde{\xi}}_c$ values for which the $c'$ branch line exponent 
curves cross zero between $k\approx 0$ and $k \approx \delta k_0\approx 0.07\pi$ in Fig. \ref{figure4} and in 
Figs. \ref{figure5} and \ref{figure6} of Appendix \ref{APA}, the exponent ${\tilde{\zeta}}_s (k)$ is negative in corresponding 
intervals $k \in ]-k_F + k_{Fs}^{*}, k_F - k_{Fs}^{*}[$ and thus positive for $\vert k\vert \in ](k_F - k_{Fs}^{*}),k_F]$. 
Here $k_{Fs}^{*}$ is a function of $n_e$, $u$, and ${\tilde{\xi}}_c$ and $k = \pm (k_F - k_{Fs}^{*})$ are the two 
momentum values at which ${\tilde{\zeta}}_{s} (k)$ vanishes. 

The predicted location at $k \in ]-k_F + k_{Fs}^{*}, k_F - k_{Fs}^{*}[$ of the ARPES MDC peaks associated with the 
$s$ branch line cannot be confirmed from the available experimental data. Indeed and as mentioned in Sec. \ref{Criteria}, 
it is not possible to extract from such data the dispersion of that line. However, the corresponding energy intervals 
$\vert\omega\vert\in [\vert{\tilde{\omega}}_{s} (k_F - k_{Fs}^{*})\vert,{\tilde{W}}_s]$ are consistent with the available 
experimental data from the EDCs in Fig. \ref{figure3}(d). Here $\vert\omega\vert ={\tilde{W}}_s =  \vert{\tilde{\omega}}_{s} (0)\vert$
is the bottom of the $s$ branch line energy, as estimated in Sec. \ref{BiARPES1} from the interplay of the kinematical constraints, 
Eq. (\ref{equ30}), and the ARPES MDC shown in Fig. \ref{figure3}(c) for $\vert\omega\vert = 0.05$ eV.

\subsubsection{Third type of agreement}
\label{BiARPES3}

From the above results we see that for $l = 6-12$ agreement with the experimentally observed high-energy
ARPES MDC and EDC peaks in Figs. \ref{figure3}(e) and (f) is reached by the exponents curves referring 
to ${\tilde{\xi}}_c$ and $\alpha$ values belonging to the small intervals reported in Table \ref{table5}. The overlap 
of the subintervals obtained for each $l=6-12$ given in that table then leads to the
theoretical predictions ${\tilde{\xi}}_c \in [0.66,0.69]$ and $\alpha \in [0.610-0.700]$. 

Table \ref{table5} also provides the corresponding intervals of the effective range $R_{\rm eff}$ in units
of the lattice spacing that refer to first and second types of agreements. The effective range dependence on 
the bare charge parameter $\xi_c = \xi_c (n_e,u)$, renormalized charge parameter ${\tilde{\xi}}_c$, and integer
quantum number $l>5$ values at which such agreements have been reached is defined by combining 
Eqs. (\ref{equ10}) and (\ref{equ26}). That table also provides the values of the length scale $2r_l$ in the same 
units whose dependence on $l$ is given in Eq. (\ref{equ29}). Upon increasing $l$ from $l=6$ to $l=12$, the 
effective range $R_c^{\rm eff}$ values for which there is agreement with the experiments change from 
$R_c^{\rm eff}\approx 5r_l$ to $R_c^{\rm eff}\approx r_l$, respectively. 

According to the analysis of Sec. \ref{BiARPES2},
agreement with the experimentally observed high-energy $(k,\omega)$-plane ARPES MDC and EDC 
peaks distribution has been reached for the SDS exponent range $\alpha \in [0.610-0.700]$.
The third type of agreement between theory and experiments defined in Sec. \ref{BiARPES2} is 
reached provided that such a predicted SDS exponent range agrees with the $\alpha$ values measured 
within the low-energy angle integrated photoemission 
intensity. An experimental uncertainty $\alpha = 0.65\pm 0.05$ of the SDS exponent was found for $-\omega <$ 0.1 eV 
in Ref. \onlinecite{Ohtsubo_15}. 

The remarkable quantitative agreement of the MQIM-HO predictions within the third criterion 
reported in Sec. \ref{BiARPES2} provides evidence of finite-range interactions playing an active role in the Bi/InSb(001) spectral 
properties and confirms the 1D character of its metallic states also found in Ref. \onlinecite{Ohtsubo_15}.
\begin{table}
\begin{tabular}{|c|c|c|c|c|c|} 
\hline
$l$ & ${\tilde{\xi}}_c$ & $\alpha$ & $R_{\rm eff}/a_0$ & $2r_l/a_0$ & $\sqrt{2\mu\gamma_c}$ \\
\hline
$6$ & $0.68-0.69$ & $0.610-0.633$ & $14.4-17.0$ & $6.0$ & $50.1$ \\
\hline
$7$ & $0.67-0.68$ & $0.640-0.662$ & $6.9-8.1$ & 6.3 & $140.5$ \\
\hline
$8$ & $0.67$ & $0.660-0.680$ & $5.0-5.8$ & $6.4$ & $377.2$ \\
\hline
$9$ & $0.66-0.67$ & $0.665-0.691$ & $4.0-4.8$ & 6.5 & $983.3$ \\
\hline
$10$ & $0.66-0.67$ & $0.670-0.695$ & $3.4-4.2$ & $6.5$ & $2.51\times 10^3$ \\
\hline
$11$ & $0.66-0.67$ & $0.672-0.699$ & $3.1-3.8$ & 6.5 & $6.29\times 10^3$ \\
\hline
$12$ & $0.66-0.67$ & $0.674-0.700$ & $2.9-3.5$ & $6.4$ & $1.56\times 10^4$ \\
\hline
\end{tabular}
\caption{The renormalized charge parameter ${\tilde{\xi}}_c$, SDS exponent $\alpha$, and
effective range $R_c^{\rm eff}$ intervals for which there is agreement between the 
$(k,\omega)$-plane regions where the theoretical branch lines display
singularities and the corresponding experimentally observed high-energy Bi/InSb(001) 
ARPES MDC and EDC peaks in Figs. \ref{figure3}(e) and \ref{figure3}(f) for 
$n_e = 0.176$, $u=0.30$, and $l=6-12$. As given in Table \ref{table4}, for each integer $l$ the 
smallest and largest ${\tilde{\xi}}_c$ value refers to the largest and smallest corresponding 
$\alpha = (2-{\tilde{\xi}}_c^2)^2/(8{\tilde{\xi}}_c^2)$ value, respectively. 
(The $\alpha$ values were derived using more digits in the ${\tilde{\xi}}_c$ values than given in the table.)
The values in units of $a_0 =1$ of the length scale $2r_l$, Eq. (\ref{equ29}), and related parameter 
$\sqrt{2\mu\gamma_c} =\sqrt{2}\,(2r_l)^{l-2\over 2}$ are also provided.}
\label{table5}
\end{table} 

\subsection{Interplay of relaxation processes with the momentum dependence of the exponents}
\label{Relaxation}

Here we discuss the physical mechanisms within the MQIM-HO that
underlie the dependence of the exponents ${\tilde{\zeta}}_c (k)$, ${\tilde{\zeta}}_{c'} (k)$, and
${\tilde{\zeta}}_{s} (k)$ on the charge parameter ${\tilde{\xi}}_c$ . These exponents are plotted in Fig. \ref{figure4} 
and in Figs. \ref{figure5} and \ref{figure6} of Appendix \ref{APA}.

In the bare charge parameter limit, ${\tilde{\xi}}_c=\xi_c$, the exponents being negative or positive just 
refers to a different type of power-law behavior near the corresponding charge and spin branch lines.
For ${\tilde{\xi}}_c < \xi_c$, this applies only to the spin $s$ branch line. It coincides with the edge of support 
of the one-electron removal spectral function that separates $(k,\omega)$-plane regions without
and with finite spectral weight. Hence conservation laws impose that, near that line, the spectral function 
remains of power-law form, Eq. (\ref{equ2}), for both intervals ${\tilde{\xi}}_c \in ]1/2,1[$ and 
${\tilde{\xi}}_c \in]1,\xi_c]$. As confirmed from an analysis of the $s$ branch-line exponents
plotted in Fig. \ref{figure4} and in Figs. \ref{figure5} and \ref{figure6} of Appendix \ref{APA}, the effect of 
decreasing the charge parameter ${\tilde{\xi}}_c$ from its initial bare value $\xi_c$ (and thus
increasing the SDS exponent $\alpha = (2-{\tilde{\xi}}_c^2)^2/(8{\tilde{\xi}}_c^2)$ from 
$\alpha_0 = (2 - \xi_c^2)/(8\xi_c^2)\in [0,1/8]$) is merely to increase the spin branch line exponent 
${\tilde{\zeta}}_{s} (k)$. Except for two regions near $-k_F$ and $k_F$ corresponding to 
$\vert k\vert \in [(k_F - k_{Fs}^{*}),k_F]$, that exponent remains negative, so that the singularities 
prevail. In the complementarily small momentum regions near $\pm k_F$ defined by 
$\vert k\vert \in [(k_F - k_{Fs}^{*}),k_F]$ where the exponent is positive, the line shape remains of 
power-law type.

Analysis of the $c$ and $c'$ branch-line exponents curves plotted in the same figures reveals that the 
situation is different for the one-electron removal spectral function in the vicinity of the charge $c$ and $c'$ 
branch lines, Eq. (\ref{equ2}). These are located in the continuum of the one-electron spectral function. The 
physics is though different for the subintervals ${\tilde{\xi}}_c \in ]1,\xi_c]$ and ${\tilde{\xi}}_c \in ]1/2,1[$, 
respectively.

Smoothly decreasing ${\tilde{\xi}}_c$ from its initial bare value $\xi_c$ to ${\tilde{\xi}}_c\rightarrow 1$, 
produces effects quite similar to those of increasing $U$ within the 1D Hubbard model to 
$U\rightarrow\infty$ \cite{Carmelo_18}. Indeed, these changes render ${\tilde{\zeta}}_c (k)$ and 
${\tilde{\zeta}}_{c'} (k)$ more negative and lead to an increase of the width of the $k$ intervals in which 
they are negative. Within the ${\tilde{\xi}}_c\in ]1,\xi_c]$ interval, a large number of ${\tilde{\xi}}_c=\xi_c$
conservation laws that are behind the factorization of the scattering $S$ matrix into two-particle 
scattering processes survive, which tend to prevent the $c$ impurity from undergoing relaxation processes.
Hence the lifetimes $\tau_{c} (k)$ and $\tau_{c'} (k)$ in Eq. (\ref{equ2}) are very large for the $k$ intervals for which
the corresponding branch line exponents are negative, so that the expression given in the equation 
for the spectral function near the $\beta = c,c'$ branch lines is nearly power-law like, 
${\tilde{B}} (k,\omega)\propto \left({\tilde{\omega}}_{\beta} (k)-\omega\right)^{{\tilde{\zeta}}_{\beta} (k)}$.

The effects of the finite-range interactions increase upon decreasing
${\tilde{\xi}}_c$ within the interval ${\tilde{\xi}}_c \in [{\tilde{\xi}}_c^{\oslash},1[$ where
${\tilde{\xi}}_c^{\oslash}=1/\xi_c = 0.805$ for $n_e = 0.176$ and $u=0.30$.
Indeed, smoothly decreasing ${\tilde{\xi}}_c$ within that interval tends to remove an increasing number of 
conservation laws, which strengthens the effects of the impurity relaxation processes. 
Such effects become more pronounced when $\Delta a/{\tilde{a}}\in ]-1,0[$ and $\tan(\Phi)>0$, 
upon further decreasing ${\tilde{\xi}}_c$ within the interval ${\tilde{\xi}}_c \in]1/2,{\tilde{\xi}}_c^{\oslash}]$. 

In the $k$ intervals for which the $\beta = c,c'$ branch line exponents ${\tilde{\zeta}}_{\beta} (k)$ 
remain negative, the lifetimes $\tau_{\beta} (k)$ in Eq. (\ref{equ2}) remain large and the
$c$ impurity relaxation processes only slightly broaden the spectral-function power-law singularities, 
as given in Eq. (\ref{equ2}). For the complementary $k$ ranges for which such exponents become positive 
upon decreasing ${\tilde{\xi}}_c$ and thus increasing $\alpha$, the high-energy singularities are rather washed 
out by the relaxation processes. 

As confirmed by analysis of the curves plotted in Fig. \ref{figure4} and in Figs. \ref{figure5} and \ref{figure6} of 
Appendix \ref{APA}, starting at $\vert k\vert = 3k_F-k_{Fc}^0$ and downwards, smoothly decreasing 
${\tilde{\xi}}_c$ from ${\tilde{\xi}}_c^{\oslash}$ first gradually enhances the $k$ domains where ${\tilde{\zeta}}_{c'} (k)$ 
is positive. Further decreasing ${\tilde{\xi}}_c$ after the $c'$ branch line singularities are fully washed out leads 
to the emergence of a $c$ branch line $k$ domain starting at $\vert k\vert=0$ and upwards in which that line 
singularities are finally fully washed out up to $\vert k\vert = k_F-k_{Fc}^0$ below a smaller ${\tilde{\xi}}_c$ value.

\section{Discussion and concluding remarks}
\label{DISCONCL}

\subsection{Discussion of other effects and properties outside the range of the MQIM-HO}
\label{OTHER}

As reported in Sec. \ref{PREP}, the ARPES data were taken at $8$ K and the angle integrations to detect 
the suppression of the photoelectron intensity were performed at $k_y = 0.2\,\textrm{\AA}^{-1}$, near the 
boundary of the ($1 \times 3$) surface Brillouin zone ($0.23\,\textrm{\AA}^{-1}$).

As shown in Fig. 2(b) of Ref. \onlinecite{Ohtsubo_15}, at $k_y = 0.2\,\textrm{\AA}^{-1}$ there is an energy 
gap between the spectral features studied in this paper within a 1D theoretical framework and a bulk valence 
band. Due to that energy gap, the coupling between the two problems is negligible, which justifies that the 
system studied here corresponds to 1D physics.

Smoothly changing $k_y$ from $k_y = 0.2\,\textrm{\AA}^{-1}$ to $k_y=0$ corresponds to smoothly turning 
on the coupling to the 2D physics.  As shown in Fig. 2(a) of Ref. \onlinecite{Ohtsubo_15}, at $k_y = 0$ the 
energy gap between the spectral features studied in this paper and that bulk valence band has been closed. 
The study of the microscopic mechanisms involved in the physics associated with turning on the coupling to 
the 2D physics by smoothly changing $k_y$ from $k_y = 0.2\,\textrm{\AA}^{-1}$ to $k_y=0$ is an interesting 
problem that deserves further investigation.

Another interesting open problem refers to theoretical prediction of the MDC for extended momentum intervals and 
of the EDC for corresponding energy ranges. The universal form of the spectral function near the singularities, Eq. (\ref{equ2}), 
is determined by the large $x$ behavior of the potential $V_c (x)$, Eq. (\ref{equ5}), which follows from that of the 
potential $V_e (r)$ in Eq. (\ref{equ1}), and potential sum rules, Eqs. (\ref{equ6}) and (\ref{equ7}). As reported in 
Eqs. (\ref{equ12}) and (\ref{equ13}), the value of the renormalized charge parameter ${\tilde{\xi}}_c$ behind the 
renormalization of the phase shifts in the exponents of that spectral function expression, Eq. (\ref{equ2}), is 
indeed controlled by the value of the initial bare charge parameter $\xi_c = \xi_c (n_e,u)$, the integer quantum 
number $l>5$ associated with the potential $V_c (x)$ large-$x$ behavior, Eq. (\ref{equ5}), and the zero-energy 
phase $\Phi$ determined by that potential sum rules, Eqs. (\ref{equ6}) and (\ref{equ7}). Plotting a MDC for extended 
momentum intervals and an EDC for corresponding energy ranges is a problem that involves non-universal properties 
of the one-electron removal spectral function. This would require additional information of that function in $(k,\omega)$-plane
regions where it is determined by the detailed non-universal dependence on $r$ of the specific electronic potential 
$V_e (r)$ {\it suitable} to Bi/InSb(001). 

Another interesting issue refers to the validity of the MQIM-HO to describe the Bi/InSb(001) one-electron spectral 
properties. The question is whether the interplay of one dimensionality and electron finite-range interactions 
is indeed the main microscopic mechanism behind such properties. As in all lattice electronic condensed matter systems, 
it is to be expected that there are both some degree of disorder effects and electron-electron effects in the Bi/InSb(001) 
physics. However, we can provide evidence that the interplay of latter effects with the Bi/InSb(001) metallic states 
one dimensionality is the dominant contribution to the one-electron removal spectral properties.

The first strong evidence that this is so is the experimentally observed universal power-law scaling of the 
spectral intensity $I (\omega,T)$. (Here $\omega =0$ refers to the Fermi-level energy.) For instance, at 
$\omega =0$ and finite $T$ and at $T=0$ and low $\omega$ it was found in Ref. \onlinecite{Ohtsubo_15} to 
have the following TLL behaviors for Bi/InSb(001),
\begin{equation}
I (0,T) \propto T^{\alpha} \hspace{0.30cm}{\rm and}\hspace{0.30cm}
I (\omega,0) \propto \vert\omega\vert^{\alpha} \, ,
\label{equ31}
\end{equation}
respectively, where $\alpha$ is the SDS exponent. 

If there were important effects from disorder, its interplay with electron-electron interactions 
would rather give rise in the case of 1D and quasi-1D systems to a spectral intensity $I (\omega,T)$ with 
the following behaviors \cite{Rollbuhler_01,Mishchenko_01,Bartoscha _02},
\begin{eqnarray}
I (0,T) & \propto & e^{-\sqrt{C_0^2\over 16\pi D_0 T}} \hspace{0.20cm}{\rm and}
\nonumber \\
I (\omega,0) & \propto & \vert\omega\vert^{1/2} {\sqrt{32\pi D_0}\over C_0}\,e^{-{C_0^2\over 32\pi D_0 \vert\omega\vert}} \,,
\label{equ32}
\end{eqnarray}
for $\omega \ll C_0^2/(32\pi D_0)$. Here $D_0\propto l$ is the {\it bare} diffusion coefficient and $C_0$ is a constant that 
depends on the effective electron-electron interaction and electronic density.

The behaviors in Eq. (\ref{equ32}) are qualitatively different from those reported in Eq. (\ref{equ31}), which are those 
experimentally observed in Bi/InSb(001). This holds specially for $I (0,T)$, in which case disorder effects {\it cannot} 
generate such a temperature power-law scaling. Also the experimentally found behavior 
$I (\omega,0) \propto \vert\omega\vert^{\alpha}$ disagrees with that implied by Eq. (\ref{equ32}).

Further, in the limit of low $\omega$ and $T$, the MQIM-HO describes the corresponding TLL limit in
which the universal power-law scaling of the spectral intensity $I (\omega,T)$ has the behaviors 
reported in Eq. (\ref{equ31}). Theoretically, the value of the SDS exponent $\alpha$ 
depends on those of the electronic density $n_e$, the interaction $u=U/4t$, and the renormalized 
charge parameter ${\tilde{\xi}}_c$. Within the MQIM-HO phase shifts constraints, its values span the 
intervals $\alpha \in [\alpha_0,1/8[\,;]1/8,49/32[$. 

The theoretically predicted $\alpha$ value has been determined from agreement of the $T=0$ one-electron 
spectral function, Eq. (\ref{equ2}), with the ARPES peaks location in the $(k,\omega)$-plane. The quantitative agreement
then reached refers to the experimental value $\alpha =0.65\pm 0.05$ obtained for $I (\omega,0) \propto \vert\omega\vert^{\alpha}$
at $\vert\omega\vert<0.1$ eV in Ref. \onlinecite{Ohtsubo_15}. For low temperatures and $\omega =0$ the MQIM-HO also
leads to the $I (0,T)$ behavior given in Eq. (\ref{equ31}). 

Finally, despite bismuth Bi, indium In, and antimony Sb being heavy elements, the present 1D surface metallic states do not show any 
detectable spin-orbit coupling effects and nor any related Rashba-split bands. In this regard, it is very important to distinguish the system studied in this paper with 1-2 monatomic layers of Bi thickness whose ARPES data were first reported in Ref. \onlinecite{Ohtsubo_15} from the system with a similar chemical name which was studied in 
Ref. \onlinecite{Kishi_17}, which refers to 5-20 monatomic layers of Bi thickness. One expects, and indeed observes, significant qualitative differences in these two systems.

\subsection{Concluding remarks}
\label{CONCL}

In this paper we have discussed an extension of the MQIM-LO used in the theoretical
studies of the ARPES in the line defects of MoSe$_2$ \cite{Ma_17}. This MQIM-type approach 
\cite{Imambekov_09,Imambekov_12} accounts only for the renormalization of the leading term in 
the effective range expansion of the charge-charge phase shift, Eq. (\ref{equ4}). As shown in 
Ref. \onlinecite{Cadez_19}, this is a good approximation if the effective range of the interactions 
of the $c$ particles and the $c$ impurity is of about one lattice spacing. 

The MQIM-HO developed in this paper accounts for the renormalization of the higher terms 
in the effective range expansion of the charge-charge phase shift, Eq. (\ref{equ4}).
It applies to a class of 1D lattice electronic systems described by the Hamiltonian, Eq. (\ref{equ1}),
which has longer range interactions. The quantum problem described by that Hamiltonian
is very involved in terms of many-electron interactions. However, we found that a key simplification is
the unitary limit associated with the scattering of the fractionalized charged particles by the $c$ impurity. 
We have shown a theory based on the MQIM-HO with finite-range interactions, Eq. (\ref{equ1}), applies 
to the study of {\it some} of the one-electron spectral properties of Bi/InSb(001) measured at $y$ 
momentum component $k_y = 0.2\,\textrm{\AA}^{-1}$ and temperature $8$ K.

Consistent with the relation of the electron and $c$ particle representations
discussed in Appendix \ref{RECP}, the form of the attractive potential $V_c (x)$ associated with the 
interaction of the $c$ particles and the $c$ impurity at a distance $x$ is determined by that of 
the electronic potential $V_e (r)$ in Eq. (\ref{equ1}). The universal behavior of the spectral function near the singularities 
given in Eq. (\ref{equ2}) whose $(k,\omega)$-plane location refers to that of the experimentally observed high-energy 
ARPES peaks, is determined by the large $x$ behavior of $V_c (x)$ shown in Eq. (\ref{equ5}) and sum rules, 
Eqs. (\ref{equ6}) and (\ref{equ7}). Otherwise the spectral function expression in the continuum is not universal.

Despite the limited available experimental information about the ARPES peaks located on the spin branch line,  
we have shown that all the three criteria associated with the different types of agreement between theory and 
experiments considered in Sec. \ref{Criteria} are satisfied. This provides further evidence to that given in Ref. 
\onlinecite{Ohtsubo_15} for the interplay of one dimensionality and finite-range interactions playing an important 
role in the one-electron spectral properties of the metallic states in Bi/InSb(001).

\acknowledgements

J. M. P. C. acknowledges the late Adilet Imambekov for discussions that were helpful in producing
the theoretical part of the paper. Y. O. and S.-i. K. thank illuminating discussions to obtain and analyze 
the ARPES dataset by Patrick Le F\`evre, Fran\c{c}ois Bertran, and Amina Taleb-Ibrahimi. J. M. P. C. would like to 
thank Boston University's Condensed Matter Theory Visitors Program for support and the hospitality of MIT. 
J. M. P. C. and T. \v{C}. acknowledge the support from Funda\c{c}\~ao para a Ci\^encia e
Tecnologia (FCT) through the Grants UID/FIS/04650/2013, 
PTDC/FIS-MAC/29291/2017, and SFRH/BSAB/142925/2018, NSAF U1530401 and computational resources 
from Computational Science Research Center (CSRC) (Beijing), and the National Science Foundation of
China (NSFC) Grant 11650110443.

\appendix

\section{Useful quantities}
\label{APA}

The spectra of the $\gamma =s,c,c'$ branch lines in the expressions for the one-electron removal spectral function 
in Eq. (\ref{equ2}) have for the MQIM-HO the same general form as for the MQIM-LO\cite{Ma_17,Cadez_19} and read,
\begin{eqnarray}
{\tilde{\omega}}_{s} (k) & = & {\tilde{\varepsilon}}_s (k) = \varepsilon_{s} (k) \leq 0\hspace{0.20cm}{\rm for}
\hspace{0.20cm} k = -q' \in [-k_F,k_F] 
\nonumber \\
{\tilde{\omega}}_c (k) & = & {\tilde{\varepsilon}}_c (\vert k\vert + k_F)\leq 0
\hspace{0.20cm}{\rm for}
\nonumber \\
k & = & k_c = -{\rm sgn}\{k\} k_F - q \in [-k_F,k_F] 
\nonumber \\
{\tilde{\omega}}_{c'} (k) & = & {\tilde{\varepsilon}}_c (\vert k\vert - k_F) \leq 0
\hspace{0.20cm}{\rm for}
\nonumber \\
k & = & k_{c'} = {\rm sgn}\{k\} k_F - q \in [-3k_F,3k_F] \, .
\label{equA1}
\end{eqnarray}
Here ${\tilde{\varepsilon}}_s (q')$ and ${\tilde{\varepsilon}}_c (q)$ are the $s$ and
$c$ particle energy dispersions, respectively. For the $c$ and $s$ band momentum intervals
at which the $c$ and $s$ impurities, respectively, are created under one-electron
excitations the energy dispersions and corresponding group velocities read,
\begin{eqnarray}
{\tilde{\varepsilon}}_c (q) & = & \left(1 + \beta_c\right)\varepsilon_c (q)\hspace{0.20cm}{\rm for}\hspace{0.20cm} 
q \in [-2k_F,2k_F]
\nonumber \\
{\tilde{\varepsilon}}_s (q') & = & \varepsilon_s (q') \hspace{0.20cm}{\rm for}\hspace{0.20cm} 
q' \in [-k_F,k_F] 
\nonumber \\
{\tilde{v}}_{\beta} (q) & = & {\partial {\tilde{\varepsilon}}_{\beta} (q)\over\partial q}
\, ,\hspace{0.20cm}
v_{\beta} (q) = {\partial\varepsilon_{\beta} (q)\over\partial q}\, ,\hspace{0.20cm}\beta = c, s \, .
\label{equA2}
\end{eqnarray}

Here the bare ${\tilde{\xi}}_c = \xi_c$ energy dispersions $\varepsilon_c (q)$ and $\varepsilon_s (q')$ are defined below
and $\beta_ c = \sqrt{1 + \alpha_c} -1$ where $\alpha_c$ is given in Eq. (\ref{equC6}) of Appendix \ref{RECP}. 
It reads $\beta_ c = (\xi_c^2 - {\tilde{\xi}}_c^2)/{\tilde{\xi}}_c^2$ for ${\tilde{\xi}}_c>{\tilde{\xi}}_c^{\,\,\breve{}}$ whereas 
$\beta_ c = (2 - \xi_c^2)/\xi_c^2$ for the range ${\tilde{\xi}}_c<{\tilde{\xi}}_c^{\,\,\breve{}}$ of most interest for our studies
where ${\tilde{\xi}}_c^{\,\,\breve{}} = \xi_c^2/\sqrt{2}$. For the latter range, its limiting behaviors are
\begin{eqnarray}
\beta_c & = & {U\over 4\pi\,t\,\sin k_F}
\hspace{0.20cm}{\rm for}\hspace{0.20cm}u\ll 1
\nonumber \\
& = & 1 - {8\,\ln 2\over \pi\,U}\,2t\sin (2k_F)\hspace{0.20cm}{\rm for}\hspace{0.20cm}u\gg 1 \, .
\nonumber 
\end{eqnarray}

The renormalization ${\tilde{\varepsilon}}_c (q) = \left(1 + \beta_c\right)\varepsilon_c (q)$ in
Eq. (\ref{equA2}) is related to the expression $\lim_{k\rightarrow 0}{\cal{V}}_c (k) = {\pi\over 2}\,\alpha_c\,v_c (2k_F)$, 
Eq. (\ref{equC6}) of Appendix \ref{RECP}, and
the corresponding ratio ${\tilde{v}}_c (2k_F)/v_c (2k_F)=\sqrt{1 + \alpha_c}=(1+\beta_c)$ 
of the renormalized and bare $c$ band Fermi velocities \cite{Carmelo_TT}.
That the spin dispersion ${\tilde{\varepsilon}}_s (q')$ remains invariant under finite-range interactions whereas
the charge dispersion bandwidth ${\tilde{W}}_c^p = - {\tilde{\varepsilon}}_c (0)$ and the charge Fermi velocity 
${\tilde{v}}_c (2k_F)$ are slightly increased as the range of interactions increases is known from numerical 
studies \cite{Hohenadler_12}. (See the related charge and spin spectra in Fig. 7 of Ref. \onlinecite{Hohenadler_12} 
and the corresponding discussion.)

\begin{figure}
\begin{center}
\subfigure{\includegraphics[width=8.25cm]{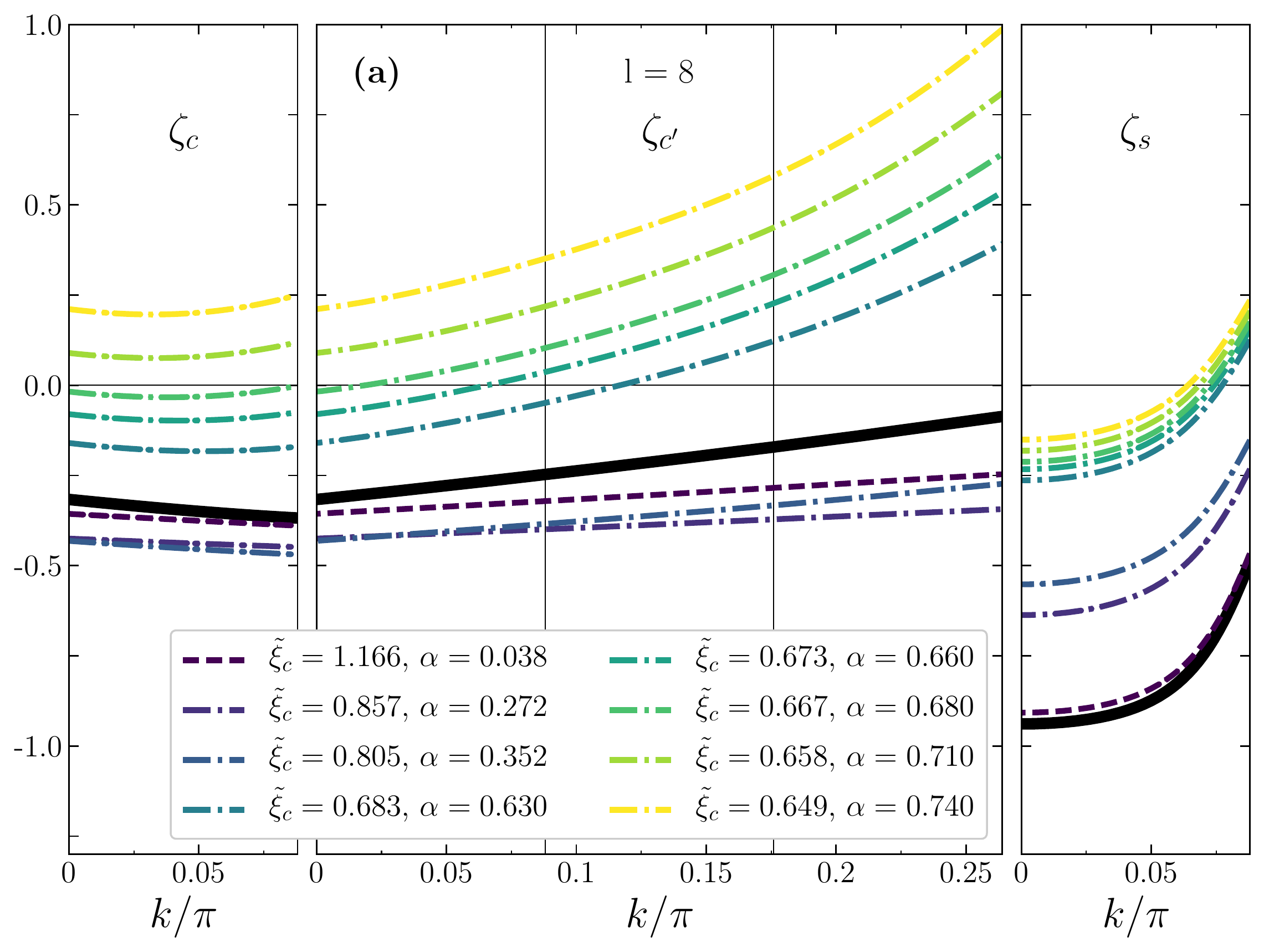}}
\subfigure{\includegraphics[width=8.25cm]{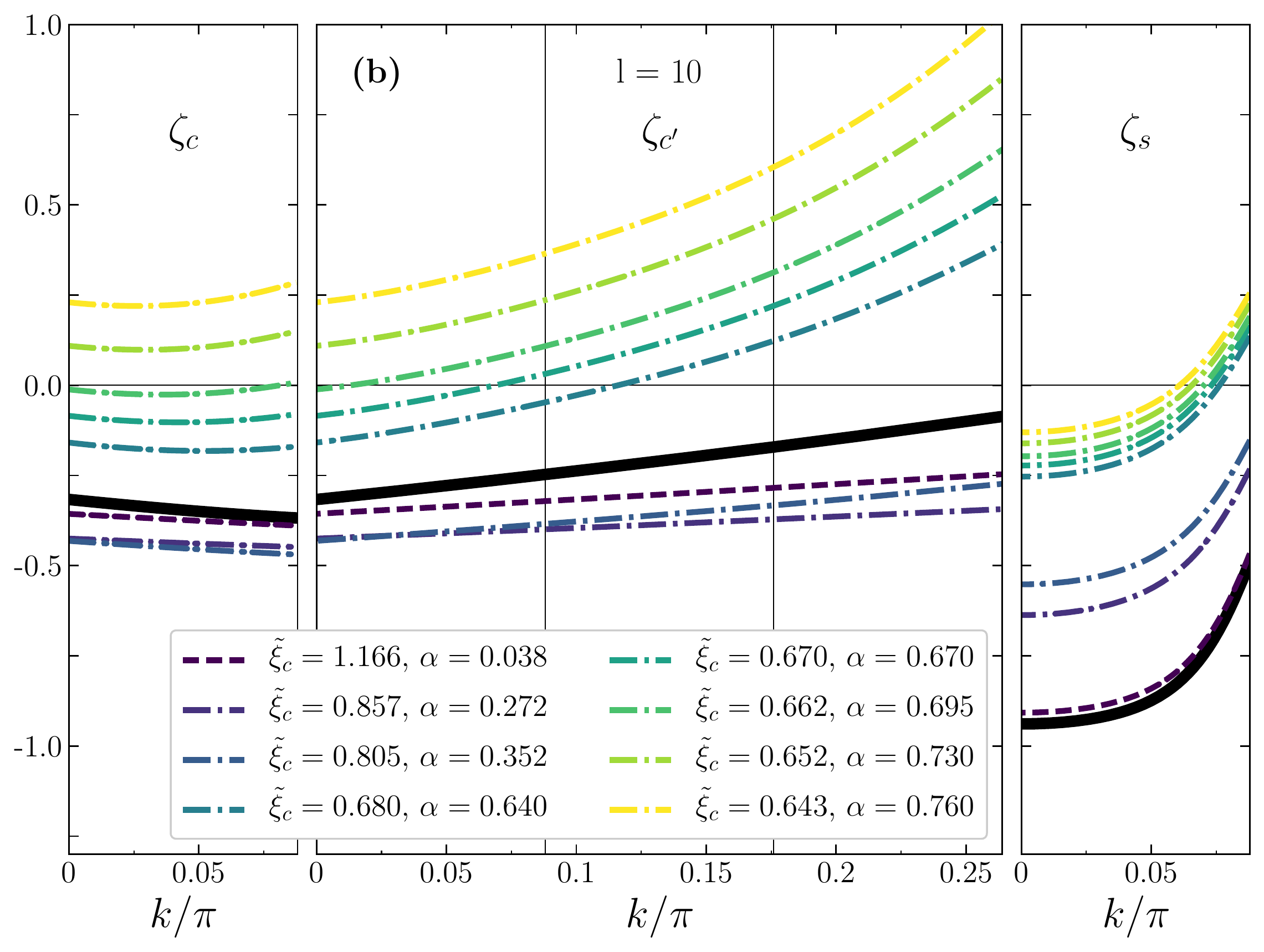}}
\caption{The same exponents as in Fig. \ref{figure4} for (a) $l=8$ and (b) $l=10$. 
The choice of the ${\tilde{\xi}}_c$ intervals corresponding to the lines whose negative 
exponents ranges agree with the experimentally observed high-energy ARPES $(k,\omega)$-plane
MDC and EDC peaks in Fig. \ref{figure3}(e) obeys the same criterion as in that figure.
For $l=8$ and $l=10$, such intervals whose limiting values are given
in Tables \ref{table4} and \ref{table5} are $\alpha = 0.660-0.680$ and $\alpha = 0.670-0.695$, respectively.}
\label{figure5}
\end{center}
\end{figure}

\begin{figure}
\begin{center}
\subfigure{\includegraphics[width=8.25cm]{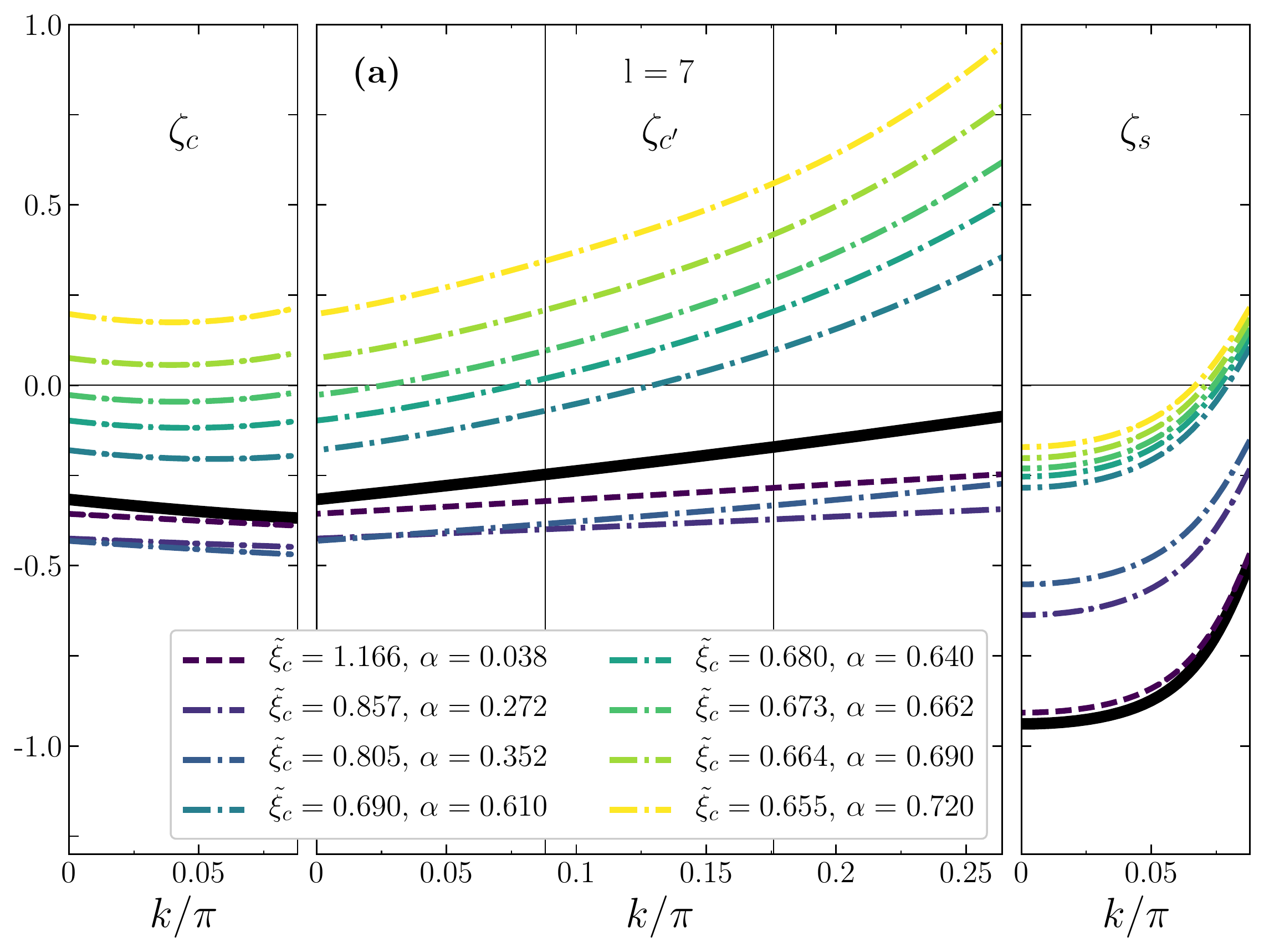}}
\subfigure{\includegraphics[width=8.25cm]{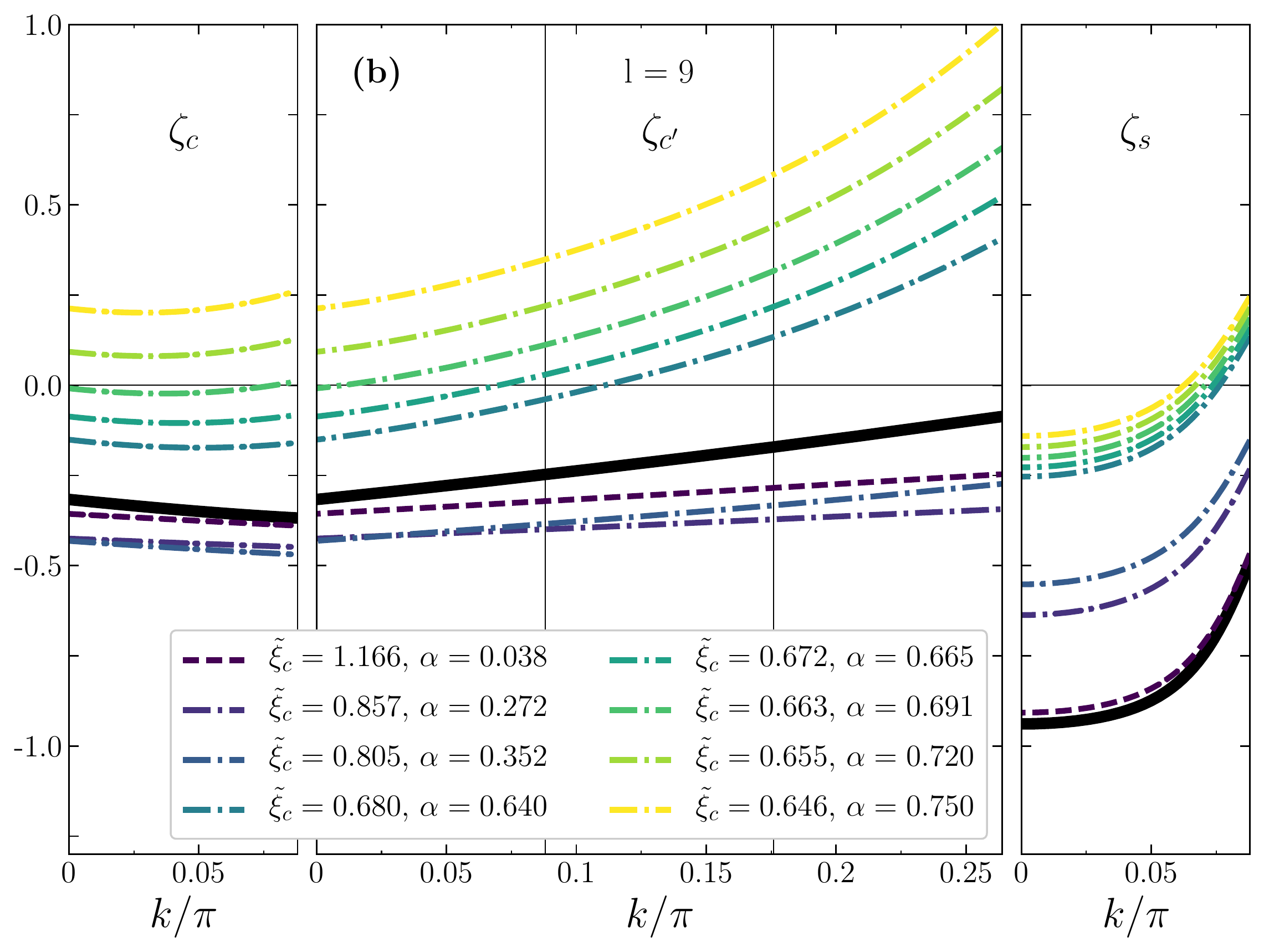}}
\subfigure{\includegraphics[width=8.25cm]{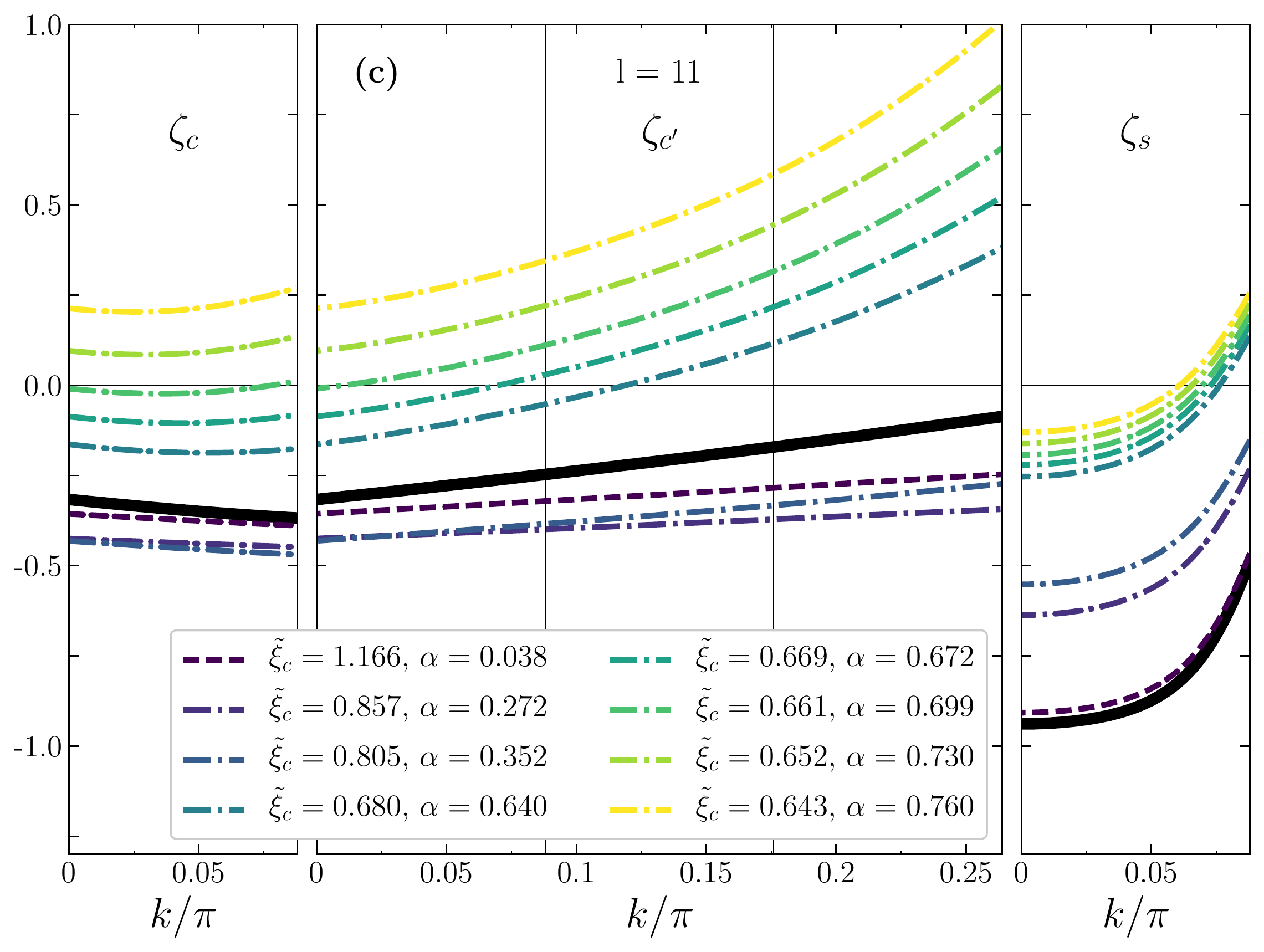}}
\caption{The same exponents as in Figs. \ref{figure4} and \ref{figure5} for (a) $l=7$, (b) $l=9$, and (c) $l=11$.
The choice of the ${\tilde{\xi}}_c$ intervals corresponding to the lines whose negative 
exponents ranges agree with the experimentally observed high-energy ARPES $(k,\omega)$-plane
MDC and EDC peaks in Fig. \ref{figure3}(e) obeys the same criterion as in Fig. \ref{figure4}.
For  $l=7$, $l=9$, and $l=11$ such intervals whose limiting values are given
in Tables \ref{table4} and \ref{table5} are $\alpha = 0.640-0.662$, $\alpha = 0.665-0.691$, and $\alpha = 0.672-0.699$, 
respectively.}
\label{figure6}
\end{center}
\end{figure}

In the MQIM-HO, the momentum dependent $\gamma = c,c',s$ exponents ${\tilde{\zeta}}_{\gamma} (k)$ in the 
expressions for the one-electron removal spectral function, Eq. (\ref{equ2}), also have the same form as for the 
MQIM-LO. However, some of the quantities in their following expressions have additional MQIM-HO terms,
\begin{eqnarray}
{\tilde{\zeta}}_c (k) & = & -{1\over 2} + \sum_{\iota=\pm1}\left({{\tilde{\xi}}_c\over 4} - {\tilde{\Phi}}_{c,c}(\iota 2k_F,q)\right)^2  
\hspace{0.20cm}{\rm where}
\nonumber \\
k & = & {\rm sgn}\{q\} k_F - q \in [-k_F+k_{Fc}^0,k_F-k_{Fc}^0] 
\nonumber \\
q & = & - {\rm sgn}\{k\} k_F - k \in [-2k_F+k_{Fc}^0,-k_F] \hspace{0.20cm}{\rm and}
\nonumber \\
& = & - {\rm sgn}\{k\} k_F - k \in [k_F,2k_F-k_{Fc}^0] 
\nonumber \\
{\tilde{\zeta}}_{c'} (k) & = & -{1\over 2} + \sum_{\iota=\pm1}\left({{\tilde{\xi}}_c\over 4} - {\tilde{\Phi}}_{c,c}(\iota 2k_F,q)\right)^2  
\hspace{0.20cm}{\rm where}
\nonumber \\
k & = & - {\rm sgn}\{q\} k_F - q \in [-3k_F+k_{Fc}^0,3k_F-k_{Fc}^0] 
\nonumber \\
q & = & {\rm sgn}\{k\} k_F - k \in [-2k_F+k_{Fc}^0,k_F] \hspace{0.20cm}{\rm and} 
\nonumber \\
& = & {\rm sgn}\{k\} k_F - k \in [-k_F,2k_F-k_{Fc}^0] 
\nonumber \\
{\tilde{\zeta}}_{s} (k) & = & -1 + \sum_{\iota=\pm1}\left(- {\iota\over 2{\tilde{\xi}}_c} - {\tilde{\Phi}}_{c,s}(\iota 2k_F,q')\right)^2  
\hspace{0.20cm}{\rm where}
\nonumber \\
k & = & - q' \in [-k_F+k_{Fs}^0,k_F-k_{Fs}^0] 
\nonumber \\
q' & = & -k \in  [-k_F+k_{Fs}^0,k_F-k_{Fs}^0] \, .
\label{equA3}
\end{eqnarray}
These exponents are plotted in Figs. \ref{figure4}(a) and \ref{figure4}(b) as a function of the momentum $k$ within the 
MQIM-HO for $u=0.30$, $n_e=0.176$, and even values $l=6$ and $l=12$, respectively. In this
appendix they are plotted in Figs. \ref{figure5} and \ref{figure6} for even values $l=8,10$ and
odd values $l=7,9,11$, respectively.

The phase shifts $2\pi{\tilde{\Phi}}_{c,s} (\pm 2k_F,q')$ and $2\pi{\tilde{\Phi}}_{c,c} (\pm 2k_F,q)$ that
in Eq. (\ref{equA3}) appear in units of $2\pi$ are defined in Eqs. (\ref{equ3}) and (\ref{equ14}), respectively.
The bare phase shifts $2\pi\Phi_{c,s} (\pm 2k_F,q')$ and $2\pi\Phi_{c,c} (\pm 2k_F,q)$  
in the latter equations are defined below. The MQIM-HO phase shift term $2\pi{\tilde{\Phi}}_{c,c}^{R_{\rm eff}} (k_r)$ 
in Eq. (\ref{equ14}) accounts for effects of finite-range interactions beyond the MQIM-LO through the
spectral function exponents ${\tilde{\zeta}}_c (k)$ and ${\tilde{\zeta}}_{c'} (k)$ in Eq. (\ref{equA3}).

The small momentum $k_{Fs}^0$ in Eq. (\ref{equA3}) such that $k_{Fs}^0/k_F\ll 1$ is in general smaller than 
the momentum $k_{Fs}^*$ considered in the discussions of Sec. \ref{BiARPES2}. It has the same role for the $s$ band
as $k_{Fc}^0$ for the $c$ band, concerning the crossover between the low-energy TLL and high-energy regimes.
As discussed below in Appendix \ref{RTLL}, for physical momenta $k$ associated with creation of the $c$ impurity 
at $q$ in the small $c$ band momentum in the absolute value interval $\vert q\vert \in [2k_F-k_{Fc}^0,2k_F]$ and 
creation of the $s$ impurity at $q'$ in the small $s$ band momentum absolute value interval 
$\vert q'\vert \in [k_F-k_{Fs}^0,k_F]$ corresponding to the low-energy TLL regime the expressions for the 
exponents, Eq. (\ref{equA3}), are not valid. 

The bare energy dispersions $\varepsilon_c (q)$ and $\varepsilon_s (q')$ in Eq. (\ref{equA2}) are defined as follows,
\begin{eqnarray}
\varepsilon_c (q) & = & {\bar{\varepsilon}_c} (k (q)) 
\hspace{0.20cm}{\rm and}\hspace{0.20cm} \varepsilon_{s} (q') = {\bar{\varepsilon}_{s}} (\Lambda (q')) 
\hspace{0.20cm}{\rm where}
\nonumber \\
{\bar{\varepsilon}_c} (k) & = & \int_Q^{k}dk^{\prime}\,2t\,\eta_c (k^{\prime}) 
\nonumber \\
{\bar{\varepsilon}_{s}} (\Lambda) & = & \int_{\infty}^{\Lambda}d\Lambda^{\prime}\,2t\,\eta_{s} (\Lambda^{\prime}) \, .
\label{equA4}
\end{eqnarray}
The distributions $2t\,\eta_c (k)$ and $2t\,\eta_{s} (\Lambda)$ appearing here are solutions of the coupled integral equations,
\begin{equation}
2t\,\eta_c (k) = 2t\sin k + \frac{\cos k}{\pi\,u} \int_{-\infty}^{\infty}d\Lambda\,{2t\,\eta_{s} (\Lambda)\over 1 +  \left({\sin k - \Lambda\over u}\right)^2} \, ,
\label{equA5}
\end{equation}
and
\begin{eqnarray}
2t\,\eta_{s} (\Lambda) & = & {1\over\pi\,u}\int_{-Q}^{Q}dk\,{2t\,\eta_c (k)\over 1 +  \left({\Lambda-\sin k\over u}\right)^2} 
\nonumber \\
& - & \frac{1}{2\pi\,u} \int_{-\infty}^{\infty}d\Lambda^{\prime}\,{2t\,\eta_{s} (\Lambda^{\prime})\over 1 +  \left({\Lambda -
\Lambda^{\prime}\over 2u}\right)^2} \, .
\label{equA6}
\end{eqnarray}

The rapidity distribution functions $k (q)$ and $\Lambda (q')$ for the $c$ and $s$ impurity occupancies 
$q \in [-2k_F,2k_F]$ and $q' \in [-k_F,k_F]$, respectively, in the arguments of the auxiliary dispersions
${\bar{\varepsilon}_c}$ and $ {\bar{\varepsilon}_{s}}$ in Eq. (\ref{equA4}) are defined in terms of their inverse 
functions $q = q (k)$ where $k \in [-Q,Q]$ and $q' = q' (\Lambda)$ where $\Lambda \in [-\infty,\infty]$,
respectively. The latter are defined by the equations,
\begin{eqnarray}
q (k) & = & k + \frac{1}{\pi} \int_{-\infty}^{\infty}d\Lambda\,2\pi\sigma (\Lambda)\, \arctan \left({\sin k -
\Lambda\over u}\right) 
\nonumber \\
& & {\rm for}\hspace{0.20cm} k \in [-Q,Q] 
\nonumber \\
q' (\Lambda) & = & {1\over\pi}\int_{-Q}^{Q}dk\,2\pi\rho (k)\, \arctan \left({\Lambda-\sin k\over u}\right) 
\nonumber \\
& - & \frac{1}{\pi} \int_{-\infty}^{\infty}d\Lambda^{\prime}\,2\pi\sigma (\Lambda^{\prime})\, \arctan \left({\Lambda -
\Lambda^{\prime}\over 2u}\right) \nonumber \\
& & {\rm for}\hspace{0.20cm} \in [-\infty,\infty] \, .
\label{equA7}
\end{eqnarray}
The parameter $Q$ in Eqs. (\ref{equA4}), (\ref{equA6}), and (\ref{equA7}) is defined by the relations,
\begin{equation}
Q = k (2k_F) \hspace{0.20cm}{\rm and}\hspace{0.20cm} q (Q) = 2k_F \, .
\label{equA8}
\end{equation}
Furthermore, the distributions $2\pi\rho (k)$ and $2\pi\sigma (\Lambda)$ in Eq. (\ref{equA7}) are the solutions of 
the coupled integral equations,
\begin{equation}
2\pi\rho (k) = 1 + \frac{\cos k}{\pi\,u} \int_{-\infty}^{\infty}d\Lambda\,{2\pi\sigma (\Lambda)\over 1 +  \left({\sin k - \Lambda\over u}\right)^2} \, ,
\label{equA9}
\end{equation}
and
\begin{eqnarray}
2\pi\sigma (\Lambda) & = & {1\over\pi\,u}\int_{-Q}^{Q}dk\,{2\pi\rho (k)\over 1 +  \left({\Lambda-\sin k\over u}\right) ^2} 
\nonumber \\
& - & \frac{1}{2\pi\,u} \int_{-\infty}^{\infty}d\Lambda^{\prime}\,{2\pi\sigma (\Lambda^{\prime})\over 1 +  \left({\Lambda -
\Lambda^{\prime}\over 2u}\right)^2} \, .
\label{equA10}
\end{eqnarray}

In the $u\rightarrow 0$ and $u\gg 1$ limits the solution and the use of Eqs. (\ref{equA4})-(\ref{equA10})
leads to the following analytical expressions for the dispersions $\varepsilon_c (q)$ and $\varepsilon_s (q')$,
\begin{eqnarray}
\varepsilon_c (q) & = & - 4t\left(\cos \left({q\over 2}\right) - \cos k_{F}\right)
\hspace{0.20cm}{\rm for}\hspace{0.20cm}q\in [-2k_{F},2k_F] 
\nonumber \\
\varepsilon_{s} (q') & = & -2t\left(\cos q' - \cos k_{F}\right)\hspace{0.20cm}
{\rm for}\hspace{0.20cm} q' \in [-k_{F},k_{F}] 
\nonumber
\end{eqnarray}
and
\begin{eqnarray}
\varepsilon_c (q) & = & -2t(\cos q - \cos 2k_F)
\nonumber \\
& & - {2t\,n_e\ln 2\over u}(\sin^2 q - \sin^2 2k_F)
\nonumber \\
& & {\rm for}\hspace{0.20cm}q \in [-2k_F,2k_F]  
\nonumber \\
\varepsilon_{s} (q') & = & - {\pi n_e\,t\over 2u}\left(1 - {\sin 2\pi n_e\over 2\pi n_e}\right)
\cos\left({q\over n_e}\right) 
\nonumber \\
& & {\rm for}\hspace{0.20cm}q \in [-k_F,k_F]  \, ,
\nonumber
\end{eqnarray}
respectively. 

The bare phase shifts $2\pi\Phi_{c,s} (\pm 2k_F,q')$ and $2\pi\Phi_{c,c} (\pm 2k_F,q)$ in 
the expressions of the phase shifts $2\pi{\tilde{\Phi}}_{c,s} (\pm 2k_F,q')$ and $2\pi{\tilde{\Phi}}_{c,c} (\pm 2k_F,q)$ 
provided in Eqs. (\ref{equ3}) and (\ref{equ14}), respectively, are given by,
\begin{eqnarray}
2\pi\Phi_{c,c}\left(\pm 2k_F,q\right) & = & 2\pi\bar{\Phi }_{c,c} \left(\pm {\sin Q\over u},{\sin k (q)\over u}\right) 
\nonumber \\
2\pi\Phi_{c,s}\left(\pm 2k_F,q'\right) & = & 2\pi\bar{\Phi }_{c,s} \left(\pm{\sin Q\over u},{\Lambda (q')\over u}\right) 
\label{equA11}
\end{eqnarray}
where the quantities $2\pi\bar{\Phi }_{c,c}\left(\pm r_Q,r'\right)$ and $2\pi\bar{\Phi }_{c,s}\left(\pm r_Q,r'\right)$ where
$r_Q = {\sin Q\over u}$ are particular cases of the rapidity dependent auxiliary phase shifts $2\pi\bar{\Phi }_{c,c}\left(r,r'\right)$ 
and $2\pi\bar{\Phi }_{c,s}\left(r,r'\right)$. Those are the solution of the following integral equations,
\begin{eqnarray}
& & 2\pi{\bar{\Phi }}_{c,c}(r,r') = D_0 (r-r') 
\nonumber \\
& & + \int_{-{\sin Q\over u}}^{{\sin Q\over u}}dr''\,D (r-r'')\,2\pi{\bar{\Phi }}_{c,c}(r'',r') \, ,
\label{equA12}
\end{eqnarray}
and
\begin{eqnarray}
& & 2\pi{\bar{\Phi }}_{c,s}(r,r') = - \arctan\Bigl(\sinh\Bigl({\pi\over 2}(r-r')\Bigr)\Bigr) 
\nonumber \\
& & + \int_{-{\sin Q\over u}}^{{\sin Q\over u}}dr''\,D (r-r'')\,2\pi{\bar{\Phi}}_{c,s1}(r'',r') \, ,
\label{equA13}
\end{eqnarray}
respectively, where,
\begin{eqnarray}
D_0 (r) & = & - 2\int_{0}^{\infty} d\omega{\sin (\omega\,r)\over \omega
(1+e^{2\omega})} 
\nonumber \\
& = & {i\over 2\pi} \ln {\Gamma \Bigl({1\over 2}+i{r\over 4}\Bigr)\,\Gamma
\Bigl(1-i{r\over 4}\Bigr)\over \Gamma \Bigl({1\over 2}-i{r\over 4}\Bigr)\,\Gamma
\Bigl(1+i{r\over 4}\Bigr)} \, , 
\label{equA14}
\end{eqnarray}
\begin{eqnarray}
D (r) & = & - {1\over 2\pi}{d D_0 (r)\over dr} = {1\over\pi}\int_{0}^{\infty} d\omega{\cos (\omega\,r)\over 1+e^{2\omega}} 
\nonumber \\
& = & {i\over 2\pi} {d\over dr}\ln {\Gamma \Bigl({1\over 2}+i{r\over 4}\Bigr)\,\Gamma
\Bigl(1-i{r\over 4}\Bigr)\over \Gamma \Bigl({1\over 2}-i{r\over 4}\Bigr)\,\Gamma
\Bigl(1+i{r\over 4}\Bigr)} \, , 
\label{equA15}
\end{eqnarray}
and $\Gamma (z)$ is the usual $\Gamma$ function.

In the $u\rightarrow 0$ and $u\gg 1$ limits the solution and the use of Eqs. (\ref{equA11})-(\ref{equA15})
leads to the following analytical expressions for the bare phase shifts $2\pi\Phi_{c,c}(\pm 2k_F,q)$ and 
$2\pi\Phi_{c,s}(\pm 2k_F,q')$,
\begin{eqnarray}
2\pi\Phi_{c,c}(\pm 2k_F,q) & = & \mp {\pi\over\sqrt{2}}  
\hspace{0.20cm}{\rm for}\hspace{0.20cm}q \in ]-2k_F,2k_F[
\nonumber \\
2\pi\Phi_{c,s}(\iota 2k_F,q') & = & \mp {\pi\over\sqrt{2}}  
\hspace{0.20cm}{\rm for}\hspace{0.20cm}q' \in ]-k_F,k_F[ \, ,
\nonumber
\end{eqnarray}
and
\begin{eqnarray}
&& 2\pi\Phi_{c,c}(\pm 2k_F,q) = {\ln 2\over u}(\mp\sin 2k_F + \sin q) 
\nonumber \\
& & \hspace{3.25cm}{\rm for}\hspace{0.20cm}q \in ]-2k_F,2k_F[
\nonumber \\
&& 2\pi\Phi_{c,s}(\pm 2k_F,q') = {q'\over n_e} \mp {\pi\over 2u}\sin 2k_F\,\cos\left({\pi\,q'\over 2k_F}\right)
\nonumber \\
& & \hspace{1.50cm} +\,q'\,{\ln 2\over u}{\sin 2k_F\over 2k_F}\hspace{0.20cm}{\rm for}\hspace{0.20cm}q' \in ]-k_F,k_F[ \, ,
\nonumber
\end{eqnarray}
respectively

The dependence on the electronic density $n_e \in ]0,1[$ and interaction $u=U/4t$ of the bare charge 
parameter $\xi_c$ is defined by the following relation and equation,
\begin{eqnarray}
\xi_c & = & \xi_c \left({\sin Q\over u}\right)
\hspace{0.20cm}{\rm where}\hspace{0.20cm}\xi_c (r)
\hspace{0.20cm}{\rm is}\hspace{0.20cm}{\rm the}\hspace{0.20cm}{\rm solution}\hspace{0.20cm}{\rm of}
\nonumber \\
&& {\rm the}\hspace{0.20cm}{\rm integral}\hspace{0.20cm}{\rm equation,}
\nonumber \\
\xi_c (r) & = & 1 + \int_{-{\sin Q\over u}}^{\sin Q\over u} d r' D (r-r')\,\xi_c (r') \, ,
\label{equA16}
\end{eqnarray}
where $D (r)$ is given in Eq. (\ref{equA15}). Its limiting behaviors are,
\begin{eqnarray}
\xi_c & = & \sqrt{2}\left(1 - {u\over 2\pi\sin Q}\right) \hspace{0.20cm}{\rm for}\hspace{0.20cm} u\ll 1
\nonumber \\
& = & 1 + {\ln 2\over \pi\,u}\sin 2k_F \hspace{0.20cm}{\rm for}\hspace{0.20cm} u\gg 1 \, ,
\nonumber
\end{eqnarray}
where $\lim_{u\rightarrow 0}Q=k_F$ for $n_e \in ]0,1[$.

Finally, the extended domain of relative fluctuation $\Delta a/{\tilde{a}}$, Eq. (\ref{equ11}), is 
briefly discussed. For the physical processes of interest for the problem studied in this paper,
$\xi_c$ in Eq. (\ref{equA16}) varies in the domain $\xi_c \in ]1,\sqrt{2}[$.
The relative fluctuation $\Delta a/{\tilde{a}}$ in Eq. (\ref{equ11}) where $\Delta a = a - {\tilde{a}}$
applies though to all finite negative values $]-\infty,0[$ of $a/L$ and ${\tilde{a}}/L$. This refers to an
extended domain $\xi_c \in ]1,2[$. 

Its new subinterval $\xi_c \in [\sqrt{2},2[$ corresponds to electronic potentials of the same general form, 
$V_e (0) = U/2$ and $V_e (r) = U\,F_e (r)/r$ for $r>0$, as those in Eq. (\ref{equ1}) but for which $U\in ]-\infty,0[$. 
In the general case there are $\xi_c\rightarrow {\tilde{\xi}}_c$ transformations
within which the function $F_e (r)$ is smoothly turned on from $F_e (r)=0$ at ${\tilde{\xi}}_c =\xi_c$ to (i) positive 
$F_e (r)>0$ and (ii) negative $F_e (r)<0$ values, respectively. 
Considering all such processes and $U\in ]-\infty,\infty[$ values leads to charge
parameters that vary in the intervals $\xi_c \in ]1,2[$ and ${\tilde{\xi}}_c\in ]1/2,2[$, respectively, for which 
the relations given in Eqs. (\ref{equ12}) and (\ref{equ13}) remain valid. The scattering length symmetry relations, 
$a (\xi_c)=a (1/\xi_c)$ and $\tilde{a} (\tilde{\xi}_c) = \tilde{a} (1/\tilde{\xi}_c)$,
then confirm that both $a/L\in ]-\infty,0[$ and $\tilde{a}/L\in ]-\infty,0[$ in $\Delta a/{\tilde{a}}$.

Within the new subinterval ${\tilde{\xi}}_c \in [\sqrt{2},2[$ 
the SDS exponent $\alpha$ varies in the physically irrelevant range $\alpha \in [0,1/8]$.
$\alpha=1/8$ refers here to $\tilde{\xi}_c\rightarrow 2$.
Only $\xi_c\rightarrow {\tilde{\xi}}_c$ transformations for which both $\xi_c\in ]1,\sqrt{2}[$ and $F_e (r)>0$ 
refer to processes contributing to the physical problem studied in this paper. 

\section{Relation to the TLL regime and crossover to it}
\label{RTLL}

Both the MQIM-LO and the MQIM-HO also apply to the low-energy TLL regime whose spectral-function 
exponents near the $c,c',s$ branch lines are different from those given in Eq. (\ref{equA3}).
In the high energy regime whose spectral function expression, Eq. (\ref{equ2}), was
used in this paper to predict the $(k,\omega)$-plane location of the 
high-energy Bi/InSb(001) ARPES peaks, the velocity of the $c$ or $s$ impurity
is different from the velocity at the $c$ or $s$ band Fermi points, respectively.

In contrast, in the TLL regime the (i) $c$ or (ii) $s$ impurity 
is created in its band at a momentum in one of the intervals (i) $q\in [-2k_F,-2k_F+k_{Fc}^0]$ and
$q\in [2k_F-k_{Fc}^0,2k_F]$ or (ii) $q'\in [-k_F,-k_F+k_{Fs}^0]$ and
$q'\in [k_F-k_{Fs}^0,k_F]$. (Here both $k_{Fc}^0/2k_F\ll 1$ and 
$k_{Fs}^0/k_F\ll 1$.) The group velocity of that impurity thus becomes that of the low-energy 
particle-hole excitations near the corresponding Fermi point (i) $-2k_F$ and $2k_F$ or
(ii) $-k_F$ and $k_F$, respectively. Hence they loses their identity, as they cannot be
distinguished from the $c$ or $s$ holes (usual holons and spinons) in such excitations. 

The exponents in Eq. (\ref{equA3}) can be rewritten as,
${\tilde{\zeta}}_{\gamma} = -1 + \sum_{\iota =\pm 1} (2{\tilde{\Phi}}_c^{\iota} + 2{\tilde{\Delta}}_s^{\iota})$. 
Here $2{\tilde{\Phi}}_c^{\iota} = (-\iota/2{\tilde{\xi}}_c - {\tilde{\Phi}}_{c,s}(\iota 2k_F,q'))^2$ 
and $2{\tilde{\Delta}}_s^{\iota} = 0$ for the one-electron removal $s$ branch line and
$2{\tilde{\Phi}}_c^{\iota} = ({\tilde{\xi}}_c/4 - {\tilde{\Phi}}_{c,c}(\iota 2k_F,q))^2$ and
$2{\tilde{\Delta}}_s^{\iota} = {1\over 8}(1 + \iota)^2$ for both the one-electron removal $c$ and $c'$ 
branch lines, which correspond to different intervals of the $c$ band momentum $q$ in
${\tilde{\Phi}}_{c,c}(\iota 2k_F,q)$.

It turns out that in the TLL regime the expressions for the $\gamma =c,c',s$ exponents in the above equation lose 
one of the four $2{\tilde{\Delta}}_{\gamma}^{\iota}$s. It is the $2{\tilde{\Delta}}_{\gamma}^{\iota}$ 
whose sign of $\iota = \pm 1$ is that of the Fermi point whose velocity is
the same as the $\gamma =c$ or $\gamma =s$ impurity velocity.
The expressions of the exponents, Eq. (\ref{equA3}), in the high-energy 
spectral function expressions Eq. (\ref{equ2}) used in the theoretical study 
of the Bi/InSb(001) high-energy ARPES peaks are thus different from those of the TLL regime.

In the case of a large finite system, there is a cross-over regime between the   
high energy regime and the low-energy TLL regime within which the above quantity 
$2{\tilde{\Phi}}_c^{\iota}$ or $2{\tilde{\Delta}}_s^{\iota}$ is gradually removed
as the energy decreases. This cross-over regime refers to $(k,\omega)$-plane
regions whose momentum and energy widths are very small or vanish in the 
thermodynamic limit. It is an interesting theoretical 
problem, but the details of its physics have no impact on the specific problems discussed in this paper.

\section{Electron and $c$ particle representations}
\label{RECP}

In the bare limit, ${\tilde{\xi}}_c=\xi_c$, the $c$ particles are directly related to {\it rotated electrons} for which double
occupancy is a good quantum number for $u>0$. Their operators,
\begin{equation}
{\tilde{c}}_{j,\sigma}^{\dag} = {\hat{U}}^{\dag}\,c_{j,\sigma}^{\dag}\,{\hat{U}}\, ,\hspace{0.20cm}
{\tilde{c}}_{j,\sigma} =
{\hat{U}}^{\dag}\,c_{j,\sigma}\,{\hat{U}} 
\, ,\hspace{0.20cm}
{\tilde{n}}_{j,\sigma} = {\tilde{c}}_{j,\sigma}^{\dag}{\tilde{c}}_{j,\sigma} \, ,
\label{equC1}
\end{equation}
and ${\tilde{n}}_{j}=\sum_{\sigma}{\tilde{n}}_{j,\sigma}$ are generated from those of the electrons by the
unitary operator $\hat{U} = e^{\hat{S}}$. It is uniquely defined in Ref. \onlinecite{Carmelo_17}
in terms of the $4^L\times 4^L$ matrix elements between all the model energy eigenstates.

The present quantum problem is defined in a subspace without rotated-electron double occupancy.
Hence $\hat{U} = e^{\hat{S}}$ merely removes the corresponding electron double occupancy
from all sites around that of index $j$ at which ${\tilde{c}}_{j,\sigma}^{\dag}$ or ${\tilde{c}}_{j,\sigma}$ acts.
In that subspace, the electron double occupancy reads $D = N_e {n_e\over 4} f(n_e,u)$ where 
$f(n_e,u) = {\ln 2\over u^2}\left(1 - {\sin \left(2\pi n_e\right)\over 2\pi n_e}\right)$ for $u\gg 1$ and 
$\lim_{u\rightarrow 0}f(n_e,u) = 1$, ${\tilde{c}}_{j,\uparrow}^{\dag} =
({1\over 2} - {\tilde{S}}^{z}_{j,s})f_{j,c}^{\dag}$, and ${\tilde{c}}_{j,\downarrow}^{\dag} =
{\tilde{S}}^{+}_{j,s}\,f_{j,c}^{\dag}$. Here $f_{j,c}^{\dag}$ is the $c$ particle creation operator
and ${\tilde{S}}^{z}_{j,s}$ and ${\tilde{S}}^{+}_{j,s}$
are usual spin operators written in terms of rotated-electron operators. 

The same rotated-electron basis and corresponding $c$ particle
representation can be used for the ${\tilde{\xi}}_c<\xi_c$ renormalized Hamiltonian, Eq. (\ref{equ1}).
The difference relative to the bare limit is that the states generated from the fractionalized particles configurations
are not in general energy eigenstates but in the relevant subspaces for which the $c$ 
impurity has a large lifetime, they have overlap with single energy eigenstates.

The operator $\hat{U} = e^{\hat{S}}$ preserves the distance $r=j-j'$ between 
electrons, being the same for rotated electrons. The rotated-electron
anti-commutation relations that follow from unitarity imply that the $c$ particle operators 
obey a fermionic algebra, $\{f^{\dag}_{j,c}\, ,f_{j',c}\} = \delta_{j,j'}$ and
$\{f_{j,c}^{\dag}\, ,f_{j',c}^{\dag}\} =\{f_{j,c}\, ,f_{j',c}\} = 0$.
In the rotated-electron basis the Hamiltonian, Eq. (\ref{equ1}), has an infinite number of terms
given by the Baker-Campbell-Hausdorff formula,
\begin{equation}
{\hat{H}} = t\,\tilde{T} + {\tilde{V}} + [{\tilde{H}},{\tilde{S}}\,] + {1\over2}\,[[{\tilde{H}},{\tilde{S}}\,],{\tilde{S}}\,] + ... \, .
\label{equC2}
\end{equation}
Here ${\tilde{S}} = {\hat{U}}^{\dag}\,\hat{S}\,{\hat{U}}= \hat{S}$,
${\tilde{H}}=t\,\tilde{T} + {\tilde{V}}$ has the same expression in terms of rotated-electron operators
as ${\hat{H}}$ in terms of electron operators, and all higher terms have a kinetic nature.
Indeed, the expression of ${\tilde{S}}$ only involves the three kinetic operators 
$\tilde{T}_{0} = \sum_{j=1}^L \sum_{\iota =\pm1}(\tilde{T}_{0,j,\iota}
+ \tilde{T}_{0,j,\iota}^{\dag})$, $\tilde{T}_{+1} = \sum_{j=1}^L \sum_{\iota =\pm1}\tilde{T}_{+1,j,\iota}$, 
and $\tilde{T}_{-1} = \tilde{T}_{+1}^{\dag}$ in $\tilde{T} = \sum_d \tilde{T}_{d}$
where $d=0,\pm 1$ gives the change in the number of rotated-electron doubly occupied sites, and,
\begin{eqnarray}
\tilde{T}_{0,j,\iota} & = &  - \sum_{\sigma}\{\tilde{n}_{j,-\sigma}\,\tilde{c}_{j,\sigma}^{\dag}\,
\tilde{c}_{j+\iota,\sigma}\,\tilde{n}_{j+\iota,-\sigma} 
\nonumber \\
& + & (1-\tilde{n}_{j,-\sigma})\,\tilde{c}_{j,\sigma}^{\dag}\,\tilde{c}_{j+\iota,\sigma}\,(1-\tilde{n}_{j+\iota,-\sigma})\}
\nonumber \\
\hat{T}_{+1,j,\iota} & = & 
- \sum_{\sigma}\{\tilde{n}_{j,-\sigma}\,\tilde{c}_{j,\sigma}^{\dag}\,\tilde{c}_{j+\iota,\sigma}\,(1-\tilde{n}_{j+\iota,-\sigma}) 
\nonumber \\
& + & \tilde{n}_{j+\iota,-\sigma}\,\tilde{c}_{j+\iota,\sigma}^{\dag}\,\tilde{c}_{j,\sigma}\,(1-\tilde{n}_{j,-\sigma})\} \, .
\label{equC3}
\end{eqnarray}

Consistent with the finite-range electron interactions having their strongest effects in 
the charge-charge interaction channel, ${\tilde{V}}$ in Eq. (\ref{equC2}) can be expressed {\it solely} in terms of the 
charge $c$ particle operators as,
\begin{eqnarray}
\tilde{V} & = & \sum_{r=1}^{L/2-1}V_e (r)\sum_{j=1}^L\left(1 - f_{j,c}^{\dag}f_{j,c}\right)
\left(1- f_{j+r,c}^{\dag}f_{j+r,c}\right) 
\nonumber \\
& + & {U\over 2}\sum_{j=1}^L \left({1\over 2} - f_{j,c}^{\dag}f_{j,c}\right) \, .
\label{equC4}
\end{eqnarray}
Here $f_{j,c}^{\dag} = (f_{j,c})^{\dag} = {\tilde{c}}_{j,\uparrow}^{\dag}\,
(1-{\tilde{n}}_{j,\downarrow}) + (-1)^j\,{\tilde{c}}_{j,\uparrow}\,{\tilde{n}}_{j,\downarrow}$
for whole Hilbert space where the rotated-electron operators are related to those of the electrons in 
Eq. (\ref{equC1}). 

Importantly, despite the infinite number of terms in the Hamiltonian, Eq. (\ref{equ1}), when expressed 
in terms of the rotated-electron operators, Eq. (\ref{equC2}), its relevant term for our study is that 
in Eq. (\ref{equC4}) with $V_e (r)$ replaced by rotated-electron potential $V_{re} (r)$
renormalized by the higher kinetic terms in the expansion, Eq. (\ref{equC2}). The latter are in turn renormalized
by $V_e (r)$. $[{\tilde{H}},{\tilde{S}}\,]$ only involves the $d=0,\pm 1$ operators $\tilde{T}_{d}$
and the four operators $\tilde{J}_0^+ = [\tilde{V},\tilde{T}_0]$, 
$\tilde{J}_0^- = (\tilde{J}_0^+)^{\dag}$, and
$\tilde{J}_{\pm 1}  = [\tilde{V},\tilde{T}_{\pm 1}]$,
\begin{eqnarray}
\tilde{J}_0^+ & = & \sum_{r=1}^{L/2-1}V_e (r) \sum_{j=1}^L \sum_{\iota =\pm1}
(\tilde{T}_{0,j,\iota} - \tilde{T}_{0,j,\iota}^{\dag})
\nonumber \\
& \times & (\tilde{n}_{j+r} + \tilde{n}_{j-r} - \tilde{n}_{j+r+\iota} - \tilde{n}_{j-r+\iota})
\nonumber \\
\tilde{J}_{\pm 1} & = & \pm U\tilde{T}_{\pm 1} 
\pm \sum_{r=1}^{L/2-1}4V_e (r) \sum_{j=1}^L \sum_{\iota =\pm1}\tilde{T}_{\pm 1,j,\iota} 
\nonumber \\
& \times & (\tilde{n}_{j+r} + \tilde{n}_{j-r} + \tilde{n}_{j+r+\iota} + \tilde{n}_{j-r+\iota}) \, .
\label{equC5}
\end{eqnarray}
Higher kinetic terms also only involve the operators $\tilde{T}_{0,j,\iota}$ and $\tilde{T}_{\pm 1,j,\iota}$, Eq. (\ref{equC3}),
and $\tilde{n}_{j}$ at different relative sites.

The interaction between a $c$ particle at site $j$ and the $c$ impurity 
at site $j+r$ refers to a Hamiltoninan term of the form $-V_{re} (r)\,f_{j,c}^{\dag}f_{j,c}f_{j+r,c}f_{j+r,c}^{\dag}$. 
In terms of $V_c (x)\propto - V_{re} (r)$, it refers to suitably transformed operators ${\breve{f}}_{j,c}^{\dag}$ and corresponds
to $V_c (x)\,{\breve{f}}_{j,c}^{\dag}{\breve{f}}_{j,c}{\breve{f}}_{j+x,c}{\breve{f}}_{j+x,c}^{\dag}$ \cite{Carmelo_TT}. 
The part of the $V_{re} (r)$ renormalization by the infinite kinetic energy terms
beyond $t\,\tilde{T} + {\tilde{V}}$ in Eq. (\ref{equC2}) that contributes to the universal
properties is accounted for by the $\xi_c\rightarrow {\tilde{\xi}}_c$ transformation.
The non-universal part is within the non-universal inverse reduced mass $\mu^{-1}$ to 
which $V_c (x)$ is also proportional, $V_c (x)\propto {1\over 2\mu}$. 

The Fourier transform ${\cal{V}}_{re} (k)$ of $V_{re} (r)$ controls
the $c$ band energy dispersion ${\tilde{\varepsilon}}_c (q)$ and velocity 
${\tilde{v}}_c (q)$ renormalization. At $k=0$ it has the universal behavior \cite{Carmelo_TT},
\begin{eqnarray}
{\cal{V}}_{re} (0) & = & {\pi\over 2}\,\alpha_c\,v_c (2k_F) \hspace{0.20cm}{\rm where}
\nonumber \\
\alpha_c & = & \left({\xi_c^4 - ({\tilde{\xi}}_c^{\,\,\breve{}})^4\over ({\tilde{\xi}}_c^{\,\,\breve{}})^4}\right) 
= {4 - \xi_c^4\over \xi_c^4} 
\hspace{0.20cm}{\rm for}\hspace{0.20cm} {\tilde{\xi}}_c \leq {\tilde{\xi}}_c^{\,\,\breve{}}
\nonumber \\
& = & \left({\xi_c^4 - {\tilde{\xi}}_c^4\over {\tilde{\xi}}_c^4}\right) 
\hspace{0.20cm}{\rm for}\hspace{0.20cm} {\tilde{\xi}}_c \geq {\tilde{\xi}}_c^{\,\,\breve{}} \, .
\label{equC6}
\end{eqnarray}
Here $v_{c} (2k_F)$ is the bare $c$ band velocity $v_{c} (q)$ at $q=2k_F$ defined by Eq.  (\ref{equA2}) of 
Appendix \ref{APA} for $\beta = c$, ${\tilde{\xi}}_c^{\,\,\breve{}} = \xi_c^2/\sqrt{2} \in ]1/\sqrt{2},1[$, and 
$\alpha_c$ is related to the enhancement parameter $\beta_c$ in that equation as $\alpha_c = (1 + \beta_c)^2 -1$
and thus determines its value, $\beta_ c = \sqrt{1 + \alpha_c} -1$. The $\alpha_c$ expression in Eq. (\ref{equC6}) 
gives $\alpha_c = 0$ at $U=0$ and $\alpha_c \leq U/(2\,\pi\,t\sin k_F)$ and ${\cal{V}}_{re} (0) \leq U/2$ for $u\ll 1$, 
as required by the properties of the related potential $V_e (r) = UF_e (r)/r\propto U$ for $r>0$.

On the other hand, the Fourier transform ${\cal{V}}_c (k)$ of the potential $V_c (x)$ 
is found in Ref. \onlinecite{Carmelo_TT} to read ${\cal{V}}_c (0) = - C_{ce}\,{\cal{V}}_{re} (0)$ 
at $k=0$ where the coefficient is given by $C_{ce} = 2\,(\xi_c^4-{\tilde{\xi}}_c^4)/[{\tilde{\xi}}_c^2\,\xi_c^2 (4-\xi_c^4)]$
for ${\tilde{\xi}}_c \leq {\tilde{\xi}}_c^{\,\,\breve{}}$ and $C_{ce}=1$ for ${\tilde{\xi}}_c \geq {\tilde{\xi}}_c^{\,\,\breve{}}$.
The quantity ${\cal{V}}_c (0)$ is related to the phase-shift and matrix-element renormalization.


\end{document}